\documentstyle[12pt,equations]{article}
\input psfig

\def\MPL #1 #2 #3 {Mod.~Phys.~Lett.~{\bf#1},\  #2 (#3)}
\def\NPB #1 #2 #3 {Nucl.~Phys.~{\bf#1},\  #2 (#3)}
\def\PLB #1 #2 #3 {Phys.~Lett.~{\bf#1},\  #2 (#3)}
\def\PR #1 #2 #3 {Phys.~Rep.~{\bf#1},\ #2 (#3)}
\def\PRD #1 #2 #3 {Phys.~Rev.~{\bf#1},\  #2 (#3)}
\def\PRL #1 #2 #3 {Phys.~Rev.~Lett.~{\bf#1},\  #2 (#3)}
\def\RMP #1 #2 #3 {Rev.~Mod.~Phys.~{\bf#1},\  #2 (#3)}
\def\ZP #1 #2 #3 {Z.~Phys.~{\bf#1},\  #2 (#3)}
\def\IJMP #1 #2 #3 {Int.~J.~Mod.~Phys.~{\bf#1},\  #2 (#3)}
\def\taup{\tau^+}
\def\taum{\tau^-}
\def\lam{\lambda}
\def\br{BF}
\def\tauptaum{\tau^+\tau^-}
\def\mbb{m_{b\anti b}}
\def\sprime{{s^\prime}}
\def\rtsprime{\sqrt{\sprime}}
\def\shat{{\hat s}}
\def\rtshat{\sqrt{\shat}}
\def\gam{\gamma}
\def\sigrts{\sigma_{\tiny\rts}^{}}
\def\sigrtssq{\sigma_{\tiny\rts}^2}
\def\sigrtsprime{\sigma_{E}}
\def\nsigrts{n_{\sigrts}}
\def\betao{{\beta_0}}
\def\rhoo{{\rho_0}}
\def\etal{{\it et al.}}

\def\sighbar{\overline \sigma_{\h}}
\def\sighlbar{\overline \sigma_{\hl}}
\def\sighhbar{\overline \sigma_{\hh}}
\def\sighabar{\overline \sigma_{\ha}}
\def\anti{\overline}
\def\epem{e^+e^-}
\def\zstar{Z^\star}
\def\wstar{W^\star}
\def\zstarp{Z^{(\star)}}
\def\wstarp{W^{(\star)}}
\def\mupmum{\mu^+\mu^-}
\def\lplm{\ell^+\ell^-}
\def\brwweff{\br_{WW}^{\rm eff}}
\def\brzzeff{\br_{ZZ}^{\rm eff}}
\def\mstar{M^{\star}}
\def\mstarmin{M^{\star\,{\rm min}}}

\def\rts{\sqrt s}
\def\ie{{\it i.e.}}
\def\eg{{\it e.g.}}
\def\eps{\epsilon}
\def\anti{\overline}
\def\wp{W^+}
\def\wm{W^-}
\def\mw{m_W}
\def\mz{m_Z}
\def\h{h}
\def\mh{m_{\h}}
\def\gamh{\Gamma_{\h}^{\rm tot}}
\def\hsm{h_{SM}}
\def\mhsm{m_{\hsm}}
\def\gamhsm{\Gamma_{\hsm}^{\rm tot}}
\def\tanb{\tan\beta}
\def\hl{h^0}
\def\mhl{m_{\hl}}
\def\gamhl{\Gamma_{\hl}^{\rm tot}}
\def\ha{A^0}
\def\mha{m_{\ha}}
\def\gamha{\Gamma_{\ha}^{\rm tot}}
\def\hh{H^0}
\def\mhh{m_{\hh}}
\def\gamhh{\Gamma_{\hh}^{\rm tot}}
\def\fbi{~{\rm fb}^{-1}}

\def\mev{~{\rm MeV}}
\def\gev{~{\rm GeV}}
\def\tev{~{\rm TeV}}
\def\stop{\widetilde t}
\def\mstop{m_{\stop}}
\def\mt{m_t}
\def\mb{m_b}

\def\overlay#1#2{\ifmmode \setbox 0=\hbox {$#1$}\setbox 1=\hbox to\wd 0{\hss
$#2$\hss }\else \setbox 0=\hbox {#1}\setbox 1=\hbox to\wd 0{\hss #2\hss }\fi
#1\hskip -\wd 0\box 1}
\def\case#1/#2{{\textstyle{#1\over#2}}}
\def\9{\phantom 0}      %%% for lining up numbers in columns
\renewcommand\linebreak{\unskip\break} %% breaks line & still justifies
\newcommand{\alt}{\mathrel{\raisebox{-.6ex}{$\stackrel{\textstyle<}{\sim}$}}}
\newcommand{\agt}{\mathrel{\raisebox{-.6ex}{$\stackrel{\textstyle>}{\sim}$}}}
\def\lsim{\alt}
\def\gsim{\agt}
\catcode`@=11
\newcount\@tempcntc
\def\@citex[#1]#2{\if@filesw\immediate\write\@auxout{\string\citation{#2}}\fi
  \@tempcnta\z@\@tempcntb\m@ne\def\@citea{}\@cite{\@for\@citeb:=#2\do
    {\@ifundefined
       {b@\@citeb}{\@citeo\@tempcntb\m@ne\@citea\def\@citea{,}{\bf ?}\@warning
       {Citation `\@citeb' on page \thepage \space undefined}}%
    {\setbox\z@\hbox{\global\@tempcntc0\csname b@\@citeb\endcsname\relax}%
     \ifnum\@tempcntc=\z@ \@citeo\@tempcntb\m@ne
       \@citea\def\@citea{,}\hbox{\csname b@\@citeb\endcsname}%
     \else
      \advance\@tempcntb\@ne
      \ifnum\@tempcntb=\@tempcntc
      \else\advance\@tempcntb\m@ne\@citeo
      \@tempcnta\@tempcntc\@tempcntb\@tempcntc\fi\fi}}\@citeo}{#1}}
\def\@citeo{\ifnum\@tempcnta>\@tempcntb\else\@citea\def\@citea{,}%
  \ifnum\@tempcnta=\@tempcntb\the\@tempcnta\else
   {\advance\@tempcnta\@ne\ifnum\@tempcnta=\@tempcntb \else \def\@citea{--}\fi
    \advance\@tempcnta\m@ne\the\@tempcnta\@citea\the\@tempcntb}\fi\fi}
\catcode`@=12

\renewenvironment{thebibliography}[1]
 {\begin{list}{\arabic{enumi}.}
    {\usecounter{enumi} \setlength{\parsep}{0pt}
     \setlength{\itemsep}{3pt} \settowidth{\labelwidth}{#1.}
     \sloppy
    }}{\end{list}}

\def\mm{\mu^+\mu^-}
\def\ee{e^+e^-}
\def\rta{\rightarrow}
\def\tanb{\tan\beta}

\begin{document}
\thispagestyle{empty}

\newlength{\captsize} \let\captsize=\small % use \let\normalsize=\captsize
%\newlength{\captwidth}                     % just before \caption{  ...

%\preprint{
%
\font\fortssbx=cmssbx10 scaled \magstep2
\hbox to \hsize{
%
%\special{psfile=uwlogo.ps
% hscale=8000 vscale=8000
% hoffset=-12 voffset=-2}
%\hskip.5in \raise.1in
%
$\vcenter{
\hbox{\fortssbx University of California - Davis}
\hbox{\fortssbx University of Wisconsin - Madison}
}$
\hfill
$\vcenter{
\hbox{\bf UCD-96-6} 
\hbox{\bf MADPH-96-930} 
\hbox{\bf IUHET-328}
\hbox{January 1996}
}$
}
%}

\begin{center}
{\large\bf
Higgs Boson Physics in the {\protect\boldmath$s$}-channel
at {\protect\boldmath$\mu^+\mu^-$} Colliders
%\footnotemark
}\\[.1in]
\small
V.~Barger$^a$, M.S.~Berger$^b$, J.F.~Gunion$^c$, T.~Han$^c$,
\\[.1in]
\small\it
$^a$Physics Department, University of Wisconsin, Madison, WI 53706,
USA\\
$^b$Physics Department, Indiana University, Bloomington, IN 47405,
USA\\
$^c$Physics Department, University of California,  Davis, CA 95616,  
USA\\
\end{center}

%\footnotetext{
%Report of Physics Goals Working Group presented by V. Barger at the {\it 2nd
%Workshop on Physics Potential and Development of $\mu^+\mu^-$ Colliders},
%Sausalito, California, Nov.~1994. Working group leaders V.~Barger and
%J.F.~Gunion.}

\vspace{.5in}

\begin{abstract}
Techniques and strategies for discovering and measuring the 
properties of Higgs bosons via $s$-channel production at 
a $\mm$ collider, and the associated requirements for 
the machine and detector, are discussed in detail. The 
unique feature of $s$-channel production is that, 
with good energy resolution, the mass, total width and partial widths 
of a Higgs boson can be directly measured with remarkable 
accuracy in most cases. For the expected machine parameters and 
luminosity the Standard Model (SM) Higgs boson $\hsm$, with 
mass $\lsim 2\mw$, the light $\hl$ of the minimal supersymmetric Standard 
Model (MSSM), and the heavier MSSM Higgs bosons (the CP-odd 
$\ha$ and the CP-even $\hh$) can all be studied in the $s$-channel,
with the  heavier states accessible up to the maximal $\sqrt s$ 
over a large fraction of the MSSM parameter space. 
In addition, it may be possible to discover the $\ha$ and 
$\hh$ by running the collider at full energy and observing 
excess events in the bremsstrahlung tail at lower energy. 
The integrated luminosity, beam resolution and 
machine/detector features required to distinguish between 
the $\hsm$ and $\hl$ are delineated.

\end{abstract}

\newpage
\setcounter{page}{1}
\pagenumbering{roman}
\tableofcontents

\newpage
\setcounter{page}{1}
\pagenumbering{arabic}

%\begin{center}
%{\large\bf\uppercase{I. Introduction}}
%\end{center}
%\vspace{-.1in}

\section{Introduction}
\indent

Despite the extraordinary success of the Standard Model (SM) in
describing particle physics up to the highest energy available today,
the mechanism responsible for electroweak symmetry-breaking (EWSB)
has yet to be determined. In particular,
the Higgs bosons predicted in the minimal Standard Model
and the theoretically attractive 
Supersymmetric (SUSY) Grand Unified Theory (GUT) extensions
thereof have yet to be observed. If EWSB
does indeed derive from non-zero vacuum
expectation values for elementary scalar Higgs fields,
then one of the primary goals of
constructing future colliders must be to {\it completely}
delineate the associated Higgs boson sector.
In particular, it will be crucial to discover all of the physical
Higgs bosons and determine their masses, widths and couplings.

The remainder of the introduction is divided into two subsections.
In the first, we briefly review crucial properties of the Standard Model
and MSSM Higgs bosons. In the second, we outline basic features
and parameters of the proposed $\mm$ colliders, and give a first
description of how they relate to our ability to discover and
study the SM and MSSM Higgs bosons in $s$-channel $\mm$ collisions.

\subsection{Higgs bosons in the SM and the MSSM}
\indent

The EWSB mechanism in the Standard Model is
phenomenologically characterized
by a single Higgs boson $(\hsm)$ in the physical particle
spectrum. The mass of the $\hsm$ is undetermined by the
theory, but its couplings to fermions and vector bosons
are completely determined, being given
by $gm_f/(2\mw)$, $g\mw$ and $g\mz/\cos\theta_W$ for
a fermion $f$, the $W$ and the $Z$, respectively.
Although the SM Higgs sector is very simple, it leads to problems
associated with naturalness and mass hierarchies
which suggest that the SM is simply an effective low-energy theory.
Recent summaries of the phenomenology of
the SM Higgs sector can be found in Refs.~\cite{dpflighthiggs,hhg}.

The most attractive extensions of the SM that solve the naturalness
and hierarchy problems are those based on supersymmetry.
The Higgs sector of a supersymmetric model 
must contain at least two Higgs doublet fields in order
to give masses to both up and down quarks and to be free of anomalies.
If it contains two, and only two, Higgs doublet fields, then
the strong and electroweak coupling constants all unify reasonably well at a  GUT scale of order $10^{16}$~GeV. 
Thus, the minimal supersymmetric Standard Model, defined as having
exactly two Higgs doublets, is especially attractive.
The resulting spectrum of physical Higgs fields includes
three neutral Higgs bosons, the CP-even $\hl$ and $\hh$ and
the CP-odd $\ha$.
At tree-level the entire Higgs sector is completely determined
by choosing values for the parameters $\tanb=v_2/v_1$
(where $v_2$ and $v_1$ are the vacuum expectation values
of the neutral members of the Higgs doublets responsible for
up-type and down-type fermion masses, respectively) and
$\mha$ (the mass of the CP-odd $\ha$). For a summary,
see Refs.~\cite{dpflighthiggs,hhg}.

\begin{figure}[htbp]
\let\normalsize=\captsize   %%%% changes the font to "\small"
\begin{center}
\centerline{\psfig{file=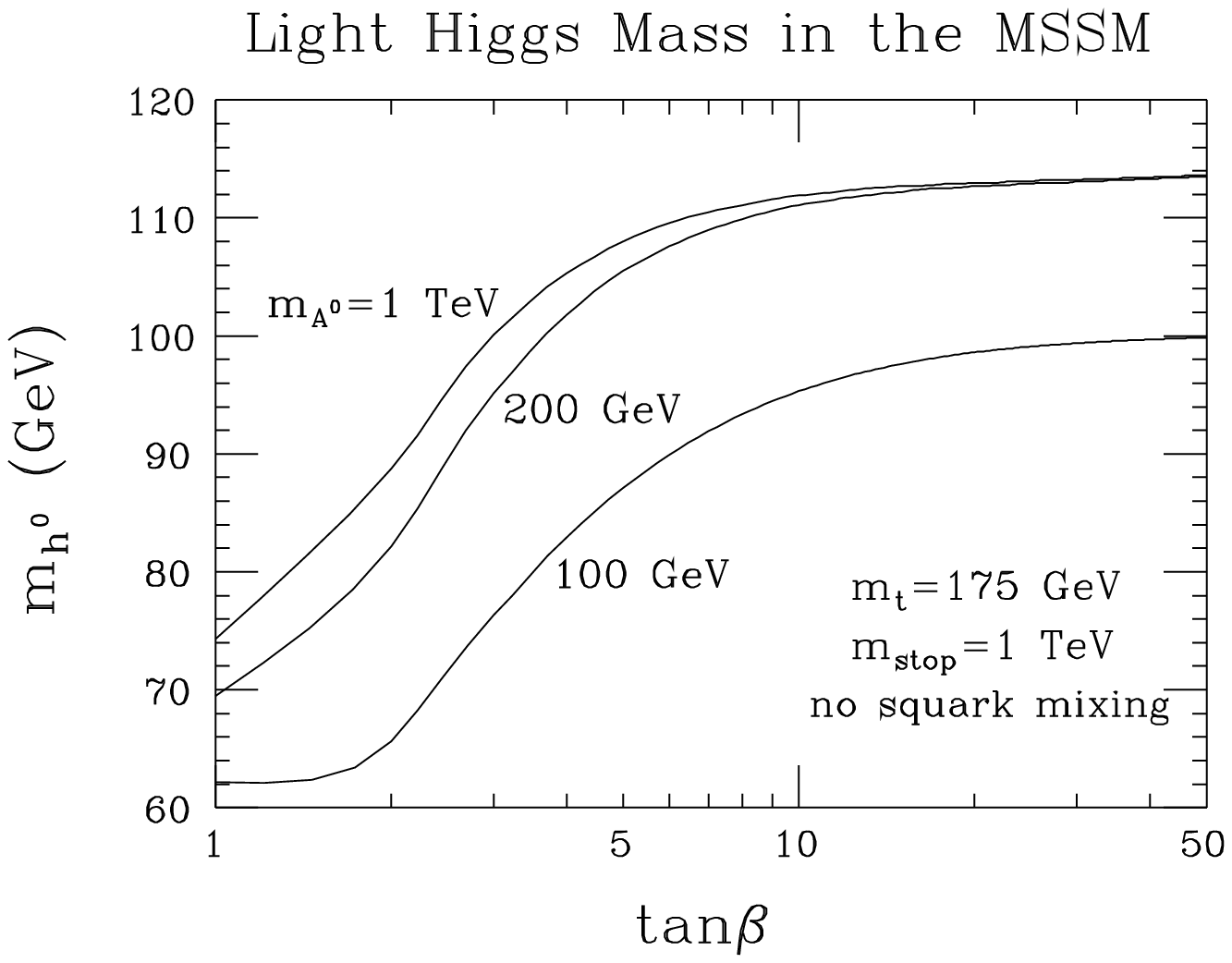,width=12.2cm}}
\begin{minipage}{12.5cm}       %%%% reduces width of caption to 12.5cm
\caption{$\mhl$ versus $\tanb$ for $\mha=100$, $200$ and 1000  
GeV. Two-loop/RGE-improved radiative corrections are included,
see Refs.~\protect\cite{habertwoloop,carenatwoloop},
taking $\mt=175\gev$, $\mstop=1\tev$ and neglecting squark mixing.}
\label{mhlvstanb}
\end{minipage}
\end{center}
\end{figure}

\begin{figure}[htbp]
\let\normalsize=\captsize   %%%% changes the font to "\small"
\begin{center}
\centerline{\psfig{file=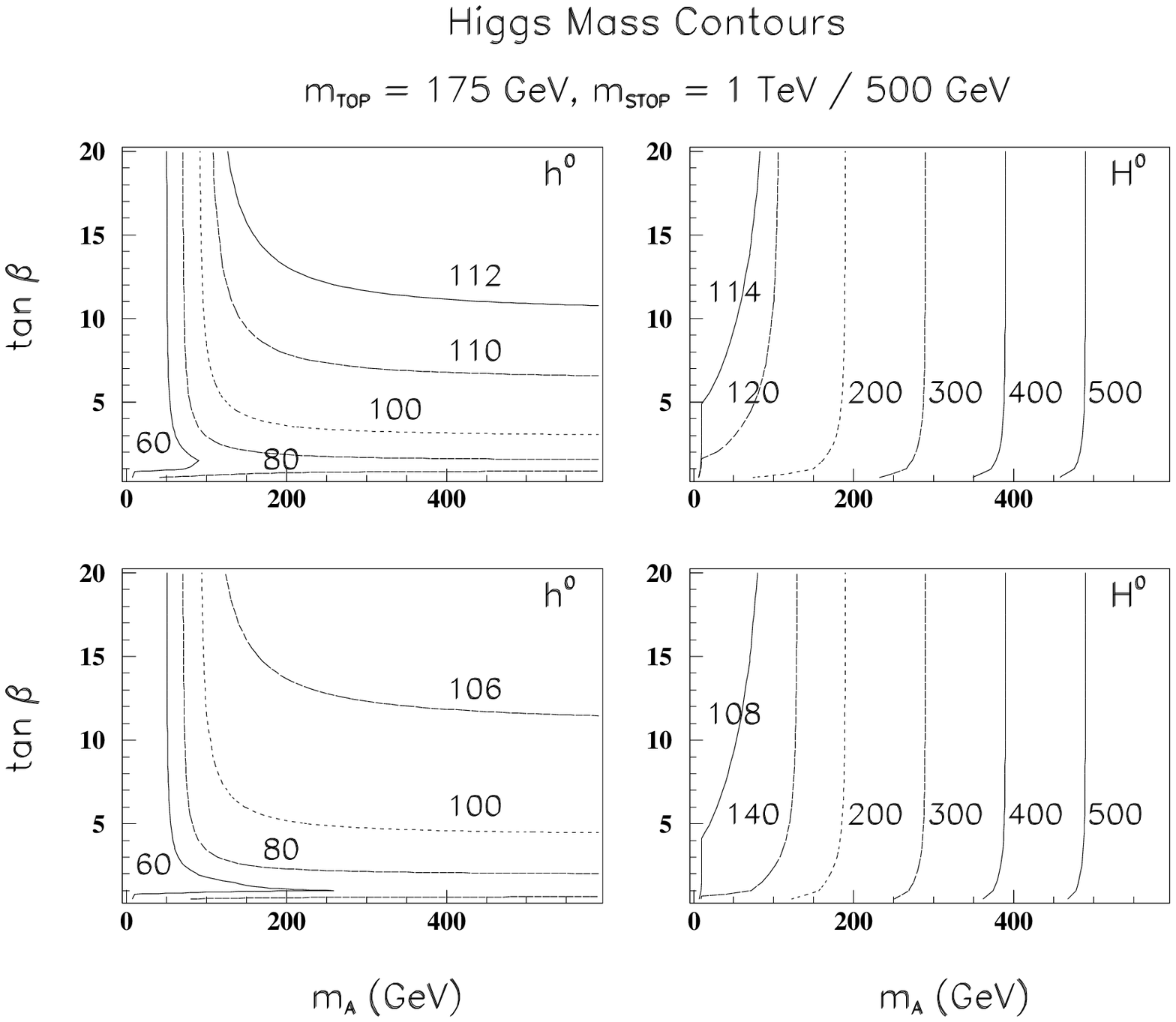,width=12.2cm}}
\begin{minipage}{12.5cm}       %%%% reduces width of caption to 12.5cm
\caption{Contours for the $\hl$ and $\hh$ masses
in $(\mha,\tanb)$ parameter space. Results include two-loop/RGE-improved
radiative corrections computed for $\mt=175\gev$,
with $\mstop=1\tev$ (upper plots) and $\mstop=500\gev$ (lower plots),
neglecting squark mixing.}
\label{rcmasses}
\end{minipage}
\end{center}
\end{figure}

In the MSSM there is a theoretical upper bound on the mass of the  
lightest state $\hl$ \cite{higgs1,higgs2} which is approached
at large $\mha$ and large $\tanb$.
After including two-loop/RGE-improved radiative corrections
\cite{habertwoloop,carenatwoloop} the bound
depends upon the top quark ($t$) and top squark ($\stop$) masses
and upon parameters associated with squark mixing. 
Assuming $\mt=175\gev$ and  $\mstop\lsim 1\tev$, the maximal mass is
\begin{equation}
\mhl^{\rm max} \sim 113\mbox{ to }130\gev\,,
\label{mhlbound}
\end{equation}
depending upon the amount of squark mixing. The 113~GeV
value is obtained in the absence of squark mixing.
Figure~\ref{mhlvstanb} illustrates the mass of the $\hl$ versus the  
parameter $\tanb$ for $\mha = 100$, $200$ and 1000 GeV.
Mass contours for the MSSM Higgs bosons are illustrated in
Fig.~\ref{rcmasses} in the conventional 
$\mha,\tanb$ parameter plane.
Both these figures include two-loop/RGE-improved radiative
corrections to the Higgs masses computed for $\mt=175\gev$, $\mstop=1\tev$
and neglecting squark mixing.

The Higgs sector of the MSSM can be extended to include extra singlet fields
without affecting any of its attractive features.  
A general supersymmetric model bound of
\begin{equation}
\mhl \alt 130  \sim 150 \; {\rm GeV}\,
\label{generalbound}
\end{equation}
applies for such non-minimal extensions of the MSSM, assuming
a perturbative renormalization group (RGE) evolved grand unified  
theory (GUT) framework.

The couplings of the MSSM Higgs bosons to fermions and vector bosons
are generally proportional to the couplings of the SM Higgs
boson, with the constant of proportionality being determined
by the angle $\beta$ (from $\tanb$) and the mixing angle
$\alpha$ between the neutral Higgs states
($\alpha$ is determined by $\mha$, $\tan\beta$, $\mt$, $\mstop$,
and the amount of stop mixing). Those couplings of 
interest in this report are \cite{gh}
\begin{equation}
\begin{array}{lcccc}
& \mm, b\anti b             & t\anti t             & ZZ,W^+W^-          
&  Z\ha
\\
\hl & -\sin\alpha/\cos\beta & \cos\alpha/\sin\beta  
&\sin(\beta-\alpha) &
\cos(\beta-\alpha)\\
\hh & \cos\alpha/\cos\beta  & \sin\alpha/\sin\beta &  
\cos(\beta-\alpha)&
-\sin(\beta-\alpha)\\
\ha & -i\gamma_5\tan\beta   & -i\gamma_5/\tan\beta & 0                  
& 0
\end{array}
\label{couplings}
\end{equation}
times the Standard-Model factor of $g m_f/(2m_W)$  in the case of  
fermions (where
$m_f$ is the relevant fermion mass), or $gm_W,gm_Z/\cos\theta_W$
in the case of the $W,Z$, and $g(p_A-p_h)^\mu/2\cos\theta_W$
in the case of $Z\ha$, where $p_A(p_h)$ is the outgoing momentum
of $\ha(\hl,\hh)$. 

An important illustrative limit is $\mha\agt 2\mz$,
since this is typical of SUSY GUT models \cite{gut}.  In this limit,
$\alpha\approx\beta-\pi/2$, $\mha\sim \mhh$, $\mhl$ approaches its
upper limit for the given value of $\tanb$,  and
the coupling factors of the Higgs bosons are approximately
\begin{equation}
\begin{array}{lcccc}
& \mm, b\anti b& t\anti t& ZZ,W^+W^-& Z\ha \\
\hl & 1 & 1 & 1 & 0\\
\hh & \tan\beta & -1/\tan\beta & 0 & -1\\
\ha & -i\gamma_5\tan\beta & -i\gamma_5/\tan\beta & 0 &  0
\end{array}
\label{couplingseq}
\end{equation}
%
%times the Standard-Model factor of $g m/(2m_W)$  in the case of fermions
%%(where
%$m$ is the relevant fermion mass), or $gm_W,gm_Z/\cos\theta_W$
%in the case of the $W,Z$.
times the Standard-Model factors as given below  
Eq.~(\ref{couplings}).
Thus at large $\mha$ it is the $\hl$ which is
SM-like, while the $\hh$, $\ha$ have similar fermion couplings
and small, zero (respectively) tree-level $WW,ZZ$ couplings.
Note that the $\hh$ and $\ha$ couplings to $\mm$
and $b\anti b$ are enhanced in the (preferred) $\tanb>1$
portion of parameter space.

For $\mha\alt\mz$, the roles of the $\hl$ and $\hh$ are reversed:
in this mass range the $\hh$ becomes roughly SM-like,
while the $\hl$ has couplings
(up to a possible overall sign) roughly
like those given for $\hh$ in Eq.~(\ref{couplingseq}).
(See Refs.~\cite{hhg,vb-susyhiggs,dpflighthiggs} for details;
Ref.~\cite{dpflighthiggs} gives the corrections
that imply that the simple rules are only roughly correct
after including radiative corrections.)
It is also useful to recall~\cite{gh,vb-susyhiggs} that the $Z\ha\hh$ ($Z\ha\hl$)  
coupling is maximal ($\sim 0$) at large $\mha$, while at small $\mha$ the  
reverse is true. The following discussions
emphasize the case of large $\mha$.

\begin{figure}[htbp]
\let\normalsize=\captsize   %%%% changes the font to "\small"
\begin{center}
\centerline{\psfig{file=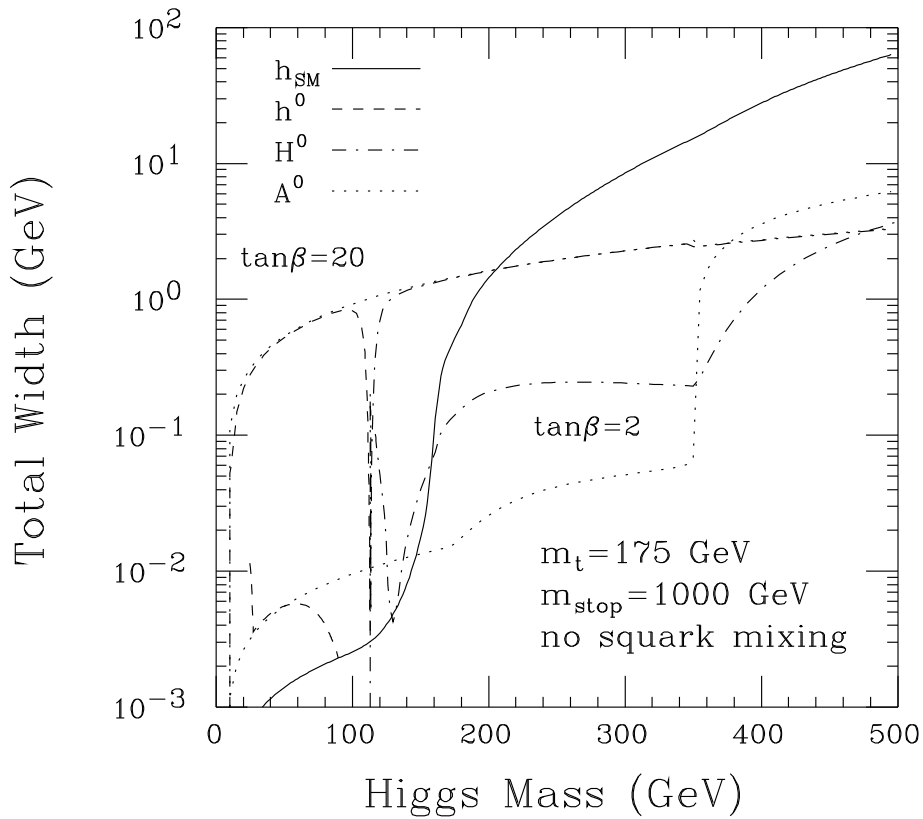,width=12.2cm}}
\begin{minipage}{12.5cm}       %%%% reduces width of caption to 12.5cm
\caption{Total width versus mass of the SM and MSSM Higgs bosons
for $\mt=175\gev$.
In the case of the MSSM, we have plotted results for
$\tan\beta =2$ and 20, taking $\mstop=1\tev$ and
including two-loop radiative corrections following
Refs.~\protect\cite{habertwoloop,carenatwoloop}
neglecting squark mixing; SUSY decay channels are assumed to be  
absent.}
\label{hwidths}
\end{minipage}
\end{center}
\end{figure}

\begin{figure}[htbp]
\let\normalsize=\captsize   %%%% changes the font to "\small"
\begin{center}
\centerline{\psfig{file=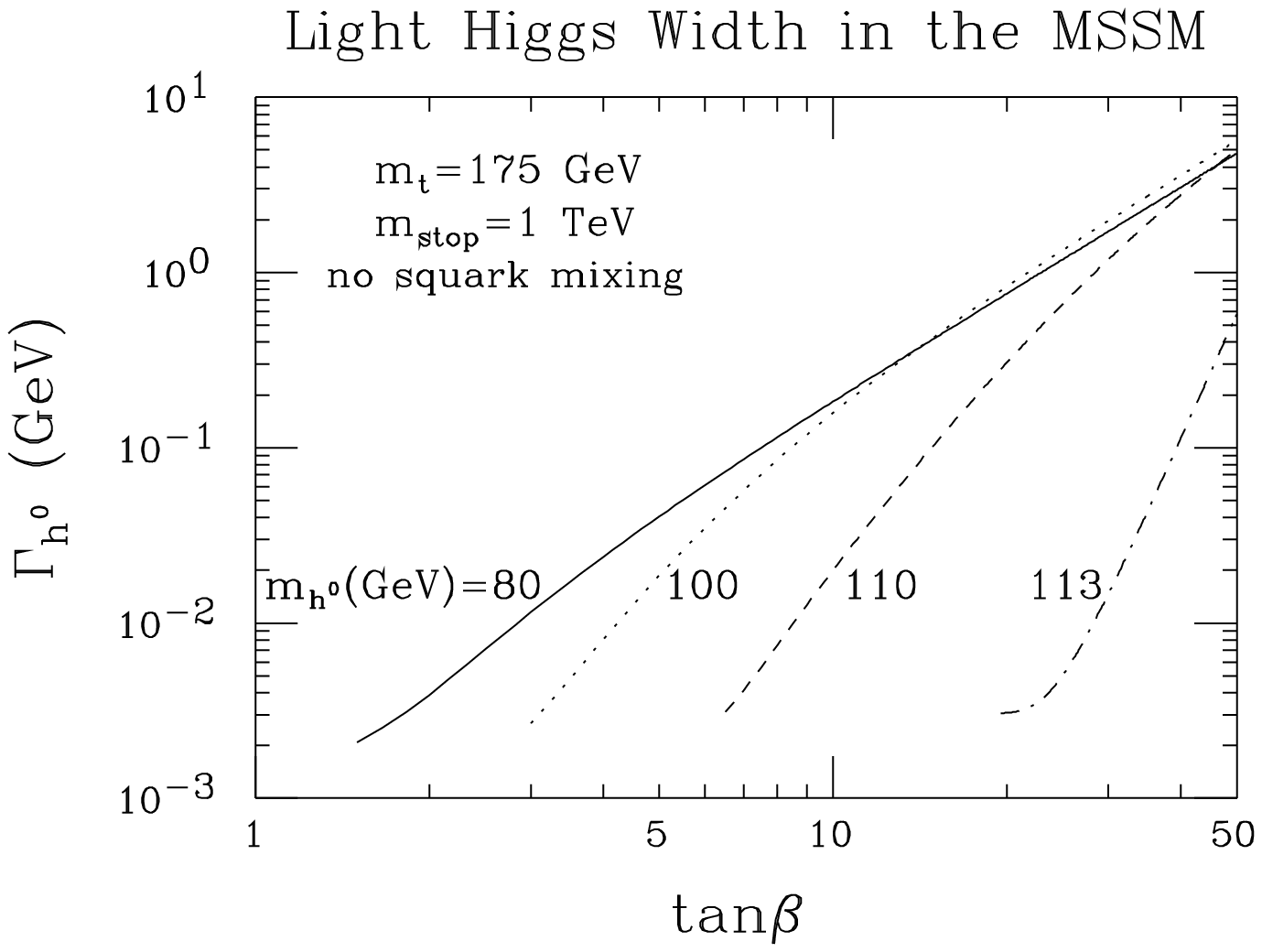,width=12.2cm}}
\begin{minipage}{12.5cm}       %%%% reduces width of caption to 12.5cm
\caption{$\gamhl$ versus $\tanb$ for $\mhl=80$, 100, 110 and 113  
GeV, assuming $\mt=175\gev$. Two-loop/RGE-improved radiative corrections 
to Higgs masses, mixing angles 
and self-couplings have been included, taking $\mstop=1\tev$
and neglecting squark mixing. SUSY decay channels are assumed to be absent.}
\label{gamhlvstanb}
\end{minipage}
\end{center}
\end{figure}

The Higgs boson widths are crucial
parameters for the searches and studies.
In particular, we shall see that the width compared
to the resolution in $\sqrt s$ of the machine is a crucial issue.
Widths for the Standard Model Higgs $\hsm$
and the three neutral Higgs bosons $\hl$, $\hh$, $\ha$ of the MSSM
are illustrated in Fig.~\ref{hwidths}; for the MSSM Higgs bosons,
results at $\tanb = 2$ and 20 are shown.
As a function of $\tanb$, the total width of $\hl$ is plotted
in Fig.~\ref{gamhlvstanb} for $\mhl = 100$, $110$ and 120 GeV.
We note that for
masses below $\sim 130\gev$, both the $\hsm$ and a SM-like $\hl$
have very small widths (in the few MeV range); we will discover that 
these widths are often smaller than the expected resolution in $\sqrt s$.
At high $\tanb$ and large $\mha\sim\mhh$,
the $\mm$, $\tauptaum$ and $b\anti b$ couplings
of the $\hh$ and $\ha$ are greatly enhanced (being proportional to  
$\tanb$). Consequently,  $\gamhh$ and $\gamha$ are
generally large compared to the expected $\sqrt s$ resolution.

\begin{figure}[htbp]
\let\normalsize=\captsize   %%%% changes the font to "\small"
\begin{center}
\centerline{\psfig{file=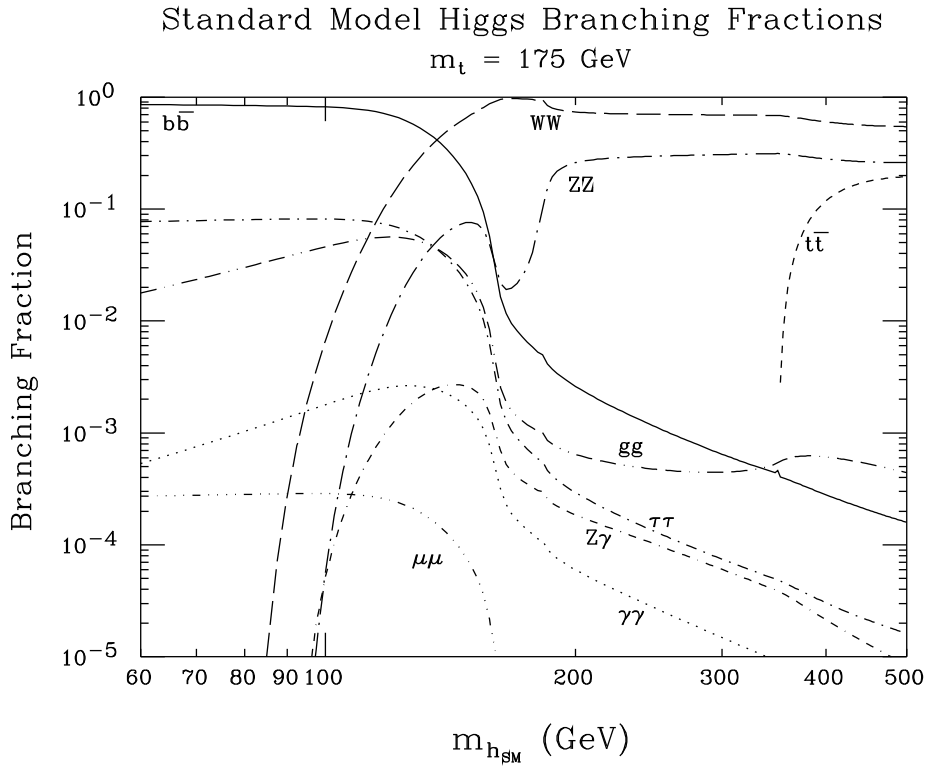,width=12.2cm}}
\begin{minipage}{12.5cm}       %%%% reduces width of caption to 12.5cm
\caption{Branching fractions for the Standard Model $\hsm$.}
\label{smbr}
\end{minipage}
\end{center}
\end{figure}

Figure~\ref{smbr} illustrates the $\hsm$
branching fractions  for the $\mm$, $b\anti b$, $W\wstarp$ and  
$Z\zstarp$ decay modes. For an $\hsm$ with $\mhsm\lsim 130\gev$, 
the $b\anti b$ branching fraction is of order 0.8--0.9,  
implying that this will be the most useful discovery channel. Once
the $W\wstarp$ and $Z\zstarp$ modes turn on ($\mhsm\gsim 2\mw$), the
$\hsm$ becomes broad and the branching fraction
$\br(\hsm\to \mm)$, which governs $s$-channel production, declines
precipitously.  Branching fractions
for the $\hl$ of the MSSM are similar to those of $\hsm$
for $\mhsm=\mhl$ when $\mha$ is large. 
At high $\tanb$ and large $\mha\sim\mhh$,
the enhancement of the $\mm$, $\tauptaum$ and $b\anti b$ couplings
implies that the $b\anti b$, $\tauptaum$ and $\mm$ branching fractions of 
the $\hh$ and $\ha$ are the only important ones, and are not unlike
those of a light $\hsm$, with relative magnitudes
determined by $\mb^2:m_{\tau}^2:m_{\mu}^2$.

Finally, it is relevant to note that
in non-minimal extensions of the MSSM,  parameter choices are  
possible such that the lightest Higgs boson 
to which the bound of Eq.~(\ref{generalbound}) applies has very weak coupling
to $ZZ$. This has been demonstrated \cite{eghrz} in
the case of the minimal non-minimal supersymmetric model (MNMSSM),  which
contains one extra singlet Higgs representation, yielding
three neutral Higgs bosons in all. However, 
for parameter choices such that the lightest Higgs decouples
from $ZZ$, there is a strong upper bound
on the mass of the least massive Higgs boson
{\it with significant $ZZ$ coupling}. 
The proof of this fact in the MNMSSM case relies on the observation
that as the lighter Higgs bosons decouple from $ZZ$,   
the upper bound on the next heaviest Higgs boson moves down. 
This result may generalize to the case of more singlets.

\subsection[$s$-channel Higgs boson physics at
$\mu^+\mu^-$ colliders]{{\protect\boldmath$s$}-channel Higgs boson physics at
{\protect\boldmath$\mu^+\mu^-$} colliders}
\indent

The ability of a new accelerator
to fully explore EWSB physics weighs heavily in its justification.
Recently, there has been much interest in the possibility of constructing
a $\mm$ collider \cite{mupmumi,saus,montauk,sanfran}, and a survey
of the physics opportunities at such a collider has been made \cite{workgr}.
It is currently anticipated that a $\mm$ collider can, at a minimum,
achieve the same integrated luminosities and energies as an $\ee$  
collider \cite{palmer,neuffersaus,npsaus}.
Further, with adequate detector segmentation the extra backgrounds resulting
from muon decays can be tamed \cite{millersaus}.  It then follows
that a $\mm$ collider can essentially
explore all the same physics that is accessible
at an $\ee$ collider of the same energy.  In particular, all
the established techniques for probing EWSB at $\ee$ colliders
are applicable at a $\mm$ collider.
In addition, should one or more Higgs boson(s) (generically denoted by $\h$)
with substantial $\mm$ coupling(s) exist,
a $\mm$ collider opens up the particularly
interesting possibility of direct $s$-channel $\mm\to\h$ production.
The SM Higgs boson, $\hsm$, is a prototypic example.
Direct $s$-channel $\hsm$ production is greatly enhanced at a $\mm$
collider compared to an $\ee$ collider because its coupling
to the incoming $\mm$ is proportional to the lepton mass.
Quantitative studies of $s$-channel Higgs production have been presented in
Refs.~\cite{workgr,mupmumprl}. With the machine energy set to the Higgs
mass ($\rts=\mh$) the $\mm\to \hsm$ rate is sufficiently large
to allow detection of the $\hsm$, provided that $\mhsm\lsim 2\mw$ (the
so-called intermediate Higgs mass region). In addition, {\it all} the Higgs
bosons of the minimal supersymmetric
model (MSSM) are produced in sufficient abundance in $s$-channel
$\mm$ collisions to allow their detection for most
of the model parameter space.

In the present report, we expand on these results and
provide the documentation underlying the discussion of  
Ref.~\cite{mupmumprl} on
precision studies of both the SM $\hsm$ and the MSSM
Higgs bosons. We find that the basic properties
of the $\hsm$ can be determined with remarkable accuracy in $\mm$  
$s$-channel
production, and that the properties of MSSM Higgs bosons can be  
detailed
over a larger fraction of model parameter space
than at any other proposed accelerator.  One particularly
important conclusion is
that $s$-channel Higgs production at a $\mm$ collider of appropriate  
design has greater potential for distinguishing between
a light SM $\hsm$ and the SM-like $\hl$ of the MSSM than other
processes/machines.
The techniques and strategies for attaining the above results,
and the associated requirements for the machine and detector,
are discussed at length.

Two possible $\mm$ machines are being actively studied  
\cite{saus,montauk,sanfran}:

\begin{itemize}

\item A first muon collider (FMC, for short) with
low c.~m.\ energy ($\rts$) between $100$ and $500\gev$ and
${\cal L}\sim2\times10^{33}\rm\,cm^{-2}\,s^{-1}$ delivering an annual
integrated integrated luminosity $L\sim 20\fbi$.

\item A next muon collider (NMC) with high $\rts \agt 4$ TeV and  
${\cal L} \sim 10^{35}\rm\,cm^{-2}\,s^{-1}$ giving
$L\sim 1000\fbi$ yearly; the extent to which
such a machine could be run at high luminosity for $\rts$ values
starting at $500\gev$ remains to be determined.

\end{itemize}

\noindent
One of our goals will be to quantify the amount of integrated luminosity
that is required to detect and study the various Higgs bosons
via $s$-channel production as the Higgs mass is varied.
For $s$-channel study of a SM-like Higgs boson,
only the lower energy machine is relevant because a SM-like Higgs can only
be detected in $s$-channel collisions if it has mass $\lsim 2\mw$,  
given the anticipated luminosity. However, higher $\rts$ will be
important if the MSSM is the correct theory.
The expected luminosity will allow
detection and study of the heavier MSSM Higgs bosons (the CP-odd $\ha$ and
the CP-even $\hh$) via $s$-channel production at the FMC
for $\mha,\mhh$ up to the maximal $\rts$. 
If the NMC can be run with
high luminosity at $\rts$ values starting at the
maximal FMC energy ($\sim500\gev$) and above,
then the ability to discover the $\ha$ and $\hh$ via $s$-channel
production would extend to correspondingly higher masses.

For $s$-channel Higgs studies, it will be important to deliver the  
maximum possible luminosity
at c.m.\ energies where Higgs bosons are either expected or observed.
Fortunately, this should be possible for the proposed FMC designs
due to the fact that the final muon storage ring(s) would comprise a
modest fraction of the overall cost \cite{palmer2}.
(The most costly component of a muon collider is the muon source ---
decays of pions produced by proton collisions.)
It is thus envisioned that multiple storage rings could
eventually be tailor-made for c.m.\ energies spanning the desired  
range. This approach could presumably also be used to
allow the high energy NMC to run with high luminosity at $\rts$
values starting at $\sim 500\gev$, where the FMC leaves off.

A crucial machine parameter for $s$-channel studies of Higgs
bosons is the energy resolution of the colliding beams.
A Gaussian shape for the energy spectrum of each beam
is expected to be a good approximation, with an rms deviation, $R$,
most naturally in the
range \cite{jackson}
$$R = 0.04\%~\mbox{to}~0.08\%$$
which could be decreased to as low as
$$R = 0.01\% $$
via additional cooling.
Excellent energy resolution is mandatory to detect and study a Higgs  
boson with a very narrow width, which is the case for
the $\hsm$ with $\mhsm\lsim 2\mw$ and the lightest MSSM Higgs boson.
The large value of the muon mass compared to the electron mass makes possible
the required energy resolution in three ways:
\begin{itemize}
\item[i)]
it is possible (albeit, probably expensive) to achieve
$R=0.01\%$;
\item[ii)]
 bremsstrahlung smearing, while non-negligible, leaves a large  
portion of the narrow central Gaussian beam energy peak intact.
\item[iii)]
 designs with small beamstrahlung are naturally achieved;
\end{itemize}
Henceforth, we neglect beamstrahlung since quantitative 
calculations of this are unavailable.

The rms spread in $\rts$ (denoted by $\sigrts$) prior
to including bremsstrahlung is given by
\begin{equation}
\sigrts = R \rts/\sqrt 2 \;,
\label{resol}
\end{equation}
where $R$ is the resolution in the energy of each beam.
A convenient formula for $\sigrts$ is
\begin{equation}
\sigrts = (7~{\rm MeV})\left({R\over 0.01\%}\right)\left({\rts\over  
{\rm
100\ GeV}}\right) \ .
\label{resolution}
\end{equation}
The critical issue is how this resolution compares to the calculated  
total widths of Higgs bosons when $\rts=\mh$.
For $R\alt 0.01\%$, the energy resolution in Eq.~(\ref{resolution})
is smaller than the Higgs widths in Fig.~\ref{hwidths}
for all but a light SM-like Higgs.  We shall demonstrate
that the smallest
possible $R$ allows the best measurement of a narrow Higgs width,
and that
the total luminosity required for discovery  by energy scanning 
when $\gamh\lsim \sigrts$
is minimized by employing the smallest possible $R$.  For a Higgs
boson with width larger than $\sigrts$, results from a fine scan
with small $R$ can be combined without any increase in
the luminosity required for discovery and width measurement.

\begin{figure}[htbp]
\let\normalsize=\captsize   %%%% changes the font to "\small"
\begin{center}
\centerline{\psfig{file=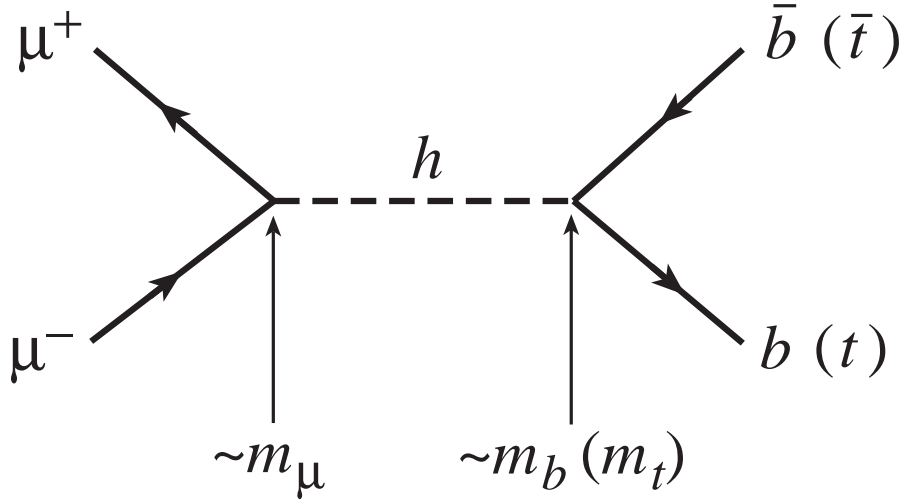,width=12.2cm}}
\begin{minipage}{12.5cm}       %%%% reduces width of caption to 12.5cm
\caption{$s$-channel diagram for production of a Higgs boson.}
\label{schannelhiggsdiagram}
\end{minipage}
\end{center}
\end{figure}

\begin{figure}[htbp]
\let\normalsize=\captsize   %%%% changes the font to "\small"
\begin{center}
\centerline{\psfig{file=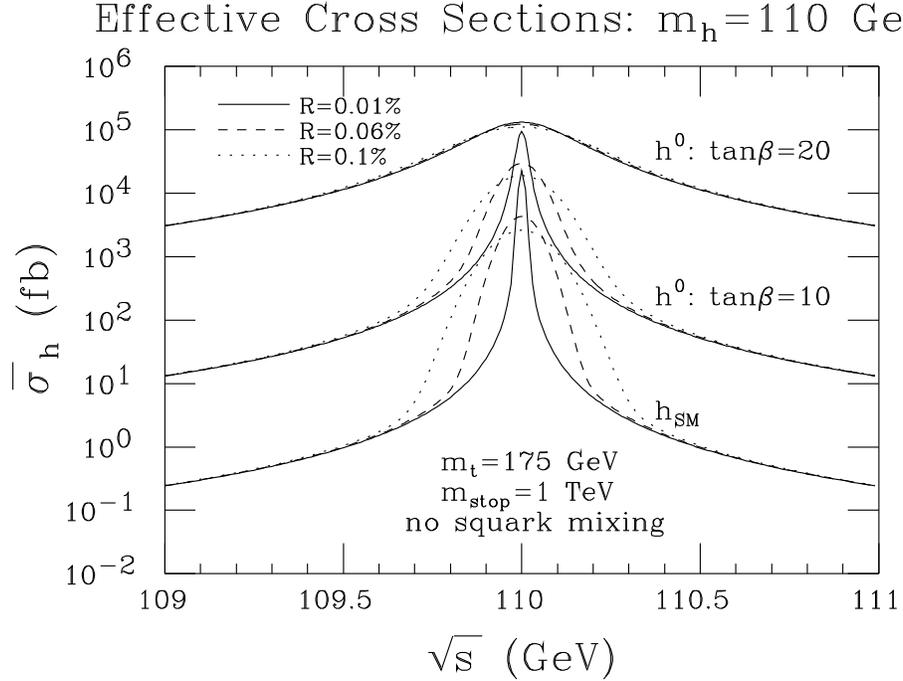,width=12.2cm}}
\begin{minipage}{12.5cm}       %%%% reduces width of caption to 12.5cm
\caption{The effective cross section, $\sighbar$, 
obtained after convoluting $\sigma_{\h}$ with 
the Gaussian distributions for $R=0.01\%$, $R=0.06\%$, and $R=0.1\%$, is
plotted as a function of $\protect\rts$ taking $\mh=110\gev$.
Results are displayed in the cases: $\hsm$,
$\hl$ with $\tanb=10$, and $\hl$ with $\tanb=20$.
In the MSSM $\hl$ cases, two-loop/RGE-improved radiative corrections
have been included for Higgs masses, mixing angles, and self-couplings
assuming $\mstop=1\tev$ and neglecting squark mixing.
The effects of bremsstrahlung are not included in this figure.
}
\label{gausssigma}
\end{minipage}
\end{center}
\end{figure}

The Feynman diagram for $s$-channel Higgs production is
illustrated in Fig.~\ref{schannelhiggsdiagram}.
The $s$-channel Higgs resonance cross section is
\begin{equation}
\sigma_{\h}(\rtshat) = {4\pi \Gamma(\h\to\mu\mu) \, \Gamma(\h\to X)\over
\left(\shat -\mh^2\right)^2 + \mh^2 [\gamh]^2} \;,
\label{basicsigma}
\end{equation}
where $\shat =(p_{\mu^+}+p_{\mu^-})^2$ is the c.~m.\ energy
squared of a given $\mm$ annihilation,
$X$ denotes a final state and $\gamh$ is the total width.\footnote{Effects
arising from implementing an energy-dependent generalization of
the $\mh\gamh$ denominator component of this simple resonance form
are of negligible importance for our studies, especially for
a Higgs boson with $\gamh\ll \mh$.}                    
The sharpness of the resonance peak is determined by $\gamh$.
Neglecting bremsstrahlung for the moment,
the effective signal cross section is obtained by convoluting
$\sigma_{\h}(\shat)$ with the Gaussian distribution in $\rtshat$
centered at $\rtshat=\rts$:
\begin{equation}
\sighbar(\rts) = \int \sigma_{\h}(\rtshat)\ {\exp\left[-(\sqrt{\shat}-\sqrt s)^2 \big/
(2\sigrtssq)\right]\over \sqrt{2\pi} \sigrts}\ d\sqrt{\shat} \; .
\label{convolutionprl}
\end{equation}
Figure~\ref{gausssigma} illustrates the effective cross section,
$\sighbar(\rts)$, as a function of $\rts$
for $\mh=110\gev$ and beam energy resolutions
of $R=0.01\%$, $R=0.06\%$, and $R=0.1\%$.  Results are given
for the cases:
$\hsm$, $\hl$ with $\tanb=10$, and $\hl$ with $\tanb=20$.
All channels $X$ are summed over.

In the case where
the Higgs width is much smaller than the Gaussian width $\sigrts$,
the effective signal cross section result for $\rts=\mh$,
denoted by $\sighbar$, is
\begin{equation}
\sighbar
={2\pi^2 \Gamma(\h\to\mu\mu)\, \br(\h\to X) \over \mh^2}\times
{1\over \sigrts \sqrt{2\pi}}
\quad \quad (\gamh \ll \sigrts)\;.
\label{narrowwidthsigma}
\end{equation}
Henceforth, we adopt the shorthand notation
\begin{equation}
G(X) = \Gamma(H\to\mu\mu) \,\br(h\to X)
\end{equation}
for the numerator of Eq.~(\ref{narrowwidthsigma}).
The increase of $\sighbar(\rts=\mh)$ with decreasing $\sigrts$ when
$\gamh\ll\sigrts$ is apparent from the $\hsm$ curves of Fig.~\ref{gausssigma}.
In the other extreme where the Higgs width is much broader than
$\sigrts$\,, then at $\rts=\mh$ we obtain
\begin{equation}
\sighbar={4\pi \br(\h\to\mu\mu)\br(\h\to X)\over \mh^2}
\quad \quad (\gamh \gg \sigrts)\;.
\label{broadwidthsigma}
\end{equation}
Note that this equation implies that if there is a large contribution
to the Higgs width from some channel other than $\mu\mu$,
we will get a correspondingly smaller total event rate due to the  
small size of $\br(\h\to\mu\mu)$. That
$\sighbar(\rts=\mh)$ is independent of the value of $\sigrts$ 
when $\gamh\gg\sigrts$
is illustrated by the $\tanb=20$ curves for the $\hl$ in Fig.~\ref{gausssigma}.
Raw signal rates (\ie\ before applying cuts
and including other efficiency factors)
are computed by multiplying $\sighbar$
by the total integrated luminosity $L$.

The basic results of Eqs.~(\ref{narrowwidthsigma}) and  
(\ref{broadwidthsigma})
are modified by the effects of photon bremsstrahlung
from the colliding muon beams.
In the case of a narrow Higgs boson,
the primary modification for $\rts=\mh$ is due to the fact
that not all of the integrated luminosity remains in the central
Gaussian peak.
These modifications are discussed in Appendix~A; to a good  
approximation, the
resulting signal rate is obtained
by multiplying $\sighbar$ of Eq.~(\ref{narrowwidthsigma})
by the total luminosity $L$ times the fraction $f$
of the peak luminosity in the Gaussian after
including bremsstrahlung relative to that before (typically $f\approx0.6$).
For a broad Higgs resonance,
the lower energy tail in the luminosity distribution due to  
bremsstrahlung
makes some contribution as well.
In the results to follow, we
avoid any approximation and numerically convolute the full
effective luminosity distribution (including bremsstrahlung)
with the Higgs cross section of Eq.~(\ref{basicsigma}).
In performing this convolution, we require that the effective  
$\mupmum$
c.m.\ energy be within $10\gev$ of the nominal value. Such a  
requirement
can be implemented by reconstructing the mass of the final
state as seen in the detector; planned detectors
would have the necessary resolution to impose the above
fairly loose limit. This invariant mass selection is imposed in order
to reduce continuum (non-resonant) backgrounds that would
otherwise accumulate from the entire low-energy bremsstrahlung tail
of the luminosity distribution.

As is apparent from Fig.~\ref{gausssigma},
discovery and study of a Higgs boson with a very narrow width
at the $\mm$ collider will require
that the machine energy $\rts$ be within $\sigrts$ of $\mh$.
The amount of scanning required to find the correct $\rts$
depends upon $R$.
From Fig.~\ref{gausssigma} it is apparent that the larger $R$ is,
the less the accuracy with which the machine energy needs to be set 
at each scan point and
the fewer the number of scan points needed. But,
small $R$ results in much greater event rate for $\rts\simeq \mh$.
{\it If $\rts$
can be \underline{rapidly} changed with an accuracy that is a small fraction
of $R$}, then we shall find that smaller $R$ implies that less total
time (and, hence, luminosity) will be required for the scan.
Further, we find that $R\sim 0.01\%$ and the ability to
set $\rts$ with an accuracy of order 1 part in $10^6$
are both required if we are to be able to measure the Higgs width
with sufficient precision to distinguish between the SM $\hsm$ and 
the MSSM $\hl$ when the latter is SM-like. Thus, 
for a $\mm$ collider to reach its full potential, it should
be designed so that $R\sim 0.01\%$ and so that it is possible
to vary $\rts$ rapidly and with great precision.  These are not
insurmountable tasks \cite{palmer2}, but careful planning
is certainly required. For Higgs bosons with a large width, 
the design demands upon the $\mm$ collider are clearly less.

Due to the bremsstrahlung tail,
it is also possible to search for a Higgs boson by running the $\mm$
collider at an energy well above the mass of the Higgs boson itself.
In some collisions, one (or both) of the muons will have radiated enough
of its initial energy that the effective $\rtshat$ of the collision
is much lower than $\rts$.
In this circumstance, detection of the Higgs boson requires   
reconstruction with good resolution of
the effective $\rtshat$ of each collision from the final state  
momenta. For a final state mass bin
centered at $\rtshat=\mh$, if $d{\cal L}/d\rtshat$
is slowly varying in the vicinity
of $\rtshat=\mh$ over an interval several times
the Higgs total width $\gamh$, the effective cross section is
\begin{equation}
\sighbar
={2\pi^2 \Gamma(\h\to\mu\mu)\, \br(\h\to X) \over \mh^2}\times
\left.{d{\cal L}\over d\rtshat}\right|_{\rtshat=\mh}\;.
\label{bremtailsigma}
\end{equation}
In exploring the possible utility of this bremsstrahlung tail
for Higgs detection, we have performed our explicit calculations using
the spectrum obtained for $R=0.1\%$.
However, we note that the bremsstrahlung
tail well away from the central Gaussian peak
is essentially independent of the beam energy resolution $R$.
If a mass resolution in the final state of $\pm 5\gev$
is possible in the $b\anti b$ final state, then
even when running the FMC at full nominal energy of $\rts=500\gev$ we  
find that it will be possible to detect a Higgs boson with $\mh$
in a broad range below $\rts$ (but not near $\mz$) provided that the
$\h\to\mm$ coupling is significantly enhanced with respect to the SM
$\hsm\to\mm$ coupling. The total integrated luminosity
required for Higgs discovery using the bremsstrahlung tail will be  
compared to that needed for discovery by scanning using a large number of  
$\rts$ machine energy settings.

Highly polarized beams may be possible since the muons are
naturally polarized from $\pi^\pm$ ($K^\pm$) decays in the parent
rest-frame. However, the luminosity for polarized beams
may be significantly reduced
during the cooling and acceleration process.
If a degree of polarization $P$
is possible for {\it both} beams, then, relative to the unpolarized
case, the $s$-channel Higgs signal is enhanced by the factor
$(1+P^2)$ while the background is suppressed by $(1-P^2)$.
High polarization $P$ of both beams would be useful
if the luminosity reduction is less than a factor of
$\left(1+P^2\right)^2 / \left(1-P^2\right)$, \ie\ the factor
which would leave the significance of the signal unchanged.
For example, $P=0.84$ would compensate a factor of 10
reduction in luminosity~\cite{parsa}. We mainly present our
results without assuming high polarization beams, but we comment on
improvements with beam polarization.

With this introduction, we now proceed with a detailed
description of the capability of a $\mm$ collider 
to detect and study different types of Higgs bosons.
In the next section,
we begin with SM-like Higgs bosons.  The following section
explores the non-SM-like Higgs bosons of the MSSM.
The final section gives our conclusions.

\section{A SM-like Higgs boson}

\indent\indent
We first review the prospects for discovering and studying
a SM-like Higgs boson without $s$-channel production
at a $\mm$ collider. We then turn to the role of $s$-channel  
$\mm\to\h$
production, emphasizing the prospects for precision studies
of the Higgs mass and width.

\subsection[Discovery and study without $s$-channel
production]{Discovery and study without {\protect\boldmath$s$}-channel
production}

\indent\indent
Neutral Higgs bosons that are coupled to $ZZ$ with roughly
SM-like strength can be discovered
via $\zstar\rta Z \h$ production for $\mh\alt 0.7\rts$ at either an  
$\ee$ collider or a $\mm$ collider~\cite{epem}.
This discovery reach applies to both the $\hsm$ and to the $\hl$ of the
MSSM in the large-$\mha$ portion of parameter space where it is  
SM-like in its couplings.
The stringent upper bound on $\mhl$, Eq.~(\ref{mhlbound}),
in the MSSM implies that
even a $\rts=300\gev$ machine is guaranteed to find the $\hl$ if it exists.

As described in the Introduction,
we can also consider adding extra singlets to the MSSM two-doublet
Higgs sector. In the MNMSSM model, containing one singlet Higgs field,
we noted that even if the lightest Higgs boson has small $ZZ$ coupling, 
there is always a CP-even Higgs boson with substantial $ZZ$ coupling
and modest mass. Refs.~\cite{eghrz} demonstrate
that at least one of the CP-even Higgs bosons of the MNMSSM model
will be detected in the $Z\h$ mode at a machine with c.m.\ energy
$\rts=500\gev$. Since it appears that this result may
generalize to the case of more than one additional singlet,
we regard it as relatively certain that any supersymmetric theory
in the SUSY GUT context will contain at least one CP-even Higgs
boson that will be discovered in the $Z\h$ mode at a machine with  
$\rts=500$~GeV, and 
its mass will be in the intermediate mass range ($\lsim 2\mw$).

Assuming that a SM-like $\h$ is discovered in the $Z\h$ mode,
an important question for $s$-channel production and study
of the $\h$ in $\mm$ collisions is the accuracy with which its mass
can be measured \'a priori via $Z\h$ production. The better
this accuracy, the easier it will be to set $\rts$ of the $\mm$
collider to a value centered on $\mh$ within the rms spread  
$\sigrts$.
Another critical question bearing on the importance
of the $s$-channel $\mm\to\h$ production mode
is whether the $Z\h$ mode is useful for measurement of the $\h$ width.
We find that it is not.

\begin{figure}[htbp]
\let\normalsize=\captsize   %%%% changes the font to "\small"
\begin{center}
\centerline{\psfig{file=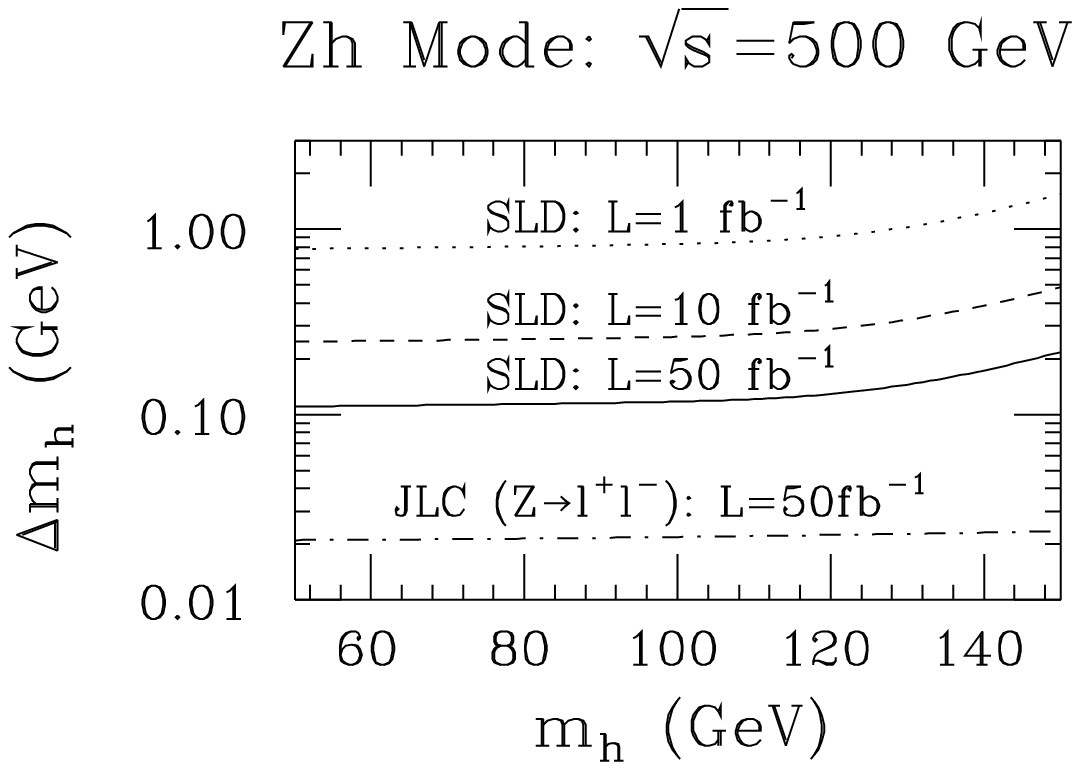,width=10.2cm}}
\begin{minipage}{12.5cm}       %%%% reduces width of caption to 12.5cm
\caption{The uncertainty $\Delta\mh$ in the
determination of $\mh$ for a SM-like Higgs boson using $Z\h$  
production and a $\pm 4\gev$ single event mass resolution,
as for an SLD-type detector, and for a $\pm 0.3\gev$ single event
mass resolution in the $Z(\to\lplm)\h(\to {\rm any})$ mode, as for
the super-LC detector.}
\label{massresolution}
\end{minipage}
\end{center}
\end{figure}

Generally speaking, the accuracy of the Higgs boson mass
measurements depends on the detector performance and
the signal statistics. As a general guide, we consider
two examples for the uncertainty on $\mh$ in the
mass range $\mh<2\mw$ (\ie\ below where $W$-pair decays become
important)
\begin{eqnarray}
\label{EQ:SLD}
\Delta m_h^{} &\simeq& 4.0\gev/\sqrt N \; \quad ({\rm SLD}),  \\
              &\simeq& 0.3\gev/\sqrt N \; \quad ({\rm  
super-LC}).
\label{EQ:JLC}
\end{eqnarray}
where our notation will always be that $\Delta X$ represents
the absolute magnitude of the $1\sigma$ error on the quantity
$X$; that is the $1\sigma$ limits on $X$ are $X\pm\Delta X$.
Equation~(\ref{EQ:SLD}) results for performance typified by the SLD  
detector
\cite{barklowpc}, where 4~GeV is the single event resolution
and $N$ is the number of events in the $Z(\to q\anti q) h(\to b\anti  b)$, 
$Z(\to q\anti q) h(\to\tau\anti\tau)$,
plus $Z(\to\ell^+\ell^-) h(\to {\rm any})$ modes.
For a SM-like Higgs, these modes have an effective final state
branching fraction that varies between about 70\% and 50\% as
$\mh$ varies from low masses up to 140 GeV.  We plot $\Delta \mh$ in
Fig.~\ref{massresolution}
according to Eqs.~(\ref{EQ:SLD}) and (\ref{EQ:JLC}),
with $N=\eps L\sigma(Z\h)\br(\rm effective)$,
assuming detection efficiencies of $\eps=0.9$ [$\eps=0.5$]
for the $Z(\to\ell^+\ell^-) h(\to {\rm any})$ 
[$Z(\to q\anti q) h(\to b\anti  b)$, $Z(\to q\anti q) h(\to\tau\anti\tau)$]
modes and assuming a fixed $\rts=500\gev$.
For SLD detector performance, results for luminosities
of $L=1$, $10$, and $50\fbi$ are shown;
with these integrated luminosities,
$\mh$ (for $\mh\lsim150$~GeV)
will be determined to an accuracy of at least 1.4, 0.5, 0.21 GeV 
(respectively).

Equation~(\ref{EQ:JLC}) is applicable for  a ``super'' performance
Linear Collider detector (hereafter referred to as
the super-LC detector) \cite{jlci,kawagoe},
the special features of which include excellent momentum resolutions
and high $b$-tagging efficiency.
%
%They assume that the machine will be run at the optimal
%$\sqrt s\sim \mz+\mhsm+ 20/30\gev$, and
%a that luminosity of $L=30\fbi$ is accumulated.
%Under these conditions, $\mhsm$ can be measured to $\sim 0.1\%$
%using direct $\mhsm$ reconstruction in
%the $Z\hsm\rta \nu\anti\nu jj,\lplm jj,jjjj$ channels. As above, this estimate
%is based on a roughly $4\gev$ uncertainty per event
%for reconstructing the mass from the detector information, divided
%by $\sqrt N$, where $N$ is the total number of accepted events.
%However, t
%
For this detector, the best determination of $\mhsm$ is obtained
by examining the recoil mass peak in $Z\hsm$ production. For
$Z\to \lplm$ events, the resolution for the recoil mass
is expected to be of order $0.3\gev$ per event. 
A measurement of $\mhsm$
to $\pm 0.3\gev/\sqrt N\sim \pm 20\mev$ would 
be possible for $\mhsm\lsim 140\gev$ and $L=50\fbi$,
as illustrated in Fig.~\ref{massresolution},
assuming detection efficiency of $\eps=0.9$ 
for the $Z(\to\ell^+\ell^-) h(\to {\rm any})$ mode.
The total width $\gamhsm$ could also be measured down to $\sim$200~MeV using 
the $Z\hsm$ recoil mass distribution.
However, this latter sensitivity is not likely to be useful since  
$\Gamma_{\hsm}\lsim 10~{\rm MeV}$ for $\mhsm\lsim 140\gev$ 
(see Fig.~\ref{hwidths}).

It could happen that there is no $\ee$ collider at the time the $\mm$  
collider
is built but that the LHC has been operational
for several years.  One of the primary modes for discovery of
a SM-like Higgs boson at the LHC is the $\gamma\gamma$ mode.
Simulations by the LHC collaborations indicate that this mode is  
detectable for
$50\lsim\mh\lsim150\gev$.  For $\mh\gsim 130\gev$, discovery
will be possible in the $4\ell$ mode.  Both modes, but especially
the $\gamma\gamma$ mode, offer the possibility of a very accurate
determination of the Higgs mass.  Resolution
will be 1\% or better in the $\gamma\gamma$ mode,
and probably not much worse than $1\%$
in the $4\ell$ mode.  Thus, even in the absence of an $\ee$
collider, the LHC can reasonably be expected to provide us
with a $\lsim 1\%$ determination of $\mh$ in the mass region where
the Higgs total width is small.

\subsection[$s$-channel production of a SM-like
$h$]{{\protect\boldmath$s$}-channel production of a SM-like
{\protect\boldmath$h$}}

\indent\indent
Once a SM-like Higgs boson is found in the $Z\h$ mode
at either an $\ee$ collider or the $\mm$ collider itself,\footnote{
While discovery at a $\mm$ collider is also possible
by scanning in $s$, the $Z\h$ mode is more
luminosity efficient for discovery.}  or at the LHC, it
will generally be easy to also produce and detect it via direct
$s$-channel production at a $\mm$ collider \cite{mupmumprl}
if $\mh\lsim 2\mw$.  Should there be no $\ee$
collider in operation, an important question at a $\mm$ collider will
then be whether to concentrate subsequent running on $s$-channel  
production
or on $Z\h$ production, as the best means for studying the properties
of the $\h$ in detail.  Generally speaking, these two different  
processes
provide complementary information and it would
be very valuable to accumulate substantial integrated
luminosity in both modes.

The potential importance of $s$-channel production of a SM-like $\h$
is illustrated by two facts pertaining to distinguishing between
the MSSM $\hl$ and the SM $\hsm$.
\begin{itemize}
\item[(1)]
Expected experimental errors imply that the ability to
discriminate between the SM $\hsm$ and the MSSM $\hl$ on the basis
of the branching fractions and production rates
that can be measured in the $Z\h$ channel is limited
to $\mha$ values below about $300\gev$ \cite{dpflighthiggs}.
\item[(2)]
Both the total width and the production rate
(proportional to $\Gamma(\h\to\mm)$) of a SM-like $\h$
could be measured at a muon collider with sufficient accuracy so as to
distinguish the $\hl$ from the $\hsm$ in the large-$\mha$ region
$300\gev\lsim m_A\lsim600\gev$  where the $\hl$ is
approximately SM-like.
\end{itemize}

A quantitative discussion of the MSSM parameter space
region for which deviations of the total width and production rate
from SM expectations are measurable will be given later.
For now we emphasize that (2)
requires the excellent $R=0.01\%$ beam energy resolution.

\subsubsection[Choosing the right  
$\protect\sqrt{s}$]{Choosing the right  
{\protect\boldmath$\protect\sqrt{s}$}}

\indent\indent
Our proposed strategy is to first discover the SM-like $\h$ via  
$\ell^+\ell^-\to
Z\h$ or in hadron collisions in order to determine
the $\rts$ region in which $\mm\to\h$ $s$-channel production should  
be
explored. If $\gamh$ is smaller than the
rms spread $\sigrts$ in $\rts$ (as is
the case for the SM  when $\mhsm\lsim 140\gev$), then
to obtain the maximum $\mm\to\h$ production rate it is necessary to set
$\rts$ equal to $\mh$ within $\lsim \sigrts$.
The ability to do this
is assessed by comparing the errors on $\mh$ from $Z\h$
production to both the $\rts$ spread $\sigrts$  at a $\mm$
collider and to $\gamh$. As an illustration, consider 
$\h=\hsm$.
With the super-LC $L=50\fbi$ determination of $\mhsm$ to $\pm 20\mev$,
$\sigrts$ for $R=0.01\%$ will be at worst a factor of 2 or 3 smaller
than the uncertainty in $\mhsm$
and only two or three tries will be needed to set the $\mm$ collider energy
to a value equal to $\mhsm$ within the rms spread in $\rts$. If the
%$4\gev/\sqrt N$
SLD $L=50\fbi$ determination of $\mhsm$ to 210 MeV 
is all that is available, then for $\mhsm\lsim 2\mw$
two or three tries would be adequate to set $\rts\simeq\mhsm$
within $\sigrts$ only if $R=0.06\%$. The number of settings required
in the case of $R=0.01\%$ would be a factor of 6 larger.
If only SLD performance and $L=1\fbi$ is available
in the $Z\hsm$ mode, or if only a $\sim 1\%$ determination
of $\mhsm$ from the LHC is provided, both of which imply
errors on $\mhsm$ that are $\gsim 1\gev$, then even with $R=0.06\%$
one must scan over 10 to 20 $\rts$ values to determine the central  
$\rts\simeq\mhsm$ value within the rms $\rts$ error, $\sigrts$.
Later, we will compute the amount of luminosity
that must be invested at each $\rts=\mh$ choice in order to
detect a SM-like Higgs signal.

In contrast to the above narrow width situation,
for $\mhsm\agt 200$ GeV one finds
$\gamhsm\gsim\sigrts$ for $R\leq 0.06\%$.
Then, even if $\mhsm$ is only known to within $\gamhsm$, we can  
immediately set
$\rts$ for the $\mm$ collider to be within the Higgs peak.
Unfortunately, we find that the event rate in $s$-channel collisions  
is too
low to allow detection of the $\hsm$ in this case.
This situation does not arise in the case of the $\hl$ of the MSSM,
which is guaranteed to have $\mhl\lsim 130\gev$.

\subsubsection[Detecting a SM-like $\h$ in the
$s$-channel]{Detecting a SM-like {\protect\boldmath$\h$} in the
{\protect\boldmath$s$-channel}}

\begin{figure}[htbp]
\let\normalsize=\captsize   %%%% changes the font to "\small"
\begin{center}
\centerline{\psfig{file=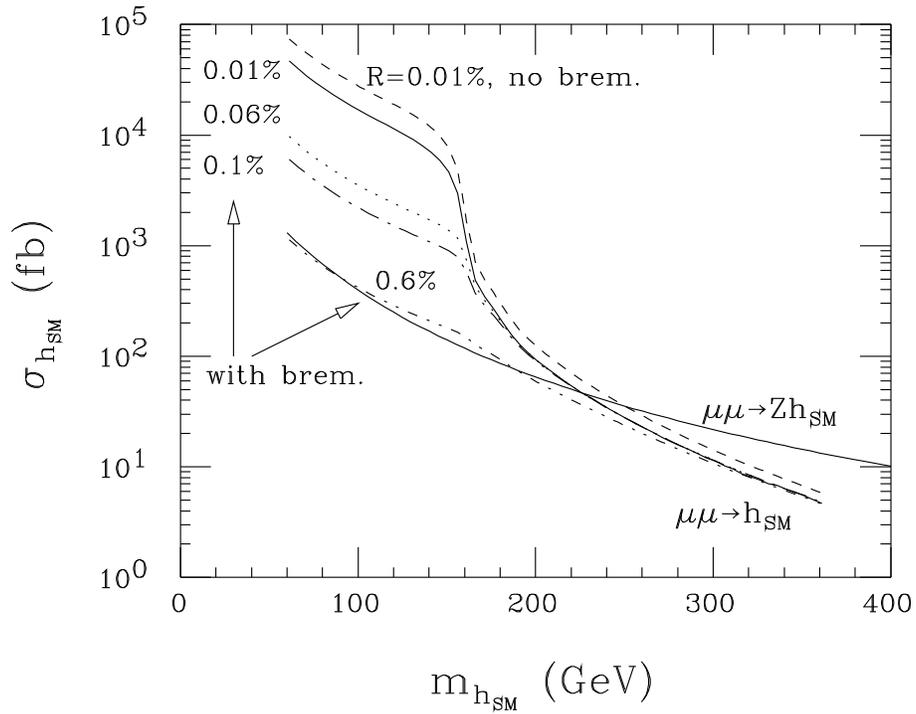,width=12.2cm}}
\begin{minipage}{12.5cm}       %%%% reduces width of caption to 12.5cm
\smallskip
\caption{
Cross sections versus $\mhsm$ for inclusive SM Higgs production: (i)  
the
$s$-channel $\sighbar$ for $\mm\to \hsm$
with $R = 0.01$\%, 0.06\%, 0.1\% and 0.6\%,
and (ii) $\sigma(\mu^+\mu^-\to Z\hsm)$
at $ \protect\sqrt s = \mz + \protect\sqrt 2 \mhsm$.
Also shown is the result for $R=0.01\%$ if bremsstrahlung effects
are not included.}
\label{smxsec}
\end{minipage}
\end{center}
\end{figure}

\indent\indent
The effective cross section, $\overline\sigma_{\hsm}(\rts=\mhsm)$ for  
inclusive SM Higgs production is given in Fig.~\protect\ref{smxsec}
versus $\rts=\mhsm$ for resolutions of 
$R = 0.01$\%, 0.06\%, 0.1\% and 0.6\%.
These results include Gaussian and bremsstrahlung smearing effects.
For comparison, the $\mm\to \zstar\to Z\hsm$ cross section
is also shown, evaluated at the energy  $\sqrt s = \mz + \sqrt 2  
\mhsm$ for
which it is a maximum. The $s$-channel $\mm\to \hsm$
cross sections for small $R$ and $\mhsm\lsim 2\mw$ are much larger
than the corresponding $Z\hsm$ cross section.
The increase in the $\mm\to\hsm$ cross section that
results if bremsstrahlung smearing is removed is illustrated in the  
most
sensitive case ($R=0.01\%$).

For a SM-like Higgs boson, the only potentially
useful final state modes $X$ are $b\anti b$, $W\wstarp$ and $Z\zstarp$, 
where the $^{(\star)}$ indicates the possibility that 
the weak boson is virtual.
The $t\anti t$ channel does not give a viable signal
for the range of luminosity that we consider.
All these channels have irreducible backgrounds from $\mm$
continuum production processes. We note that
\begin{itemize}
\item[(a)]
The light-quark backgrounds to the $b\anti b$ channel
can be rejected using $b$-tagging.  We assume a 50\% efficiency
for isolating the $2b$ final state (via tagging one of the $b$'s);  
this
efficiency is to include cuts and detector efficiencies.
\item[(b)]For the $b\bar b$ final state, we have checked that
interference between the $s$-channel signal and the backgrounds is  
never of
importance.
This is because the Higgs signal contributes to $RR$ and $LL$  
helicity
amplitudes for the incoming muons, whereas the backgrounds come
almost entirely from $RL$ and $LR$ helicity combinations (the $RR$
and $LL$ background contributions are suppressed by a factor of  
$m_\mu/E$
at the amplitude level).
\item[(c)]
For the $W\wstarp$ and $Z\zstarp$ final states the useful
channels depend upon whether or not the $\wstarp$
or $\zstarp$ is virtual. We shall find that discovery
in these channels is only possible for $\mh\lsim 2\mw$,
in which case the final states of interest are $W\wstar\to \ell \nu 2j$
with $\br^{\rm eff}_{WW}\sim 0.3$ and 
$Z\zstar\to 2\ell 2j,2\nu 2j,4\ell,2\ell 2\nu$ with $\br^{\rm eff}_{ZZ}\sim
0.42$, $4j$ final states having too large
a QCD background and mass reconstruction
of the real $W$ or $Z$ being impossible in
the $2\ell 2\nu$ or $4\nu$ final states, respectively. 
(Here, we consider only $\ell=e$ or $\mu$.)
In our analysis, we assume an overall efficiency of $50\%$
for isolating these channels.  For the $Z\zstar$, a cut requiring that
$\mstar$ (the invariant mass of the virtual $\zstar$) be greater
than a given value $\mstarmin$ is imposed.  Full details regarding
our procedures in the $W\wstarp$ and $Z\zstarp$ channels are presented
in Appendix~B.

\end{itemize}

\begin{figure}[htbp]
\let\normalsize=\captsize   %%%% changes the font to "\small"
\begin{center}
\centerline{\psfig{file=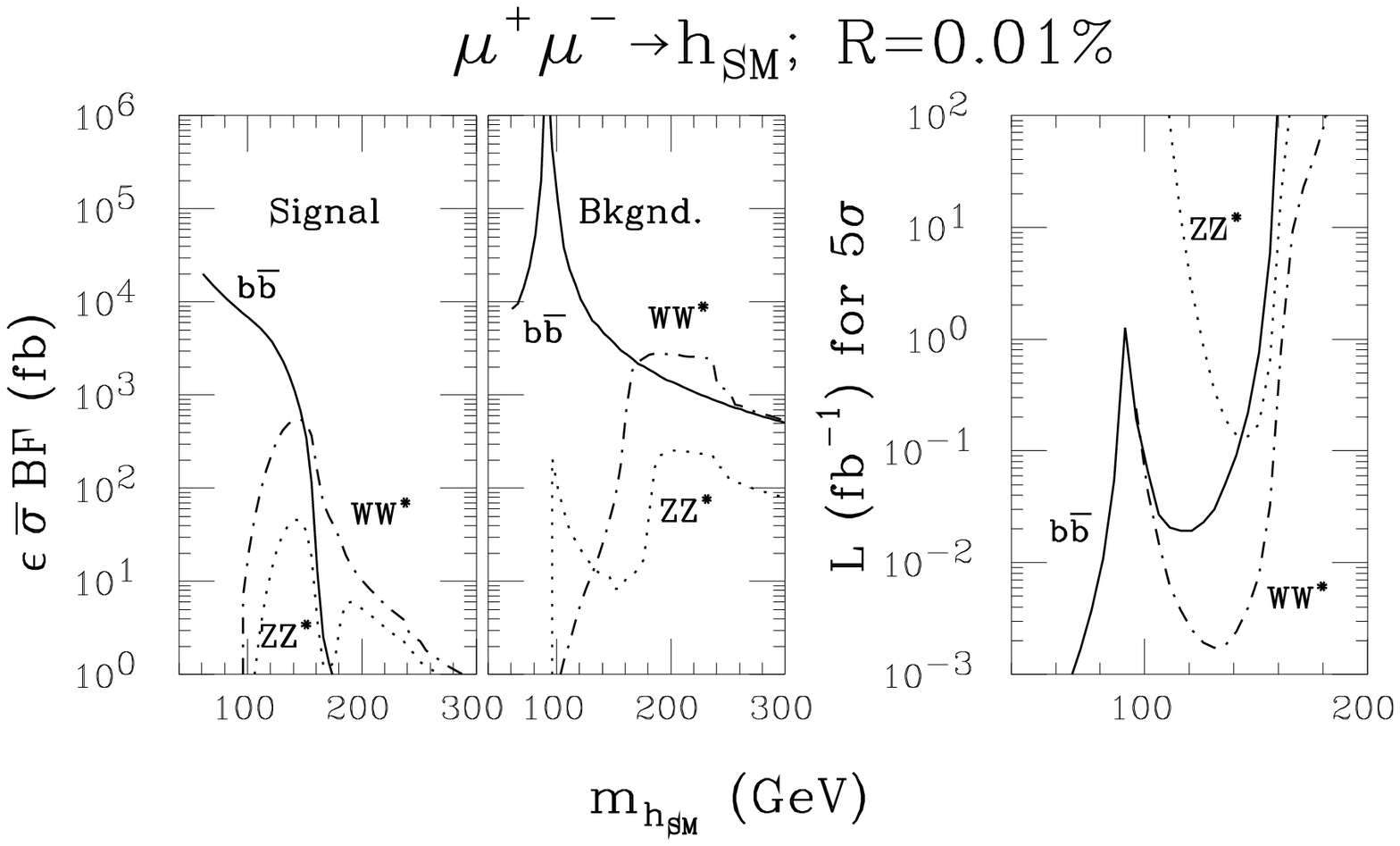,width=12.2cm}}
\vskip .3in
\centerline{\psfig{file=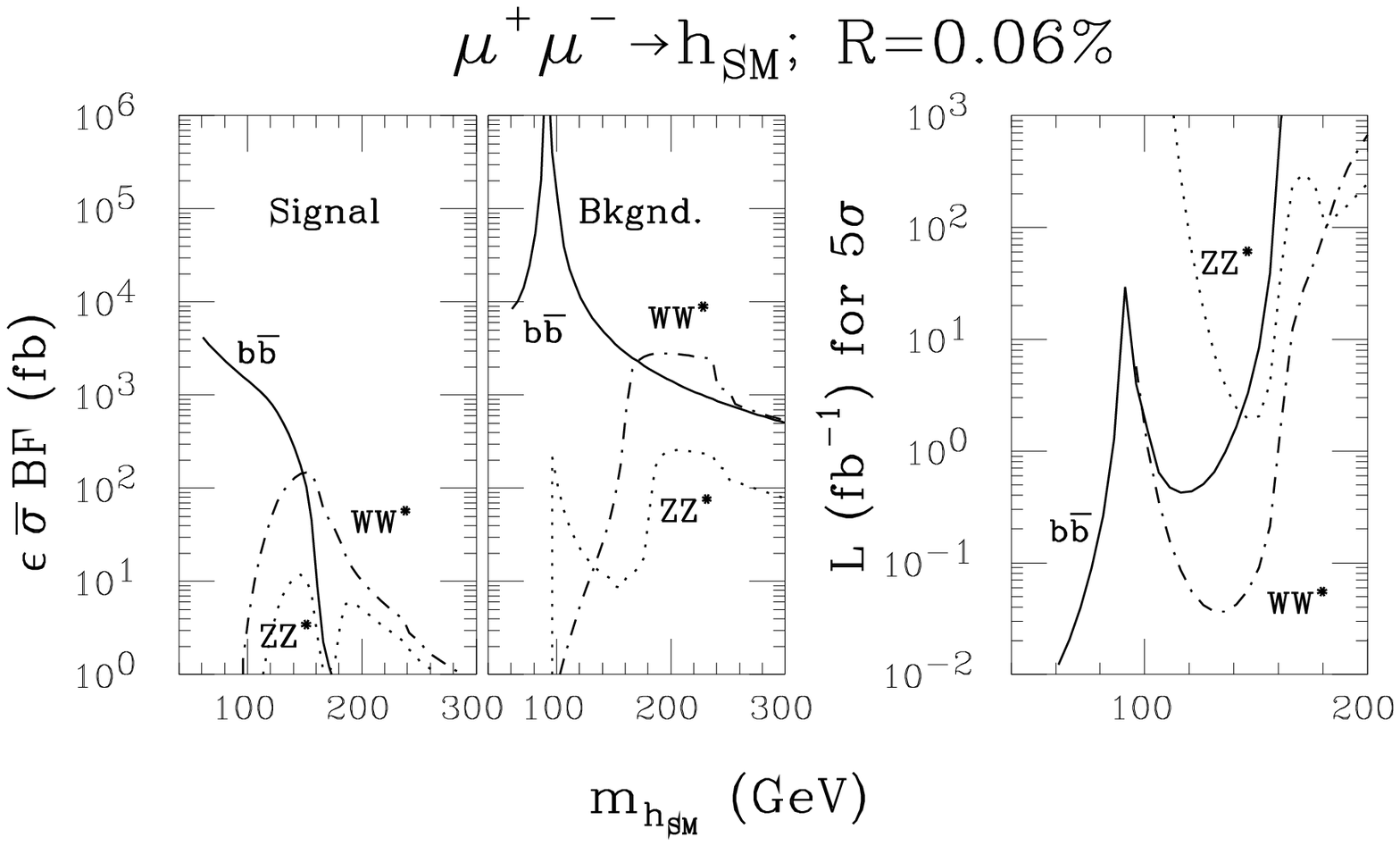,width=12.2cm}}
\begin{minipage}{12.5cm}       %%%% reduces width of caption to 12.5cm
\caption{
The (a) $\hsm$ signal and (b) background
cross sections, $\epsilon \overline\sigma \br(X)$,
for $X=b\anti b$, and useful (reconstructable, non-$4j$) 
$W\wstarp$ and $Z\zstarp$ final states
(including a channel-isolation efficiency of $\epsilon=0.5$)
versus $\mhsm$ for SM Higgs $s$-channel production.
Also shown: (c) the corresponding luminosity required for
a $S/\protect \sqrt B =5$ standard deviations signal in each
of the three channels. Results for $R=0.01\%$ and $R=0.06\%$ are  
given.}
\label{smrates}
\end{minipage}
\end{center}
\end{figure}

The $\hsm$ signal and background
cross sections, $\epsilon \overline\sigma \br(X)$,
for $X=b\anti b$, and the above $W\wstarp$ and $Z\zstarp$ final states
are presented in Fig.~\ref{smrates} (including a channel-isolation  
efficiency of $\epsilon=0.5$) as a function of
$\mhsm$ for SM Higgs $s$-channel production with
resolution $R=0.01\%$ and $R=0.06\%$.
For both resolutions, we also plot the luminosity required for
a $S/\protect \sqrt B =5\sigma$ signal in the $b\anti b$, $W\wstarp$  
and $Z\zstarp$ channels. In the case of the $W\wstarp$
final state, we give event rates only for
the mixed leptonic/hadronic final state modes;
in the case of the $Z\zstarp$ final state
we include the mixed hadronic/leptonic and (visible) purely leptonic
final state modes listed earlier.

From Fig.~\ref{smrates} we see that:
\begin{itemize}
\item
$R=0.01\%$, $L=0.1\fbi$ would  yield
a detectable $s$-channel Higgs signal
for all $\mhsm$ values between the current LEP\,I limit of 63~GeV and  $2\mw$
except in the region of the $Z$~peak; 
a luminosity $L\sim 1\fbi$ at
$\sqrt s = \mhsm$ is needed for $\mhsm\sim \mz$.
\item
For $R=0.06\%$, $5\sigma$ signals typically require about 20--30 times
the luminosity needed for $R=0.01\%$; $L=30\fbi$
would be required for a $5\sigma$ signal if $\mhsm\sim\mz$.
\end{itemize}
This argues for a $\mm$
collider design with $R$ near the $0.01\%$ level. A search for the $\hsm$
(or any Higgs with width smaller than the achievable resolution)
by scanning would be most efficient for the smallest possible $R$.
For a specific illustration, let us consider $\mhsm\sim 110\gev$ and assume
that just $L=1\fbi$ has been accumulated in the $Z\hsm$ mode
(at either an $\ee$ collider or at the $\mm$ collider itself).
Fig.~\ref{massresolution} shows that the error in the determination
of $\mhsm$ will be of order $\pm0.8\gev$ (assuming an SLD-type detector).
How much luminosity
will be required to observe the $\hsm$ in the $s$-channel by zeroing
in on $\mhsm$ within the rms resolution $\sigrts$?
The number of scan points required to cover the $1.6\gev$
mass zone at intervals of $\sigrts$, the luminosity required to observe
(or exclude) the Higgs at each point, and the total luminosity
required to zero-in on the Higgs using the scan is given
in Eq.~(\ref{smallscan}), for resolutions of $R=0.01\%$ and $0.06\%$.
\begin{equation}
\begin{array}{lrrrr}
~~R & \sigrts~~~ & \# \mbox{points} & L/\mbox{point}~~ & L_{\rm tot}~~~~ \\
0.01\% & 7\mev & 230~~ & 0.01\fbi & 2.3\fbi \\
0.06\% & 45\mev & 34~~ & 0.3\fbi & 10.2\fbi 
\end{array}
\label{smallscan}
\end{equation}

More generally, the $L$ required at each scan point
decreases as (roughly) $R^{1.7}$, whereas the number
of scan points only grows like $1/R$, implying that
the total $L$ required for the scan decreases as $\sim R^{0.7}$.
Thus, the $\mm$ collider should be constructed with the smallest 
possible $R$ value. (Note that if the Higgs resonance
is broad, using small $R$, although not necessary, is not harmful  
since the data from a fine scan can be rebinned to test for its presence.)
In the case of a narrow Higgs, a by-product of the above zeroing-in
scan will be to ascertain
if the Higgs width is in the $\alt \sigrts$ range.                 
However,  
the large number of $\rts$ settings required when conducting a scan
with small $R$ implies that it must be possible
to quickly and precisely adjust the energy of the $\mm$ collider.
For example, if the machine can deliver $50\fbi$ per year and
$R=0.01\%$, so that only $L\sim 0.01\fbi$ should be devoted to each point,
we must be able to step the machine energy in units of $\sim 7\mev$
once every hour or so. 

Let us compare the above procedure, where the $Z\h$ mode
at low luminosity is used to find the SM-like $\h$
and then $s$-channel collisions are used to zero-in on $\mh$,
to the possibility of searching directly for the $\h$ by
$s$-channel scanning without the benefit of $Zh$ data.  
The latter would be a possible alternative
if the $\mm$ collider were to be built before the light Higgs boson
is observed at either the LHC or an $\ee$ collider.
The question is whether it is most useful to employ the $Z\h$
mode or direct $s$-channel production for initial discovery. 
We shall suppose that precision radiative corrections 
pin down the mass of the SM-like Higgs boson to
a 20 GeV interval, although this may be way too optimistic. 
Let us again focus on $\mh=110$ GeV.
The number of scan points required to cover the $20\gev$
mass zone at intervals of $\sigrts$, the luminosity required to observe
(or exclude) the Higgs at each point, and the total luminosity
required to zero-in on the Higgs using the scan is given
in Eq.~(\ref{bigscan}), for resolutions of $R=0.01\%$ and $0.06\%$.
\begin{equation}
\begin{array}{lrrrr}
~~R & \sigrts~~~ & \# \mbox{points} & L/\mbox{point}~~ & L_{\rm tot}~~~~ \\
0.01\% & 7\mev & 2857~~ & 0.01\fbi & 29\fbi \\
0.06\% & 45\mev & 426~~ & 0.3\fbi & 128\fbi 
\end{array}
\label{bigscan}
\end{equation}
Thus, much greater luminosity would be required 
(not to mention the much greater demands upon the machine
for performing efficiently such a broad scan) than if
the $Z\h$ mode is employed for the initial $\h$ discovery.
Note that it is not useful to expend more than
$L\sim 1\fbi$ in the $Z\h$ mode simply to pin down the mass; however,
precision studies with $L=50\fbi$ in this mode would be useful for determining
$\sigma(Z\h)\times \br(\h\to X)$ for various different final states, $X$
\cite{dpflighthiggs}.

For $\mhsm$ above $2\mw$, $\gamhsm$ rises dramatically,
$\br(\hsm\to \mm)$ falls rapidly and, thus [see Eq.~(\ref{broadwidthsigma})
and Fig.~\ref{smxsec}], $\sighbar$ declines precipitously.
Even after combining all channels,
the luminosity requirements in the double-on-shell $WW$ and $ZZ$ final states
are such that Higgs detection in $s$-channel production will be difficult.
How severe a drawback is this?
One of the unique and most important features of
$s$-channel Higgs production
is the ability to scan with sufficient statistics to determine
the width of a narrow Higgs boson.
In the case of the $\hsm$, only below $WW$ threshold
is the Higgs so narrow that this is the only possible measurement technique.
The $\hsm$ can be detected straightforwardly in the
standard $Z\hsm$ mode and, at the super-LC detector, its width
can be measured down to $0.2\gev$ via the recoil mass spectrum
in $Z\hsm$ events with $Z\to\lplm$. Since $\gamhsm\gsim 0.2\gev$ for
$\mhsm\gsim 2\mw$, this $Z\hsm$ technique becomes viable just as
$s$-channel detection becomes difficult.
Without the super-LC detector there could, however, be a gap
between the $\mhsm\lsim 2\mw$ region where $s$-channel
measurement of $\gamhsm$ will be possible at a muon collider and the  
region $\mhsm\gsim 200\gev$ where $\gamhsm$ becomes comparable to the
event by event mass resolution of $\sim 4\gev$ (see earlier  
discussion and Fig.~\ref{hwidths}) and would become measurable at a linear  
$e^+e^-$ collider.
The high resolution for lepton momenta of the super-LC detector
could thus prove critical in avoiding a gap
in the region between about $150\gev$ and $200\gev$ where $\gamhsm$
measurement might not be possible
using either $s$-channel scanning or the $Z\hsm$ mode.

The most important conclusions of this subsection are two:
\begin{itemize}
\item[(1)] Excellent beam energy resolution is absolutely critical
to guaranteeing success in detecting a SM-like $\h$
in $\mm\to\h$ $s$-channel collisions and to our ability to perform
detailed studies once the Higgs boson mass is known. Every effort
should therefore be made to achieve excellent resolution. (It is only  
if $\mh>2\mw$ where the SM-like Higgs boson begins to become broad 
that the advantage of having small $R$ declines. But, for such masses 
$s$-channel discovery of the SM Higgs will be very difficult in any case, 
as we have discussed.)
\item[(2)] The scanning required when $R$ is small implies
that the machine design must be such that $\rts$ can be
{\it quickly} reset with a precision that is a small fraction of $\sigrts$.
\end{itemize}

\subsection[Precision measurements:
$\mh$ and $\gamh$]{Precision measurements:
{\protect\boldmath$\mh$} and {\protect\boldmath$\gamh$}}

\indent\indent
Once the machine is set to the central value of $\rts=\mh$,
one can proceed to precisely measure
the mass $\mh$ and the total width $\gamh$.
A precision determination of the total width $\gamh$ is of particular  
interest
to differentiate between the $\hsm$
and the $\hl$ of the MSSM.
Knowledge of the total width will also allow extraction
of the partial width (and associated Higgs couplings) for any channel
in which the Higgs can be observed.

\begin{figure}[htbp]
\let\normalsize=\captsize   %%%% changes the font to "\small"
\begin{center}
\centerline{\psfig{file=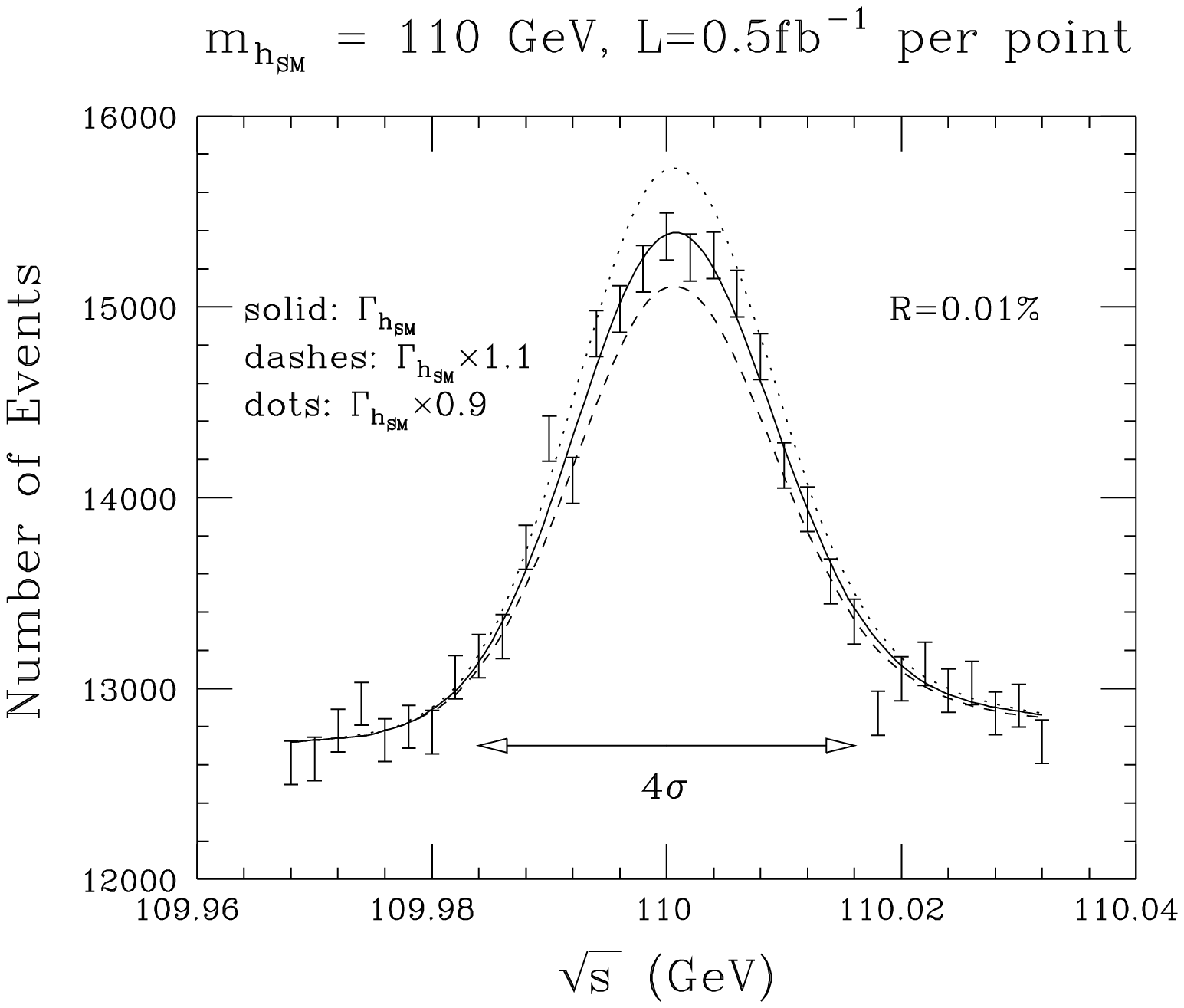,width=12.2cm}}
\begin{minipage}{12.5cm}       %%%% reduces width of caption to 12.5cm
\caption{Number of events and statistical errors in the $b\anti b$
final state as a function
of $\protect\rts$ in the vicinity of $\mhsm=110\gev$,
assuming $R=0.01\%$,
and $L=0.5\fbi$ at each data point.
The precise theoretical prediction is given by the solid line.
The dotted (dashed) curve is the theoretical prediction
if $\gamhsm$ is decreased (increased) by 10\%, {\it keeping
the $\Gamma(\hsm\to\mm)$ and $\Gamma(\hsm\to b\anti b)$
partial widths fixed at the predicted SM value.}}
\label{mhsmscan}
\end{minipage}
\end{center}
\end{figure}

A precise measurement of the Higgs mass is
possible via $s$-channel collisions. We initially focus our discussion on
$\mhsm \lsim 2\mw$, for which $\gamhsm$ is quite likely to be
smaller, perhaps much smaller, than the rms $\rts$ resolution,  
$\sigrts$. Despite this, a highly accurate determination of $\mhsm$
is still possible via a straightforward
scan in the vicinity of $\rts=\mhsm$.  In Fig.~\ref{mhsmscan} we  
illustrate sample data points (statistically fluctuated) in the case of  
$\mhsm=110\gev$, assuming $L=0.5\fbi$ is accumulated at each $\rts$ setting.
A resolution of $R=0.01\%$ is assumed. The solid curve is the  
theoretical prediction. A visual inspection reveals that $\mhsm$ can
be pinned down to within about $4\mev$ using seven scan
points centered around $\rts=\mhsm$ (involving a combined luminosity  
of $3.5\fbi$). Using somewhat more sophisticated techniques, to be described
shortly, we will find that with this same total luminosity we can do
better. These latter techniques are those needed
for a direct measurement of the total Higgs width $\gamhsm$.

If the partial widths
for $\hsm\to\mm$ and $\hsm\to b\anti b$  are regarded as  
theoretically
computable
with no systematic uncertainties (not a valid assumption in the
case of the MSSM $\hl$), then determination of $\gamhsm$
is straightforward based on Eq.~(\ref{narrowwidthsigma}).
We have plotted the theoretical
predictions for $\mhsm=110\gev$ in Fig.~\ref{mhsmscan} corresponding
to keeping the above partial widths constant while varying only
$\gamhsm$ by $\pm10\%$.  Assuming that the background can be
absolutely normalized by a combination of theory and experiment,
the height of the peak is a measure of $\gamhsm$.  The
seven central points would determine $\gamhsm$ to better than 10\%.

Since in practice we are not able to accurately pre-determine
the partial widths, a {\it model-independent} technique for
discriminating between
the total width of the SM $\hsm$ and that of some other SM-like $\h$
must be devised that does not involve a
theoretical computation of the partial widths.
Such a determination of the total width
requires measurements sensitive to
the breadth
of the spectrum illustrated in Fig.~\ref{mhsmscan}.
We outline below a procedure
by which roughly $L\sim 3\fbi$ of total luminosity
will allow a $\pm 33\%$ determination of $\gamhsm$ (for  
$\mhsm=110\gev$)
without any assumption regarding the partial widths.

The key observation is that if one adjusts the partial widths
so that the normalization of the theoretical curve at $\rts=\mhsm$
agrees with experiment, then the normalization of the wings of
the theoretical curve will be correspondingly increased or decreased  
in the case that $\gamh$ is  larger or  smaller, respectively.
Experimental measurements of sufficient precision both at
a central $\rts$ value and on the wings would thus allow
a direct measurement of $\gamhsm$ via the ratio of
the central peak cross section to the cross sections on the wings  
(the partial widths cancel out in the ratio). 
With this in mind, we define the quantity
\begin{equation}
d \equiv  | \sqrt s - \mhsm | / \sigrts
\end{equation}
and propose the following procedure:
\begin{itemize}
\item[(1)]
Perform a rough scan to determine $\mhsm$
to a precision $\sigrts d$, with $d\lsim 0.3$;
$d$~will not be known ahead of
time, but the value of $d$, and hence of $\mhsm$
will be determined by the procedure.
\item[(2)]
Then perform three measurements.
At $\rts_1=\mhsm+\sigrts d$
we employ a luminosity of $L_1$ and
measure the total rate $N_1=S_1+B_1$.
Then perform two additional measurements at
\begin{equation}
 \rts_2=\rts_1-\nsigrts\sigrts \label{measure1}
\end{equation}
 and one at
\begin{equation}
\rts_3=\rts_1+\nsigrts\sigrts \label{measure2}
\end{equation}
yielding $N_2=S_2+B_2$ and $N_3=S_3+B_3$
events, respectively, employing luminosities of
$L_2=\rho_2L_1$ and $L_3=\rho_3L_1$.
We find that
 $\nsigrts\sim 2$ and
$\rho_2=\rho_3\sim 2.5$
are optimal for maximizing sensitivity and minimizing the
error in determining $d$ (\ie\ $\mhsm$) and $\gamhsm$.
\item[(3)]
To determine $\mhsm$ and $\gamhsm$ consider the ratios
\begin{equation}
r_2\equiv (S_2/\rho_2)/S_1 = (S_2/L_2)/(S_1/L_1)  \qquad r_3\equiv
(S_3/\rho_3)/S_1 = (S_3/L_3)/(S_1/L_1)\,.
\end{equation}
The ratios $r_2$ and $r_3$ are governed by $d$  and $\gamhsm$.
 Conversely, we have implicitly
$d=d(r_2,r_3)$ and $\gamhsm=\gamhsm(r_2,r_3)$.
Determining the statistical
errors $\Delta\mhsm$ and $\Delta\gamhsm$ is then simply a matter of
computing the partial derivatives of $d$ and $\gamhsm$
with respect to the $r_{2,3}$
(we do this numerically)
and using  errors on the ratios $r_{2,3}$ implied by statistics.
The procedure is detailed in Appendix~C, as is the cross check
on its accuracy that we have used.
\end{itemize}

\begin{figure}[htbp]
\let\normalsize=\captsize   %%%% changes the font to "\small"
\begin{center}
\centerline{\psfig{file=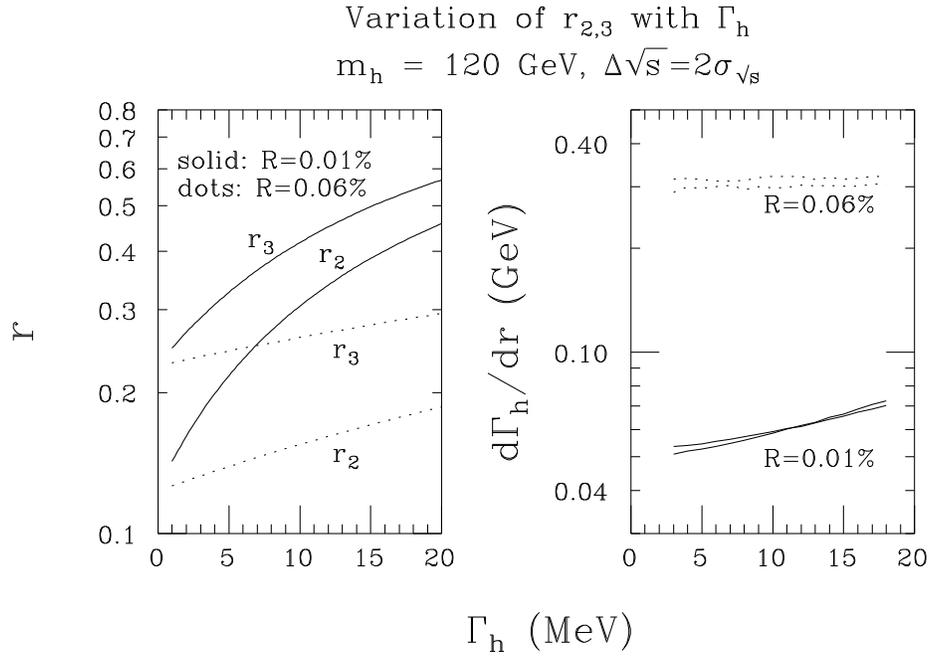,width=12.2cm}}
\begin{minipage}{12.5cm}       %%%% reduces width of caption to 12.5cm
\caption{We plot $r_2$ and $r_3$ as a function of Higgs width,  
$\gamh$,
for resolutions of $R=0.01\%$ and $R=0.06\%$, assuming that
$\protect\rts=\mh=120\gev$.  Also shown are the derivatives
$d\gamh/dr$ as a function of $\gamh$. We have taken
$\protect\nsigrts=2$ corresponding to a shift in $\protect\rts$
of $\mp 2\protect\sigrts$ in computing $r_2$ and $r_3$,  
respectively.}
\label{r2sigma}
\end{minipage}
\end{center}
\end{figure}

The utility of the ratios $r_2$ and $r_3$ is basically governed
by how rapidly they vary as $d$ and $\gamh$ are varied
in the ranges of interest. Since we are most interested
in $\gamh$ here, we illustrate the sensitivity of $r_{2,3}$
to $\gamh$ in Fig.~\ref{r2sigma} taking $\rts=\mh=120\gev$.
For this figure we employ
$\nsigrts=2$ for computing
$r_2$ and $r_3$, respectively.
Results are shown for resolutions $R=0.01\%$ and $R=0.06\%$.  Because  
of the
bremsstrahlung tail, $r_2$ is substantially larger than $r_3$.
Nonetheless, both $r_2$ and $r_3$ show rapid variation as $\gamh$
varies in the vicinity of $\gamhsm$ in the case
of $R=0.01\%$, but much less variation if $R=0.06\%$.  The error in  
the
determination of $\gamh$ is basically determined by  
$d\gamh/dr_{2,3}$.
Figure~\ref{r2sigma} shows that these derivatives are almost the
same and quite small for $R=0.01\%$.  The much larger values
of these derivatives for $R=0.06\%$ imply that determining
$\gamh$ accurately would be very difficult in this case.

\begin{figure}[htbp]
\let\normalsize=\captsize   %%%% changes the font to "\small"
\begin{center}
\centerline{\psfig{file=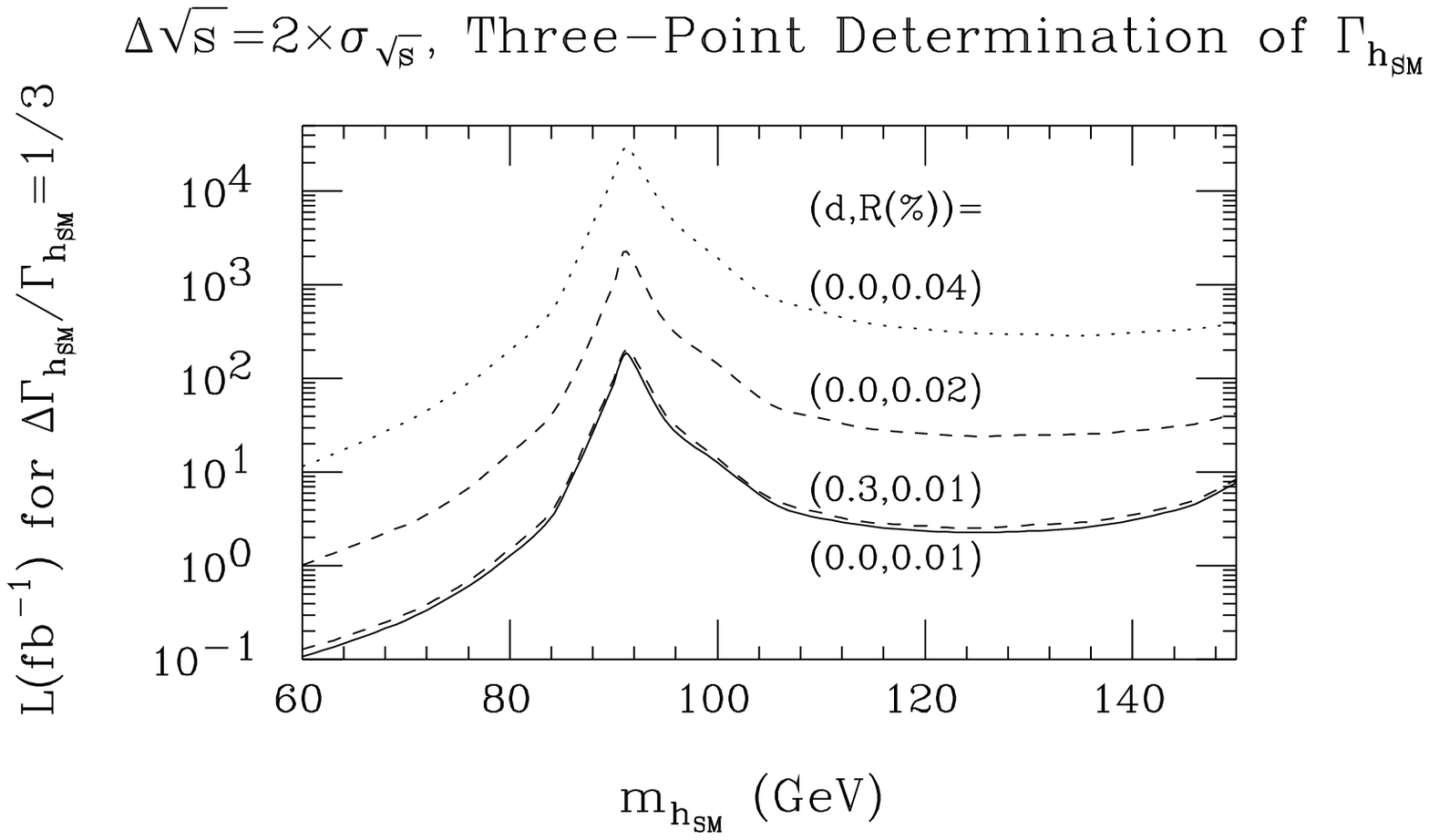,width=12.2cm}}
\begin{minipage}{12.5cm}       %%%% reduces width of caption to  12.5cm
\caption{Luminosity required for a $\Delta\gamhsm/\gamhsm=1/3$
measurement in the $b\anti b$ final state using the three
point technique described in the text.  Results for resolutions
of $R=0.01\%$, $0.02\%$ and $0.04\%$ are shown for $d=0$, where 
$d = | \protect\sqrt s - \mhsm | / \protect\sigrts$.
The result for $d=0.3$ and $R=0.01\%$ is also shown.}
\label{masswidthlum}
\end{minipage}
\end{center}
\end{figure}

In Fig.~\ref{masswidthlum},
we plot the total luminosity $L=L_1+L_2+L_3=6L_1$ required
to achieve $\Delta\gamhsm/\gamhsm= 1/3$ in the $b\anti b$
final state as a function of $\mhsm$ for several beam resolutions.
(The error scales statistically; \eg\ to achieve a 10\% measurement
would require $(10/3)^2$ as much luminosity.)  We also illustrate
the fact that the total luminosity required is rather insensitive to
the initial choice of $d$ for $d\lsim 0.3$; $d=0.3$ results
in no more than a 20\% increase in the luminosity needed
relative to $d=0$.

\begin{figure}[htbp]
\let\normalsize=\captsize   %%%% changes the font to "\small"
\begin{center}
\centerline{\psfig{file=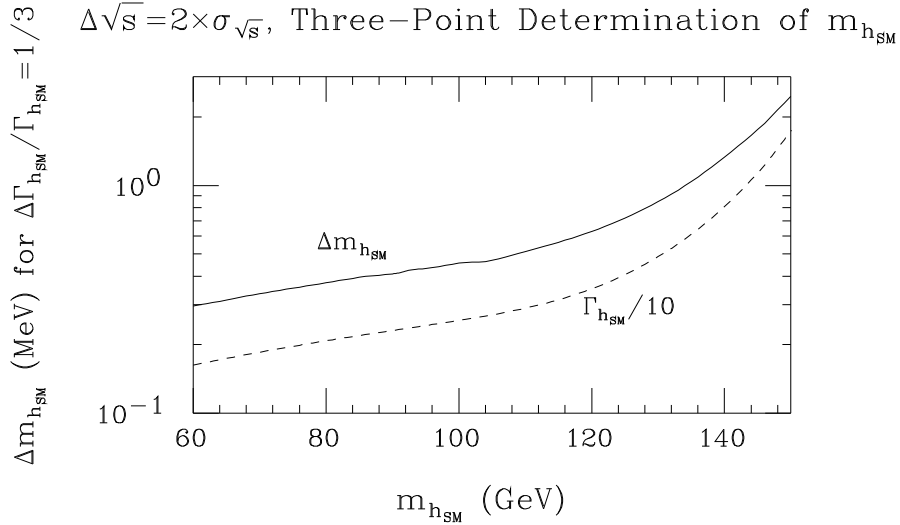,width=12.2cm}}
\begin{minipage}{12.5cm}       %%%% reduces width of caption to 12.5cm
\caption{We plot the $1\sigma$ error, $\Delta\mhsm$, in the  
determination of $\mhsm$ using the three point technique described
in the text with $R=0.01\%$ and $d=0$.
The error given is that achieved for the luminosity
that allows a $\Delta\gamhsm/\gamhsm= 1/3$
measurement in the $b\anti b$ final state. For such luminosity,
$\Delta\mhsm$ is essentially independent of $R$ and $d$.
Also shown, for comparison, is $\gamhsm/10$. }
\label{masswidthlumdelm}
\end{minipage}
\end{center}
\end{figure}

In Fig.~\ref{masswidthlumdelm},
we plot the $1\sigma$ error $\Delta\mhsm$ that results using our  
three-point technique after accumulating the luminosity required
for a $\Delta\gamhsm/\gamhsm=1/3$ measurement in the $b\anti b$
final state.  The specific result
plotted is for $R=0.01\%$ and $d=0$, but is essentially
independent of $R$ and $d$ given the stated luminosity.  Also
shown, for comparison, is $\gamhsm$ itself. We see that $\Delta\mhsm$
is of order $1.5$--$2$ times $\gamhsm/10$, \ie\ a fraction of an MeV
for $\mhsm\lsim 130\gev$. (Again, $\Delta\mhsm$ scales as $1/\sqrt L$.)

It should be stressed that the ability to precisely set the 
energy of the machine when the three measurements are taken is  
crucial for the success of the three-point technique. A misdetermination of  
the spacing of the measurements in Eqs.~(\ref{measure1}) and  
(\ref{measure2}) by just 3\% (\ie\ $\rts$ uncertainty of order
$0.25\mev$ for any one setting near $\mhsm\sim 120\gev$)
would result in an error in $\gamhsm$ of 30\%. 
For a measurement of $\gamhsm$ at the 10\% level the
$\rts$ settings must be precise at a level of better than 
one part in $10^6$. This is possible \cite{palmer2}
provided  the beam can be  partially polarized
so that the precession of the spin of the muon 
as it circulates in the final storage ring can be measured. 
From the precession  and the rotation rate the energy can be determined. 
The ability to perform this critical measurement needed for the determination of  
the total width of a narrow Higgs must be incorporated in the  
machine design.

\subsection[Precision measurements:
$\Gamma(\h\to \mm)\times \br(\h \to X)$]{Precision measurements:
{\protect\boldmath$\Gamma(\h\to \mm)\times \br(\h \to X)$}}
\indent

\begin{figure}[htbp]
\let\normalsize=\captsize   %%%% changes the font to "\small"
\begin{center}
\centerline{\psfig{file=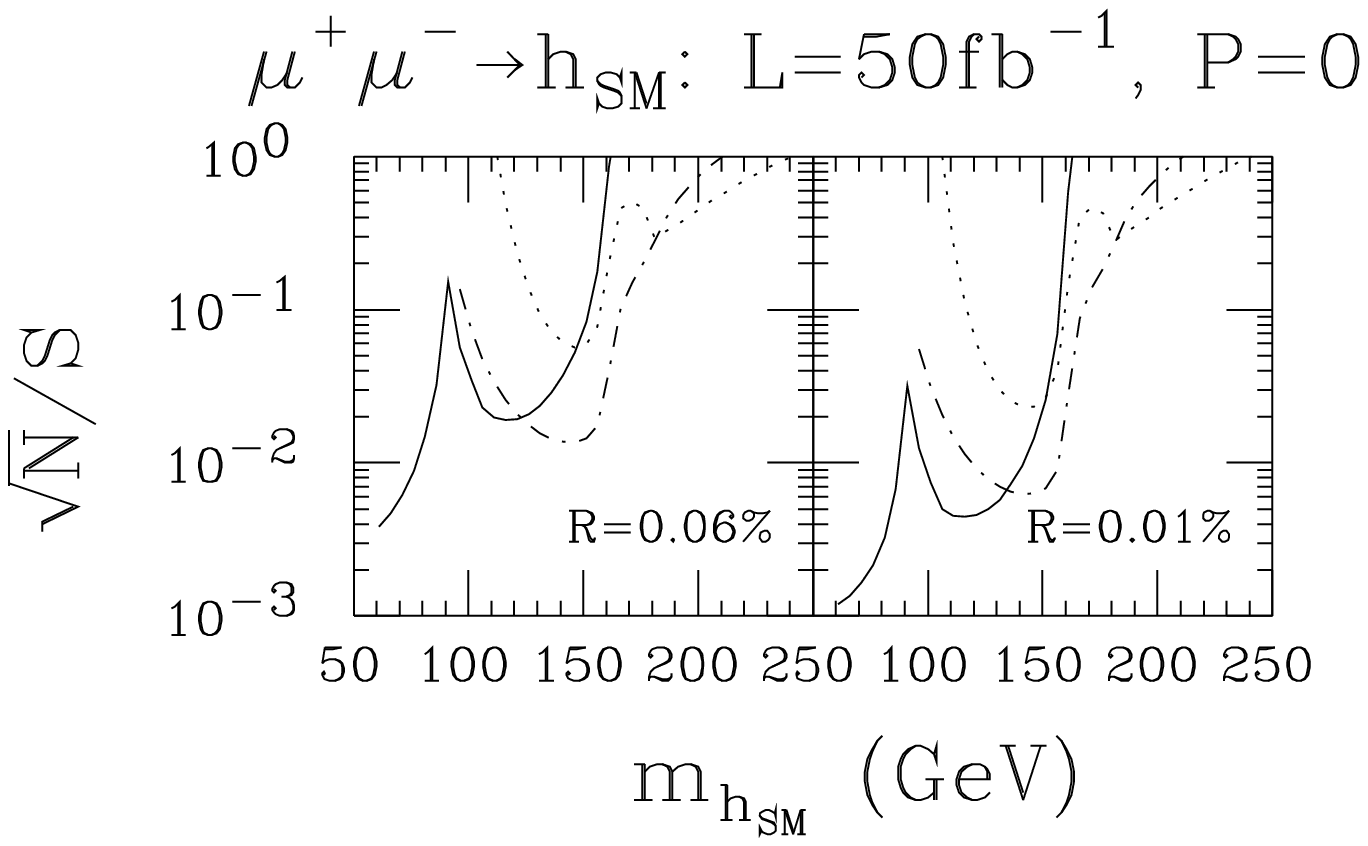,width=12.2cm}}
\begin{minipage}{12.5cm}       %%%% reduces width of caption to 12.5cm
\caption{Fractional error in determining $\Gamma(\hsm\to\mu\mu)\times
\br(\hsm\to X)$ for $X=b\anti b$ (solid), $W\wstarp$ (dotdash) and
$Z\zstarp$ (dots), assuming $L=50\fbi$.
(See text for $W\wstar$ and $Z\zstar$ final states employed.)}
\label{smerrors}
\end{minipage}
\end{center}
\end{figure}

Assuming that the Higgs width is much narrower than the rms
uncertainty in $\rts$, Eq.~(\ref{narrowwidthsigma}) shows that the  
event rate in a given channel measures 
$G(X)=\Gamma(\h\to \mm)\times \br(\h \to  X)$.
If the background can be determined precisely (either
by off-resonance measurements or theory plus Monte Carlo  
calculation), the error in the determination of this product is $\sqrt N/S$,
where $N=S+B$ and $S$, $B$ are the number of signal, background events,
respectively. The results for $\sqrt N/S$
in the case of $P=0$ and $L=50\fbi$
in the $b\anti b$, $W\wstarp$ and $Z\zstarp$ modes are shown in  
Fig.~\ref{smerrors} for  $\h=\hsm$.
For each final state, the efficiencies and procedures employed
are precisely those discussed with regard to Fig.~\ref{smrates}.
Good accuracy in this measurement is possible
for $\mhsm\alt 2\mw$ even if $\mhsm$ is near $\mz$.

%In a later section, we shall discuss  strategies for extracting
%$\h\to\mm$  and $\h\to X$ partial widths
%from the product $\Gamma(\h\to \mm)\times \br(\h \to X)$
%and how such measurements
%may allow discrimination between the $\hsm$ and the MSSM $\hl$.
%These we will compare with
%the discrimination possible in the $Z\h$ mode.
%For now, we simply note that a determination of the total
%Higgs width is a very valuable tool in this process.
%Without knowledge of $\Gamma_h$, useful information
%could still be extracted by taking the measured
%values of $\br(\h\to X)$ from $Z\h$
%production and using these to make determination(s)
%of the partial width $\Gamma(\h\to\mm)$ and corresponding
%$h\to\mu^+\mu^-$ coupling. More information could be extracted
%under certain assumptions.  In particular, if the $\mm$
%and $b\anti b$ squared couplings are assumed to the related
%to the SM squared couplings by the same factor, $f$, (as
%is the case in two doublet Higgs models of type-II ---
%the MSSM Higgs sector being an example thereof), then
%once $\Gamma(\h\to\mm)$ has been extracted as described above
%$\Gamma(\h\to b\anti b)$ can be computed for an assumed
%value of $\mb$.  The total Higgs width could then be computed
%as $\Gamma(\h\to b\anti b)/\br(\h\to b\anti b)$. This
%somewhat uncertain procedure could be avoided by  direct
%measurement of the total width.
%It is to this and the related accurate determination of the Higgs mass
%that we now turn.

%\subsection{SM-like {\protect\boldmath$\hl$}}
%
%\indent\indent
%Before turning to the $\hh$ and $\ha$, we consider $s$-channel
%$\hl$ production in more detail, assuming that $\mha\gsim 2\mz$
%for which the $\hl$ is SM-like. In general, for a SM-like $\h$
%we have seen that we can determine both
%$\Gamma(\h\to\mu\mu)\times
%\br(\h\to b\anti b)$ and the total $\h$ width
%with good accuracy, depending upon the beam resolution $R$,
%the available luminosity and the Higgs mass.  Good accuracy for 
%the $\gamh$
%measurement is especially dependent upon having the best
%possible $R$, with $R\sim 0.01\%$ being almost mandatory
%to reach the accuracy levels of interest below.

\subsection[$\hl$ or $\hsm$?]{{\protect\boldmath$\hl$} or {\protect\boldmath$\hsm$}? }
\indent

\begin{figure}[htbp]
\let\normalsize=\captsize   %%%% changes the font to "\small"
\begin{center}
\centerline{\psfig{file=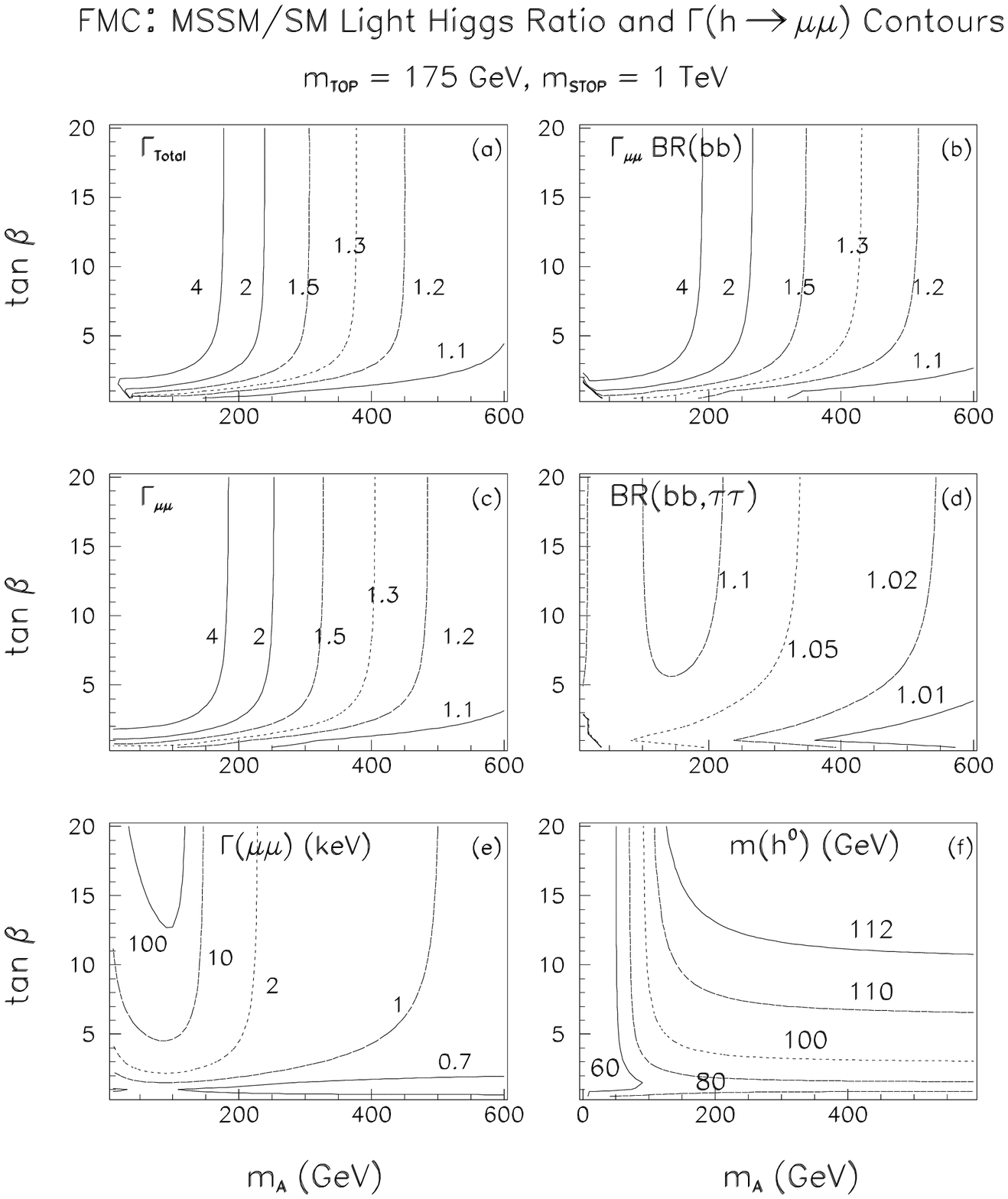,width=12.2cm}}
\begin{minipage}{12.5cm}       %%%% reduces width of caption to 12.5cm
\caption{Contours of constant MSSM/SM  ratios for $\gamh$,
$\Gamma(\h\to\mu\mu)\times \br(\h\to b\anti b)$,
$\Gamma(\h\to\mu\mu)$ and $\br(\h\to b\anti b,\tau\tau)$ in  
$(\mha,\tanb)$
parameter space. We have taken $\mt=175\gev$, $\mstop=1\tev$,
and included two-loop/RGE-improved
radiative corrections, neglecting squark mixing, for Higgs masses,
mixing angles and self-couplings.
Also shown are contours for fixed values of $\Gamma(\hl\to\mu\mu)$
using units of keV, and contours of fixed $\mhl$. 
This graph was obtained using the programs
developed for the work of Ref.~\protect\cite{ghinprogress}.
}
\label{hltohsmratios}
\end{minipage}
\end{center}
\end{figure}

We now discuss the possibility of distinguishing
the MSSM $\hl$ from the SM $\hsm$ using precision measurements of $\gamh$ and
$G(b\anti b)\equiv \Gamma(\h\to\mu\mu)\times \br(\h\to b\anti b)$.
The accuracy to which  $\gamh$ and $G(b\anti b)$ need to be determined 
can be gauged by the ratio of the $\hl$ predictions to the 
$\hsm$ predictions for these
quantities at $\mhl=\mhsm$.  Contours for
various fixed values of these ratios are plotted
in Fig.~\ref{hltohsmratios} in the standard $(\mha,\tanb)$ parameter  
space \cite{ghinprogress}. In computing results for $\gamh$ and $G(b\anti b)$
for $\hl$ we have taken $\mstop=1\tev$, $\mt=175\gev$, and included
two-loop/RGE-improved radiative corrections to the Higgs
masses, mixing angles and self-couplings, neglecting squark mixing.
The ratios for both $\gamh$ and $G(b\anti b)$ are substantially
bigger than 1, even out to fairly large $\mha$ values.
This is because the $\hl$ retains somewhat enhanced $b\anti b$,  
$\tau^+\tau^-$ and $\mupmum$ couplings until quite large $\mha$ values.
Two facts are of particular importance:

\begin{itemize}
\item
$\gamhl$ is enhanced relative to $\gamhsm$ by
virtue of the enhanced partial widths into its dominant decay  
channels, $b\anti b$ and $\tau^+\tau^-$.

\item
The enhancement in $G(b\anti b)$ derives mainly
from $\Gamma(\h\to \mu\mu)$, as can be
deduced by comparing Fig.~\ref{hltohsmratios}(b) and \ref{hltohsmratios}(c).
\end{itemize}

\noindent
This latter point is also apparent in Fig.~{hltohsmratios}(d), 
where we observe that the MSSM to SM
ratio of $\br(\h\to b\anti b)$'s is very close to 1
along the 1.1 contour of the MSSM/SM $G(b\anti b)$.
This is because the enhanced $b\anti b$ partial width in
the numerator of $\br(\h\to b\anti b)$ is largely
compensated by the extra contribution to the total width
from this same channel. Thus, in comparing the MSSM
to the SM, a measurement of $G( b\anti b)$
is most sensitive to deviations of $\Gamma(\h\to  \mu\mu)$
from SM expectations. As seen numerically in 
Fig.~\ref{hltohsmratios}(e), $\Gamma(h\to\mu\mu)$ grows rapidly
at lower $\mha$ or higher $\tanb$. For small squark mixing,
a deviation in $G(b\anti b)$ from the SM value implies
almost the same percentage deviation of $\Gamma(\h\to\mu\mu)$
from its SM value.  However, when squark mixing is large, 
this equality breaks down. In general, one must separately
determine $\Gamma(\h\to\mu\mu)$ in order to probe MSSM vs. SM
differences. The procedure for this will be discussed shortly.

\begin{figure}[htbp]
\let\normalsize=\captsize   %%%% changes the font to "\small"
\begin{center}
\centerline{\psfig{file=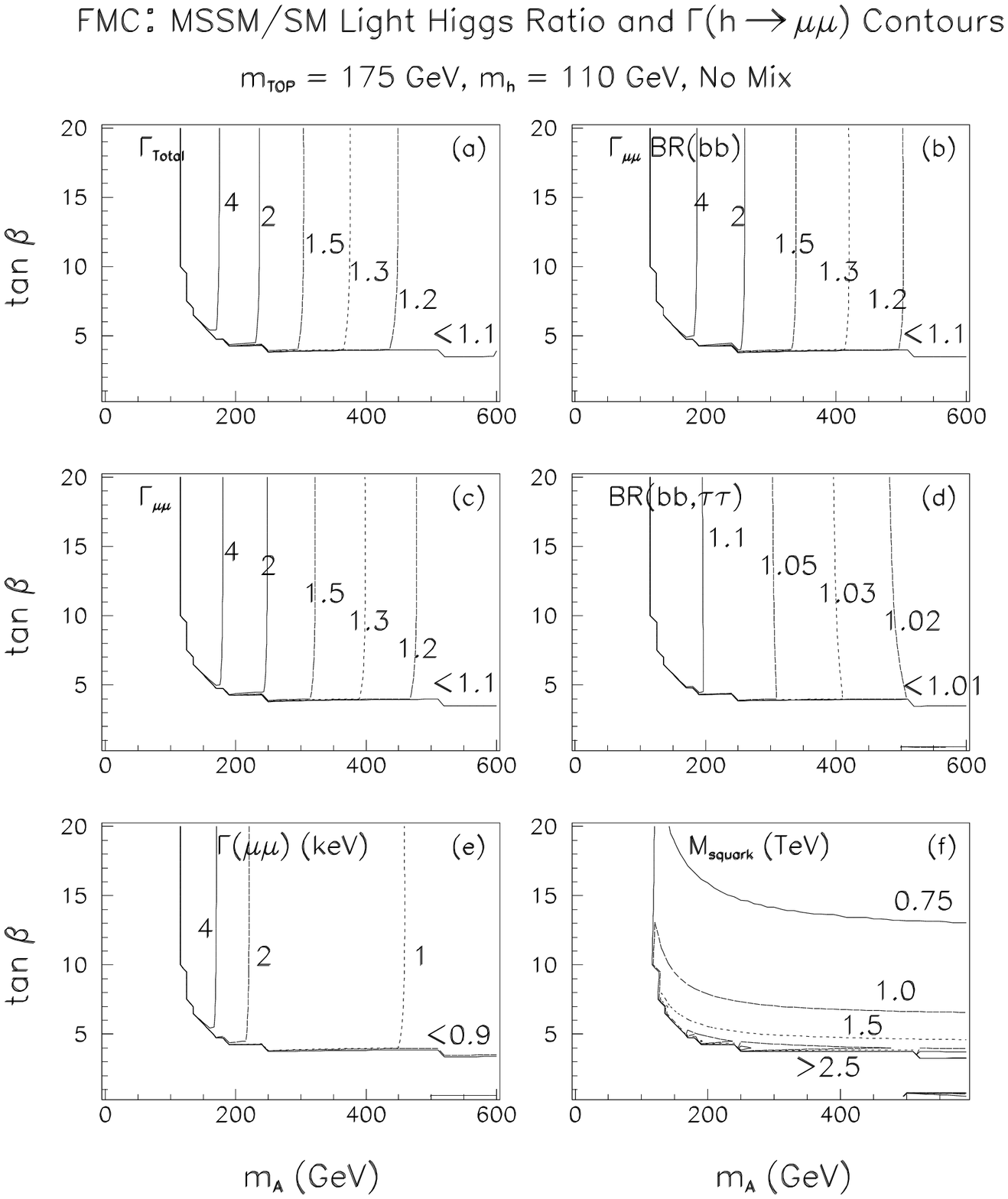,width=12.2cm}}
\begin{minipage}{12.5cm}       %%%% reduces width of caption to 12.5cm
\caption{Contours of constant MSSM/SM  ratios for $\gamh$,
$\Gamma(\h\to\mu\mu)\times \br(\h\to b\anti b)$,
$\Gamma(\h\to\mu\mu)$ and $\br(\h\to b\anti b,\tau\tau)$ in $(\mha,\tanb)$
parameter space. We have taken $\mt=175\gev$, and
we adjust $\mstop$ so as to keep a fixed value of $\mhl=110\gev$
after including two-loop/RGE-improved
radiative corrections for Higgs masses,
mixing angles and self-couplings, neglecting squark mixing.
Also shown are contours for fixed values of $\Gamma(\hl\to\mu\mu)$
in keV units, and contours for fixed values
of $\mstop$ in TeV units. This graph was obtained using the programs
developed for the work of Ref.~\protect\cite{ghinprogress}.
}
\label{hltohsmratiosfixedmhl}
\end{minipage}
\end{center}
\end{figure}

The measured value of $\mh$ provides a further constraint.
For example, suppose that a Higgs boson is observed with $\mh=110\gev$.
A fixed value for $\mh$ implies that the 
parameters which determine the radiative corrections to $\mhl$
must change as $\mha$ and $\tan\beta$ are varied.  
For example, if squark mixing is neglected, then the appropriate value of
$\mstop$ is a function of $\mha$ and $\tanb$. 
Given the assumption of no squark mixing
and the fixed value of $\mh=110\gev$, results for the same ratios
as plotted in Fig.~\ref{hltohsmratios} are given in
Fig.~\ref{hltohsmratiosfixedmhl}. Also shown are contours
of fixed $\Gamma(\hl\to \mu\mu)$ and contours of fixed $\mstop$
(as required to achieve $\mhl=110\gev$). The vertical
nature of the $\mu\mu$ ratio and partial width contours implies
that a measurement of any of these quantities could provide
a determination of $\mha$ (but would yield little information
about $\tanb$).

Contours for other mixing assumptions, can also be plotted.  The only
contours that remain essentially unaltered as the amount of squark
mixing is varied (keeping $\mh=110\gev$) are those for
the ratio $\Gamma(\hl\to\mu\mu)/\Gamma(\hsm\to\mu\mu)$
and for the $\Gamma(\hl\to\mu\mu)$ partial width itself. 
Once $\mhl\lsim 100\gev$,
even these contours show substantial variation as a function of 
the squark mixing parameters. However, it remains true that
a determination of the $\mu\mu$ partial width or partial width ratio
provides at least a rough determination of $\mha$.

In order to assess the observability of the differences between
predictions for $\gamh$, $G(b\anti b)$, and $\Gamma(\mu\mu)$
for the $\hl$ compared to the $\hsm$, we must examine
more closely the error in the experimental determination of these quantities,
and consider the theoretical uncertainties in our 
predictions for them.

\subsubsection[Interpreting a measurement of $\gamh$]{Interpreting a measurement of {\protect\boldmath$\gamh$}}

\indent\indent
Consider first the total width measurement.  Here, the experimental
error is the key issue.
%For the parameter choices of Fig.~\ref{hltohsmratios},
The $\hl$ may have a mass of order $110\gev$ in the large-$\mha$
region where it is SM-like, provided $\tanb$ is not near 1
(see Fig.~\ref{rcmasses}).
According to Fig.~\ref{masswidthlum}, $L\sim 3\fbi$
is required to measure $\gamh$ to $\pm33\%$, provided $R=0.01\%$.
A $\pm10\%$ measurement
would require $L\sim 33\fbi$ (using $\Delta\gamhsm\propto 1/\sqrt{L}$).
As seen most clearly from Fig.~\ref{hltohsmratiosfixedmhl},
this accuracy would probe MSSM/SM differences at the $3\sigma$  level
for $\mha\lsim 400\gev$ if squark mixing is small. 

Detecting a difference between the $\hl$ and $\hsm$
using $\gamh$ could prove either somewhat easier or
much more difficult than outlined above, because
the $\tanb$, $\mstop$ values and
the degree of squark mixing could very well be
different from those assumed above.
For example, if $\mhl=110\gev$, $\tanb\gsim 5$
and squark mixing is large, $\mha$ values above $400\gev$ would
be probed at the $3\sigma$ level by a 10\% measurement of $\gamh$.
On the other hand, the radiative corrections could yield a smaller
$\mhl$ value, \eg\ $\mhl\lsim 100\gev$
is quite likely if $\tanb$ is near 1 or $\mstop$ is small. 
In this range, predicted deviations from predictions
for the $\hsm$ with $\mhsm=\mhl$ are not dissimilar to those obtained
discussed above.  However,
a luminosity $L\gsim 100\fbi$ would be required for a $\pm10\%$
measurement of $\gamh$ for $80\gev\lsim\mhl\lsim 100\gev$.

Other theoretical uncertainties include:
i) extra contributions to $\gamhl$ in the MSSM model
from SUSY decay modes; ii)~the $gg$ decay width of the $\hl$ could be
altered by the presence of light colored sparticles; iii)~the $\hsm$
could have enhanced $gg$ decay width due to heavy colored
fermions (\eg\ from a fourth family).

Nonetheless, a $\mm$ collider determination of $\gamh$ will be a crucial
component in a model-independent determination of all the
properties of a SM-like $\h$, and could provide the first
circumstantial evidence for a MSSM Higgs sector prior
to direct discovery of the non-SM-like MSSM Higgs bosons.

\subsubsection[Interpreting a measurement of
$\Gamma(\h\to\mu\mu)\times \br(\h\to b\anti b)$]{Interpreting a measurement of
{\protect\boldmath$\Gamma(\h\to\mu\mu)\times \br(\h\to b\anti b)$}}

\indent\indent
How does the $\hl$--$\hsm$ discrimination power of the total width  
measurement compare to that associated with a measurement of $G(b\anti b)
\equiv \Gamma(\h\to\mu\mu)\times
\br(\h\to b\anti b)$?  Figure~\ref{smerrors}
shows that $\pm 0.4\% $ accuracy in the determination of
$G(b\anti b)$ is possible for $L=50\fbi$ and
$R=0.01\%$ in the $\mhl\sim 110$--$115\gev$
mass range predicted for $\mha\gsim 2\mz$ and larger $\tanb$ values,
assuming $\mstop\gsim 0.75\tev$ and no squark mixing.

An uncertainty
in $\br(\h\to b\anti b)$ arises from $\Gamma(\h\to b\anti b)\propto m_b^2$
due to the uncertainty in $\mb$.  Writing $\br(\h\to b\anti
b)=\Gamma_b/(\Gamma_b+\Gamma_{{\rm non}-b})$, 
the error in $\br(\h\to b\anti b)$ is given by
\begin{equation}
\Delta \br(\h\to b\anti b)={2 \Delta \mb \over \mb} \br(\h \to b\anti  b)
\br(\h \to {\rm non}-b)\;.
\label{mubproderror}
\end{equation}
Since $\br(h\to\mbox{non-}b)$ is not very large (0.1 to 0.2
in the mass range in question
for either the $\hsm$ or $\hl$), even a 10\% uncertainty in $\mb$
only leads to $\Delta \br(\h\to b\anti b)\lsim 0.05$.
Eventually $\mb$ may be known to the 5\% level, leading to $\lsim 2.5\%$
uncertainty in the branching fraction.
Comparison to Fig.~\ref{hltohsmratiosfixedmhl} 
shows that a $2.5\%$ uncertainty from $\mb$, in combination with
a still smaller statistical error,
has the potential for $\hl$--$\hsm$ discrimination at the $3\sigma$
statistical level out to large $\mha$ for $\mh=110\gev$,
if squark mixing is small.  However, as squark mixing is increased,
it turns out that the maximum $\mha$ that can potentially be probed decreases
if $\tanb$ is large.

$\br(\h\to b\anti b)$ is also subject to an
uncertainty from the total width. For example, in the MSSM
$\br(\hl\to b\anti b)$ could be smaller than the SM prediction
if $\gamhl$ is enhanced due to channels other than the $b\anti b$
channel itself (\eg\ by supersymmetric decay modes, or
a larger than expected $gg$ decay width due to loops containing
supersymmetric colored sparticle or heavy colored fermions).
Thus, a measurement of $G(b\anti b)$ alone
is not subject to unambiguous interpretation.

We note that the $L=50\fbi$
$\mm$ collider measurement of $G(b\anti b)$
is substantially more powerful than
a $L=50\fbi$ precision measurement
of $\sigma(\ee\to Z\h)\times \br(\h\to b\anti b)$ at an $\ee$ collider
\cite{dpflighthiggs}. The ratio of the
$\hl$ prediction to the $\hsm$ prediction
is essentially equal to the $\hl$ to $\hsm$ $\br(\h\to b\anti b)$
ratio and is predicted
to be within 1\% (2\%) of unity along a contour very close to
the $1.1$ ($1.2$) contour of $\Gamma(\h\to\mm)\br(\h\to b\anti b)$;
see panels (b) and (d) in Figs.~\ref{hltohsmratios}
and \ref{hltohsmratiosfixedmhl}. Since at best 5\% deviations in
$G(b\anti b)$ and $\br(\h\to b\anti b)$
can be detected at the $1\sigma$ level (after combining
a possibly small statistical error with a large theoretical error), we  
see from the 1.05 ratio contour for $\br(\h\to b\anti b)$
in Figs.~\ref{hltohsmratios} and \ref{hltohsmratiosfixedmhl} that 
the $\sigma(Z\h)\br(\h\to b\anti b)$ and $\br(\h\to b\anti b)$ ratios,
that can be determined experimentally at an $\ee$ collider,
only probe as far as $\mha\lsim 250$--300~GeV
at the $1\sigma$ significance level, with even less reach at the $3\sigma$ 
level.

We must again caution that if $\mh$ is close
to $\mz$,  there could be substantially worse experimental uncertainty
in the $G(b\anti b)$ measurement than taken above.
Pre-knowledge of $\mh$ is necessary to determine the level of  
precision that could be expected for this measurement.

\subsubsection{Combining measurements}

\indent\indent
We now discuss how the independent measurements of $\gamh$
and $G(b\anti b)$ can be combined
with one another and other experimental inputs to
provide a model-independent determination of
the properties of the $\h$. We consider three complementary  
approaches.

\smallskip\noindent (1)
A model-independent determination of $\Gamma(\h\to\mu\mu)$
can be made by combining the $s$-channel $\mm$ collider measurement of
$G(b\anti b)$ with the value
of $\br(\h\to b\anti b)$ measured in the $Z\h$ mode at
an $\ee$ collider or the $\mm$ collider.  With $L=50\fbi$
of luminosity, $\br(\h\to b\anti b)$ can potentially be
measured to $\pm 7\%$ \cite{dpflighthiggs}. From our earlier
discussion, the error on $G(b\anti b)$
will be much smaller than this if $\mh\gsim 100\gev$, and  
$\Gamma(\h\to\mu\mu)$
would be determined to roughly $\pm 8$--$10\%$.
Figures~\ref{hltohsmratios} and \ref{hltohsmratiosfixedmhl} show that this
procedure would probe the $\hl$ versus $\hsm$ differences
at the $3\sigma$ level out to $\mha\sim 400\gev$ if $\tanb$
is not close to 1 (see the 1.3 ratio contour in the figures).
This is a far superior reach to that possible at the $3\sigma$
level at either the LHC, NLC and/or $\gam\gam$ collider.
Further, we note that the $\mu\mu$ partial width
at fixed $\mh\gsim100\gev$ is relatively independent of
the squark mixing scenario and provides a rather precise determination
of $\mha$ \cite{dpflighthiggs}.

\smallskip\noindent (2)
A model-independent determination of $\Gamma(\h\to b\anti b)$
is possible by computing $\gamh \br(\h\to b\anti b)$ using
the value of $\gamh$ measured at the $\mm$ collider and
the value of $\br(\h\to b\anti b)$ measured in the $Z\h$ mode.
Taking 10\% accuracy for the former and 7\% accuracy for the latter,
we see that the error on $\Gamma(\h\to b\anti b)$ would be of order 12\%.
The ratio contours for $\Gamma(\h\to b\anti b)$ are the same
as the ratio contours for $\Gamma(\h\to \mu\mu)$.
Thus, ignoring systematics, this measurement could also probe
out to $\mha\gsim 400\gev$ at the $3\sigma$ level if $\mh\sim 110\gev$,
see Fig.~\ref{hltohsmratiosfixedmhl}. However, the $2\Delta \mb/\mb$
systematic uncertainty in the partial width is also of order
10\% for 5\% uncertainty in $\mb$, implying a total statistical
plus theoretical error of order 16\%.  This would restrict
$3\sigma$ sensitivity to $\hl$ vs.\ $\hsm$ differences to
$\mha\lsim  300\gev$.

\smallskip\noindent (3)
A third approach uses only the $\mm$ collider measurements. We note that
\begin{equation}
W\equiv\Gamma(\h\to\mu\mu)\Gamma(\h\to b\anti b)=
[\gamh]\times [\Gamma(\h\to\mu\mu) \br(\h\to b\anti b)]\, .
\label{productform}
\end{equation}
In the MSSM (or any other type-II two-Higgs-doublet model) the $\mu\mu$
and $b\anti b$ squared couplings have exactly the same factor, call it $f$,
multiplying the square of the SM coupling strength. Thus,
\begin{equation}
W=\Gamma(\h\to\mu\mu)\Gamma(\h\to b\anti b)\propto f^2\left({g\over
2\mw}\right)^4 m_\mu^2 \mb^2\, .
\end{equation}
Following our earlier discussion, in the MSSM $f^2$ would be $(1.3)^2\sim 1.7$
along the 1.3 ratio contours for $\Gamma(\h\to\mu\mu)$
in Figs.~\ref{hltohsmratios} and \ref{hltohsmratiosfixedmhl}. 
For $\mh\gsim 100\gev$, experimental
errors in $W$ of Eq.~(\ref{productform}) would be dominated
by the $\pm10\%$ error on $\gamh$. The dominant systematic error
would be that from not knowing the value of $\mb$:
$\Delta W/W=2\Delta \mb/\mb$.  Thus, a combined statistical
and theoretical $1\sigma$ error for $W$ below 20\% is entirely
possible for $\mh\gsim 100\gev$, in which case deviations in $f^2$
from unity can be probed at the $3\sigma$ level for $\mha$ values
at least as large as $\mha\sim 400\gev$. Since 
both $\Gamma(\hl\to\mu\mu)$ and $\Gamma(\hl\to b\anti b)$
are relatively independent of the squark mixing scenario for fixed
$\mhl$ and fixed $\mha$, a fairly reliable value of $\mha$ would
result from the determination of $f^2$.

By combining the strategies just discussed, one
can do even better.  Thus, a $\mm$ collider has great
promise for allowing us to measure the crucial $b\anti b$ and
$\mupmum$ couplings of a SM-like $\h$, provided $\mh$
is not within $10\gev$ of $\mz$ (nor $\gsim 2\mw$) and that  
$m_A\lsim400$~GeV. In particular,
for such masses we can distinguish the $\hl$ from the $\hsm$
in a model-independent fashion out to larger $\mha$
than at any other accelerator or combination of accelerators.

\subsubsection[The $ W\wstar$
and $ Z\zstar$ channels]{The {\protect\boldmath$ W\wstar$}
and {\protect\boldmath$ Z\zstar$ channels}}

\indent\indent
Precision measurements of $\Gamma(\h\to\mm)\br(\h\to X)$ are also
possible for $X=W\wstar$ and, to a lesser extent, $Z\zstar$,
see Fig.~\ref{smerrors}. Once again,
$\Gamma(\h\to \mm)$ can be determined in a model-independent
fashion using $\br(\h\to X)$ measured in the $Z\h$ mode, and
$\Gamma(\h\to X)$ can be computed in a model-independent
fashion as the product $\br(\h\to X)\gamh$. We will
not go through the error analysis in detail for these cases,
but clearly determination of both the $WW$ and $ZZ$ couplings
will be possible at a reasonable statistical level.
Unfortunately, the $\hl WW,\hl ZZ$ couplings are very close to the SM
values for $\mha\gsim 2\mw$ and the expected statistical errors
would not allow $\hl$ vs.\ $\hsm$ discrimination.

\section{Non-SM-like Higgs bosons in the MSSM}

\begin{figure}[htbp]
\let\normalsize=\captsize   %%%% changes the font to "\small"
\begin{center}
\centerline{\psfig{file=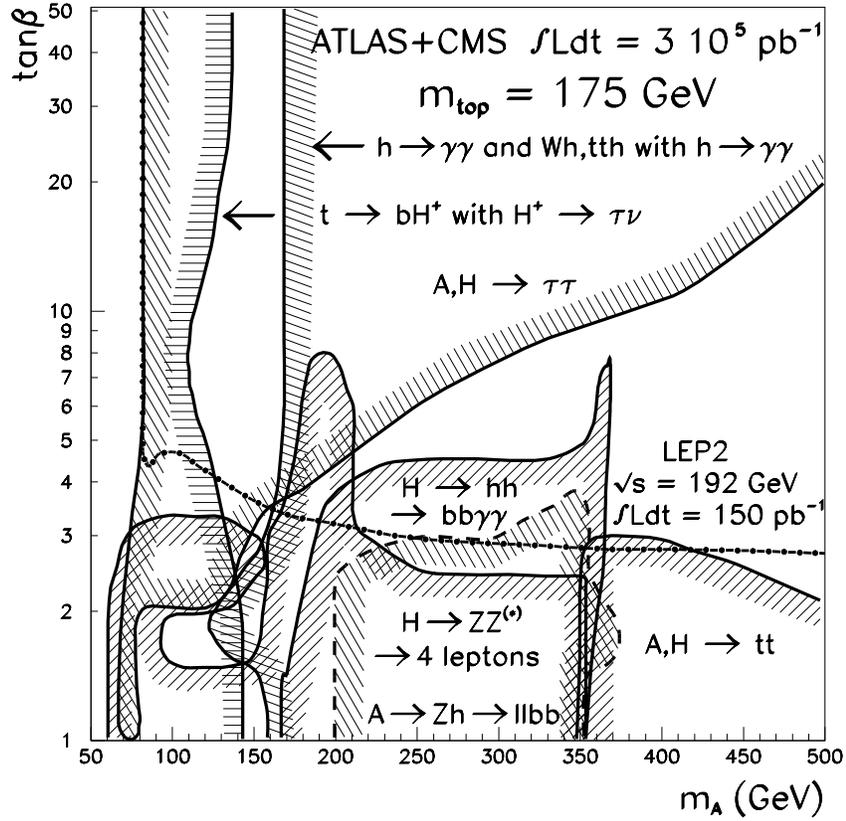,width=12.2cm}}
%%\centerline{\psfig{file=mssm_hi_froid.ps,width=12.2cm}}
\begin{minipage}{12.5cm}       %%%% reduces width of caption to 12.5cm
\caption{MSSM Higgs discovery contours ($5\sigma$)
in the parameter space of the
minimal supersymmetric model for ATLAS+CMS at the LHC: $L=300\fbi$  
per detector.  Figure
from Ref.~\protect\cite{latestplots}. Two-loop/RGE-improved
radiative corrections are included for
$\mhl$ and $\mhh$ assuming $\mstop=1\tev$ and no squark mixing.}
\label{lhcregionshilum}
\end{minipage}
\end{center}
\end{figure}

\indent\indent
In what follows, we shall demonstrate that it is possible
to observe the $\hh$ and $\ha$
in $s$-channel Higgs production for $\mha\sim\mhh> \rts/2$
over much of $(\mha,\tanb)$ parameter space.  It is this fact 
that again sets the $\mm$ collider apart from other machines.
\begin{enumerate}
\item
The LHC can only detect
the $\hh$ and $\ha$ for masses above $200$--$250\gev$ if $\tanb$
is either large or $\lsim 3$--5; a wedge of unobservability
develops beginning at $\mha\gsim 200\gev$, covering an increasingly
wide range of $\tanb$ as $\mha$ increases \cite{oldwork}. This is illustrated
in Fig.~\ref{lhcregionshilum} from Ref.~\cite{latestplots}.  

\item
At an $\ee$ collider, $Z^\star\to Z\ha,Z\hh$ production will
be negligible when $\mha > 2 \mz$.

\item
$\ee\to Z^\star\to\ha\hh$ could easily be kinematically disallowed,
especially for $\ee$ machine energies in the $\rts \sim 500\gev$ range ---
GUT scenarios often give $\mha\sim\mhh\agt 300\gev$.

\item
If an $\ee$ collider is run in the photon-photon collider mode,
discovery of  the $\hh$ and $\ha$ in the $\mha,\mhh\gsim 200\gev$ region
via $\gam\gam\to \ha,\hh$ requires extremely high luminosity ($\gsim 200\fbi$)
\cite{ghgamgam}.

\item
$s$-channel production of the $\ha$ and $\hh$
will not be significant in $\ee$
collisions due to the small size of the electron mass.
\end{enumerate}

\noindent
A $\mm$ collider can overcome the limitations 3 and 5 of an
$\ee$ collider, though
not simultaneously.  If the $\mm$ collider is run at energies
of $\sqrt s = \mha\sim\mhh$, then we shall find that $s$-channel
production will allow discovery of the $\ha$ and $\hh$
if $\tanb\gsim 3-4$. Here, the kinematical Higgs mass reach is limited only
by the maximum $\rts$ of the machine.
Alternatively, the $\mm$ collider can be designed to have $\rts \sim 4\tev$
in which case $\mha\sim\mhh$ values up to nearly $2\tev$ can be probed
via the $Z^\star\to \ha\hh$ process, a mass range that encompasses
all natural GUT scenarios.
We focus in this report on $s$-channel production and detection.
In our analysis, we will assume
that more or less full luminosity can be maintained for all $\rts$
values over the mass range of interest (using multiple storage rings,
as discussed in the introduction).

\subsection[MSSM Higgs bosons in the $s$-channel:
$\protect\rts=\mh$]{MSSM Higgs bosons in the {\protect\boldmath$s$}-channel:
{\protect\boldmath$\protect\rts=\mh$} }

\indent\indent
Here we investigate the potential of a $\mm$ collider for probing
those Higgs bosons whose couplings to $ZZ,WW$ are either suppressed
or absent at tree-level --- that is the $\ha$, the $\hh$ (at larger  
$\mha$),  or the $\hl$ (at small $\mha$).
The $W\wstarp$ and $Z\zstarp$ final states in $s$-channel production
are then not relevant.
We consider first the
$b\anti b$ and  $t\anti t$ decay modes, although we shall later
demonstrate that
the relatively background free $\hh\rta \hl\hl~{\rm or}~\ha\ha\rta  
b\anti b b\anti b$, $\hh\to Z\ha\to Zb\anti b$
and $\ha\rta Z\hl\rta Zb\anti b$ modes might also be useful.

\begin{figure}[htbp]
\let\normalsize=\captsize   %%%% changes the font to "\small"
\begin{center}
\centerline{\psfig{file=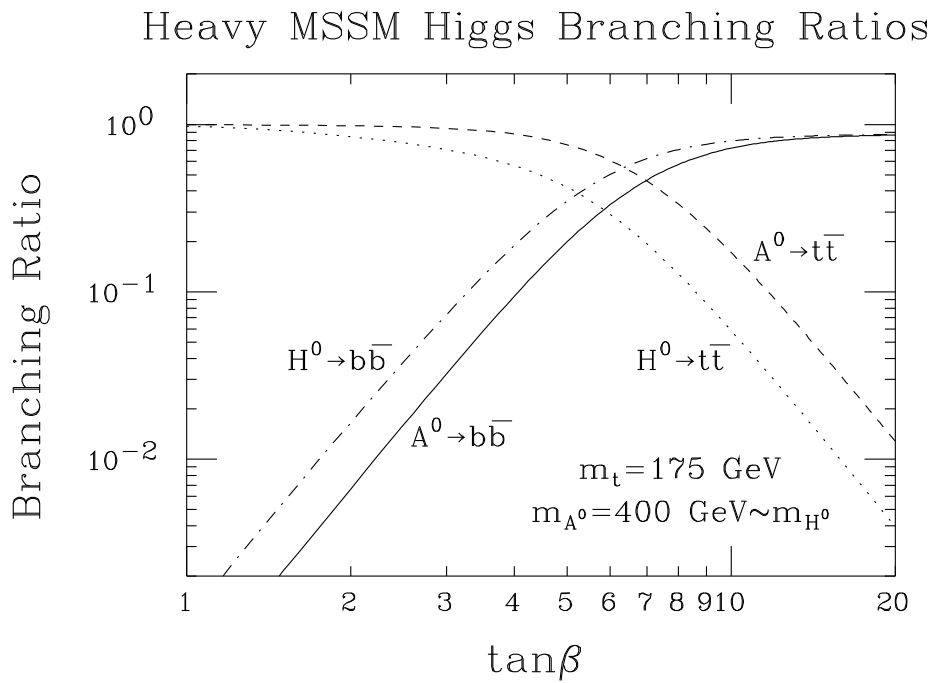,width=12.2cm}}
\begin{minipage}{12.5cm}       %%%% reduces width of caption to 12.5cm
\caption{Dependence of the $b\anti b$ and $t\anti t$
branching fractions of the heavy  supersymmetric Higgs bosons on $\tanb$.
Results are for $\mt=175\gev$ and include two-loop/RGE-improved
radiative corrections to Higgs masses, mixing angles, and self-couplings,
computed with $\mstop=1\tev$ neglecting squark mixing.}
\label{hhhabrs}
\end{minipage}
\end{center}
\end{figure}

Figure~\ref{hhhabrs}
shows the dominant branching fractions to $b\anti b$ and $t\anti t$ of
Higgs bosons of mass $\mha=400\gev\approx \mhh$ versus $\tanb$, taking
$\mt=170$~GeV. The $b\anti b$ decay mode is dominant for $\tanb>5$,
which is the region where observable signal rates are most easily  
obtained. From the figure we see that $\br(\hh,\ha\rta b\anti b)$ grows
rapidly with increasing $\tanb$ for
$\tanb\alt 5$, while $\br(\hh,\ha\rta t\anti t)$ falls slowly.

\subsubsection{Resolution compared to Higgs widths}
\indent

\begin{figure}[htbp]
\let\normalsize=\captsize   %%%% changes the font to "\small"
\begin{center}
\centerline{\psfig{file=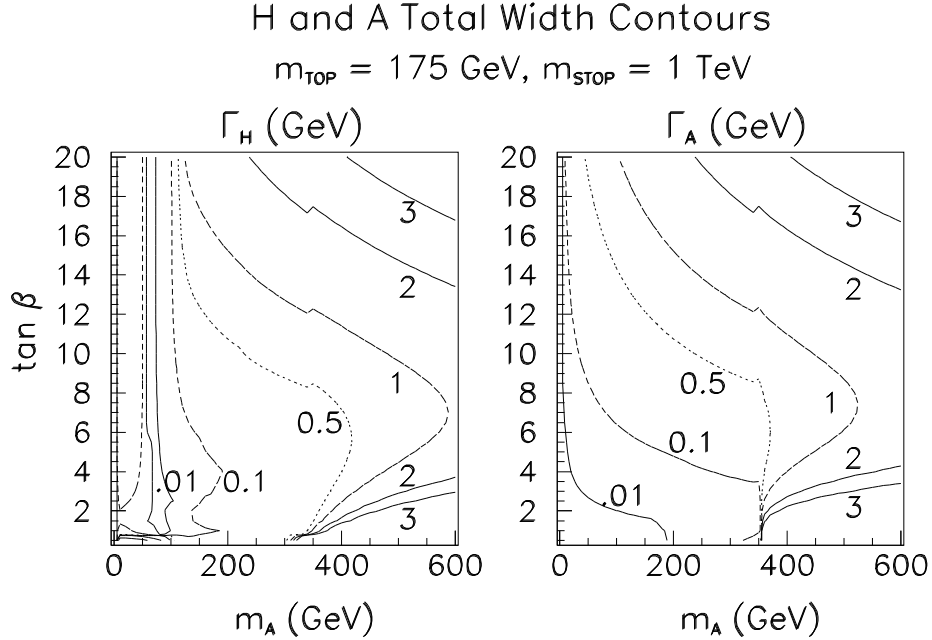,width=12.2cm}}
\begin{minipage}{12.5cm}       %%%% reduces width of caption to 12.5cm
\caption{Contours of $\hh$ and $\ha$ total widths (in GeV)
in the $(\mha,\tanb)$ parameter space. We have taken $\mt=175\gev$
and included two-loop/RGE-improved radiative corrections using $\mstop=1\tev$
and neglecting squark mixing. SUSY decay channels are assumed to
be absent.}
\label{hhhawidths}
\end{minipage}
\end{center}
\end{figure}

The first critical question is how
the resolution in $\rts$ compares to the $\hh$ and $\ha$
total widths. The calculated $\hh$ and $\ha$
widths are shown in Fig.~\ref{hwidths} versus $\mhh,\mha$ for  
$\tanb=2$ and $20$.
In Fig.~\ref{hhhawidths} we give contours of constant total widths for
the $\hh$ and $\ha$ in the $(\mha,\tanb)$ parameter space.
For $\mha, \mhh\lsim 500\gev$, the $\hh$ and
$\ha$ are typically moderately
narrow resonances ($\Gamma_{\hh,\ha}\sim 0.1$ to 6~GeV),
unless $\tanb$ is larger than 20.
For a machine energy resolution of $R=0.06\%$, and Higgs masses
in the 100 GeV to 1 TeV range, the resolution $\sigrts$ in $\rts$  
will range from roughly 0.04 GeV to to 0.4 GeV, see Eq.~(\ref{resolution}).  
Thus, Figs.~\ref{hwidths} and~\ref{hhhawidths} indicate that the $\hh$ and  
$\ha$ widths are likely to be somewhat
larger than this resolution in $\rts$. For $R=0.01\%$, this is
always the dominant situation.

When the $\rts$ resolution is smaller than the Higgs width,
then Eq.~(\ref{basicsigma}), with $\rts\sim \mh$ shows
that the cross section will behave as the product of the
$\mu\mu$ and final state branching fractions.  For low to moderate  
$\tanb$
values, $\br(\hh,\ha\to \mu\mu)$ and $\br(\hh,\ha\to b\anti b)$
grow with increasing $\tanb$, while $\br(\hh,\ha\to t\anti t)$ falls  
slowly.
Thus, the number of $\hh$ and $\ha$ events in both the $b\anti b$
and $t\anti t$ channels increases with increasing $\tanb$.
It is this growth with $\tanb$ that makes $\hh,\ha$ discovery
possible for relatively modest values of $\tanb$ larger than 1.
For higher $\tanb$ values, the $\mu\mu$
and $b\anti b$ branching fractions asymptote to constant values,
while that for $t\anti t$ falls as $1/(\tanb)^4$.  Thus,
observability in the $t\anti t$ channel does not survive to large  
$\tanb$
values.

\begin{figure}[htbp]
\let\normalsize=\captsize   %%%% changes the font to "\small"
\begin{center}
\centerline{\psfig{file=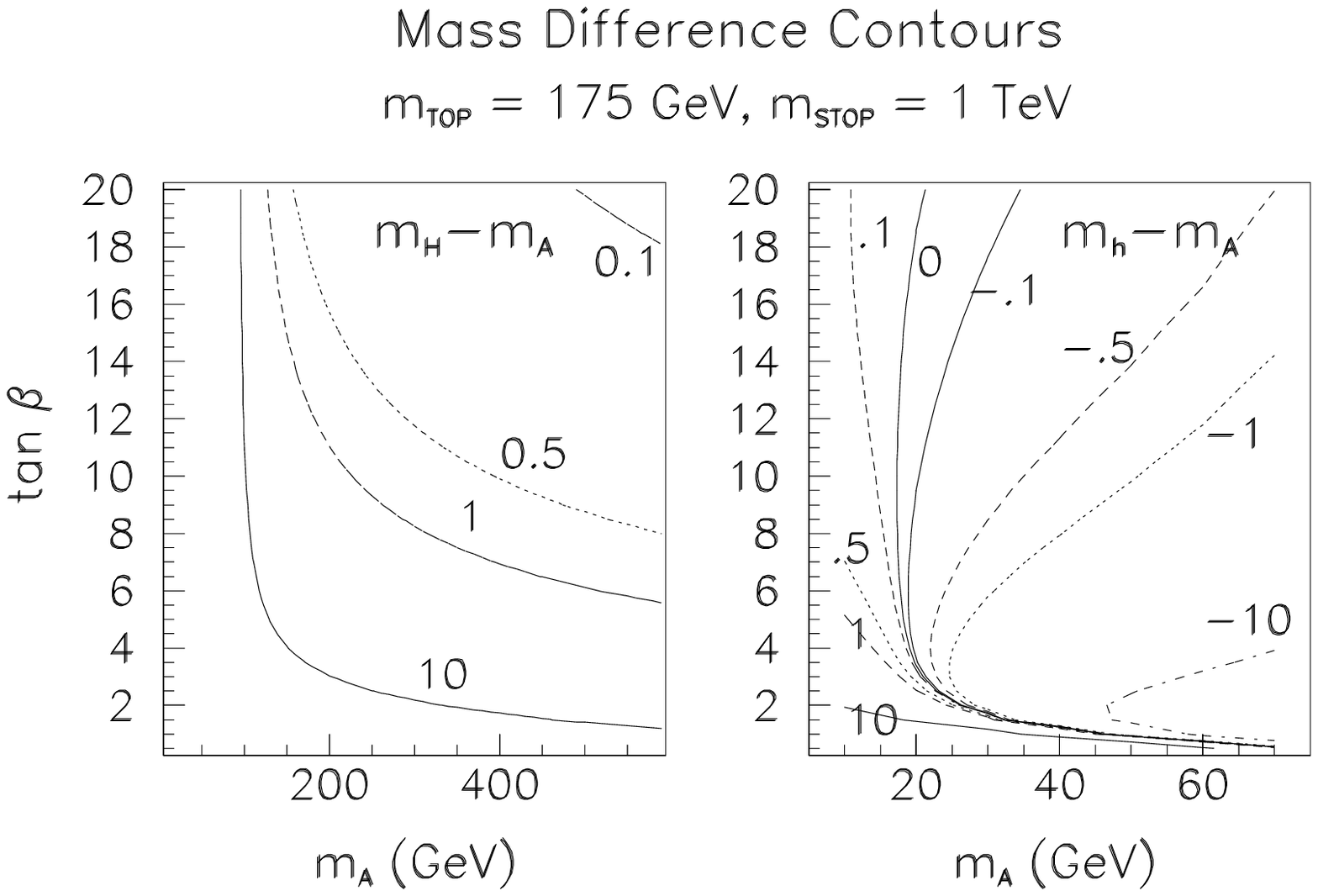,width=12.2cm}}
\begin{minipage}{12.5cm}       %%%% reduces width of caption to 12.5cm
\caption{Contours of $\mhh-\mha$ (in GeV)
in the $(\mha,\tanb)$ parameter space. Two-loop/RGE-improved
radiative corrections are included taking $\mt=175\gev$, $\mstop=1\tev$,
and neglecting squark mixing.}
\label{delmhiggs}
\end{minipage}
\end{center}
\end{figure}

\subsubsection{Overlapping Higgs resonances}
\indent

\begin{figure}[htbp]
\let\normalsize=\captsize   %%%% changes the font to "\small"
\begin{center}
\centerline{\psfig{file=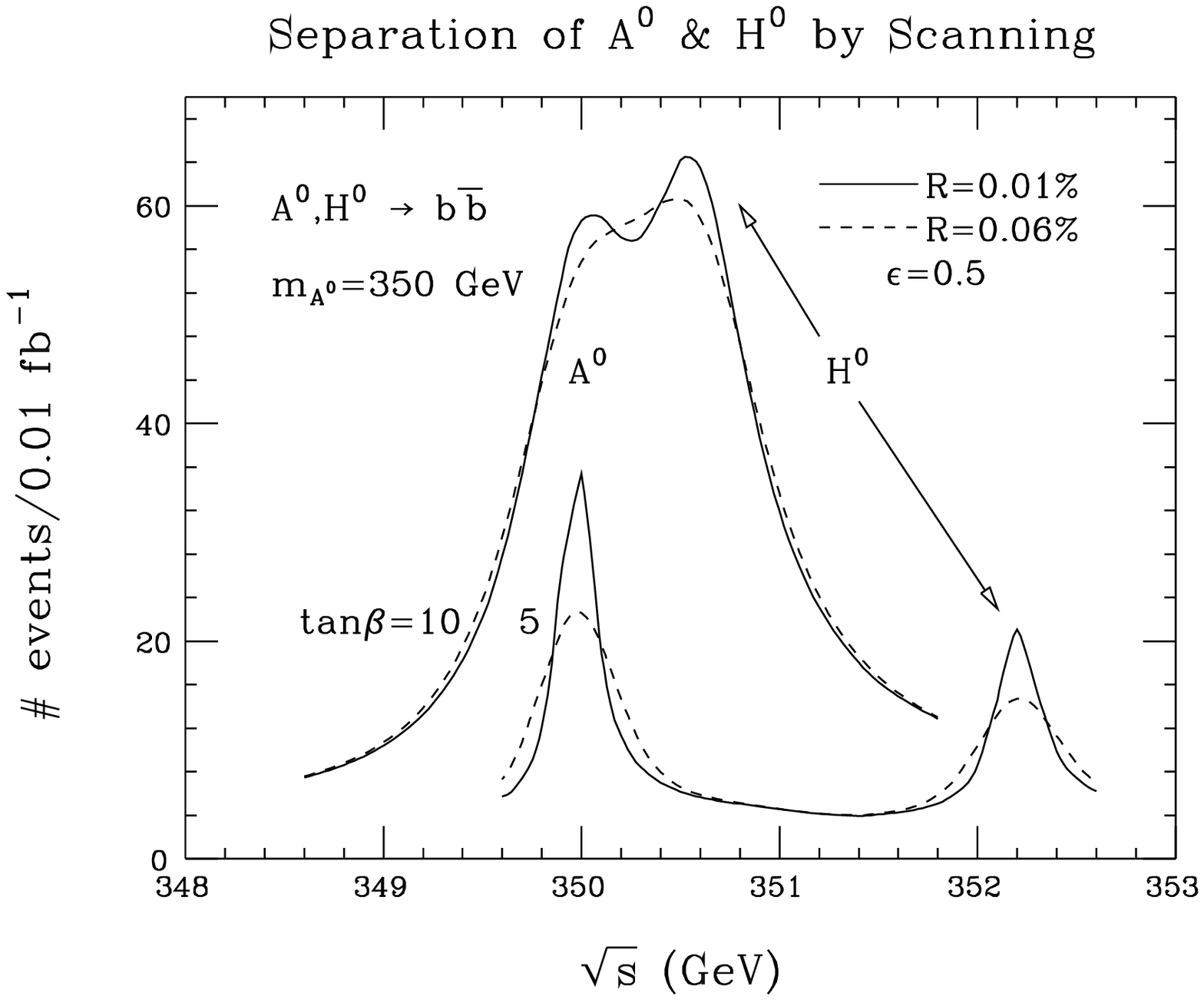,width=12.2cm}}
\begin{minipage}{12.5cm}       %%%% reduces width of caption to 12.5cm
\caption{Plot of $b\anti b$ final state
event rate as a function of $\protect\rts$ for $\mha=350\gev$,
in the cases $\tanb=5$ and 10, resulting from 
the $\hh,\ha$ resonances and the $b\anti b$ continuum background.
We have taken $L=0.01\fbi$ (at any given $\protect\rts$), 
$\eps=0.5$, $\mt=175\gev$, and
included two-loop/RGE-improved radiative corrections 
to Higgs masses, mixing angles and self-couplings using $\mstop=1\tev$
and neglecting squark mixing. SUSY decays are assumed to be absent.
Curves are given for two resolution choices: $R=0.01\%$ and $R=0.06\%$}
\label{hhhasusyrtsscan}
\end{minipage}
\end{center}
\end{figure}

The Higgs widths are a factor in the observability of a signal
in that approximate Higgs mass degeneracies are not unlikely. For larger
$\mha$, $\mha\sim\mhh$,
while at smaller $\mha$ values, $\mhl\sim\mha$ at larger $\tanb$,
as illustrated in Fig.~\ref{delmhiggs}, where the plotted mass
difference should be compared to the Higgs widths in  
Figs.~\ref{hwidths} and \ref{hhhawidths}. 
At large $\mha$ and $\tanb$,
there can be significant overlap of the $\ha$ and $\hh$ resonances.
To illustrate the possibilities,
we show in Fig.~\ref{hhhasusyrtsscan} the event rate in the $b\anti b$
channel as a function of $\rts$ (assuming $L=0.01\fbi$
and event detection/isolation efficiency $\epsilon=0.5$)
taking $\mha=350\gev$ in the cases $\tanb=5$ and $10$.
Continuum $b \anti b$ background is included. Results are plotted
for the two different resolutions, $R=0.01\%$ and $R=0.06\%$.
For $R=0.01\%$, at $\tanb=5$
the resonances are clearly separated and quite narrow, whereas at
$\tanb=10$ the resonances have become 
much broader  and much more degenerate, resulting in
substantial overlap; but, distinct resonance peaks are still visible.
For $R=0.06\%$, at $\tanb=5$ the resonances are still separated, but
have been somewhat smeared out, while at $\tanb=10$ the $\hh$ and $\ha$
peaks are no longer separately visible.  The $R=0.06\%$ smearing
does not greatly affect the observation of a signal, but would
clearly make separation of the $\hh$ and $\ha$ peaks and precise
determination of their individual widths much more difficult.

In the following section,
we perform our signal calculations by centering $\rts$ on  
$\mha$, but including any $\hh$ signal tail, and vice versa.
At small $\mha$, there is generally only small overlap between
the $\ha$ and $\hl$ since their widths are small, but
we follow a similar procedure there.
We also mainly employ
the optimistic $R=0.01\%$ resolution that is highly
preferred for a SM-like Higgs boson. Since the MSSM Higgs bosons
do not have especially small widths, results for $R=0.06\%$ are  
generally quite similar.

\subsubsection[Observability for $\hl, \; \hh$
and $\ha$]{Observability for {\protect\boldmath$\hl, \; \hh$} 
and {\protect\boldmath$\ha$}}
\indent

\begin{figure}[htbp]
\let\normalsize=\captsize   %%%% changes the font to "\small"
\begin{center}
\centerline{\psfig{file=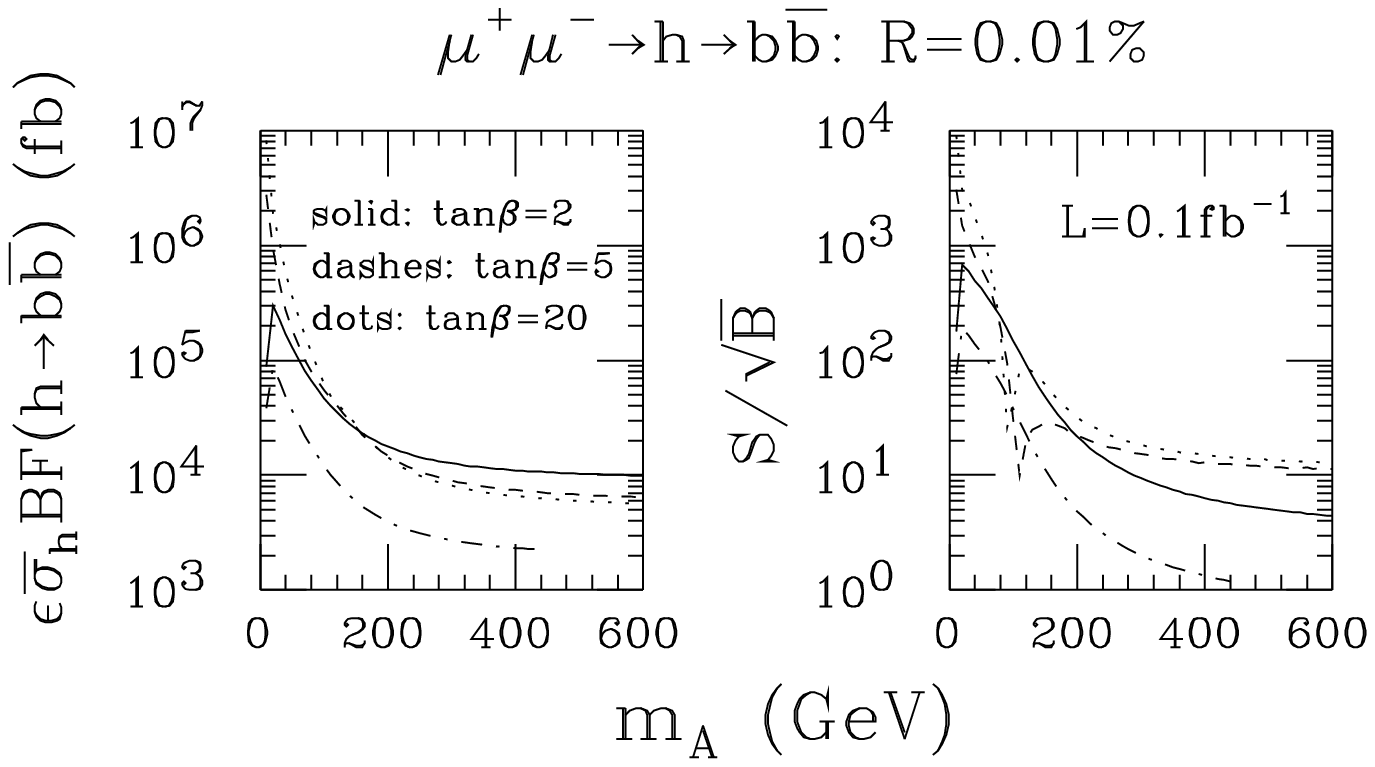,width=12.2cm}}
\begin{minipage}{12.5cm}       %%%% reduces width of caption to 12.5cm
\caption{Plot of
$\eps\sighlbar \br(\hl\to b\anti b)$ versus $\mha$ for $\tanb=2$, 5  
and 20. Also shown is the corresponding
$S/\protect\sqrt B$ for $L=0.1\fbi$.
We have taken $R=0.01\%$, $\eps=0.5$, $\mt=175\gev$, and
included two-loop/RGE-improved radiative corrections 
to Higgs masses, mixing angles and self-couplings using $\mstop=1\tev$
and neglecting squark mixing. SUSY decays are assumed to be absent
in computing $\br$. Also shown as the dot-dashed curve
are the $R=0.06\%$ results at $\tanb=2$ in the $b\anti b$ channel.}
\label{susytanbsliceshl}
\end{minipage}
\end{center}
\end{figure}

\begin{figure}[htbp]
\let\normalsize=\captsize   %%%% changes the font to "\small"
\begin{center}
\centerline{\psfig{file=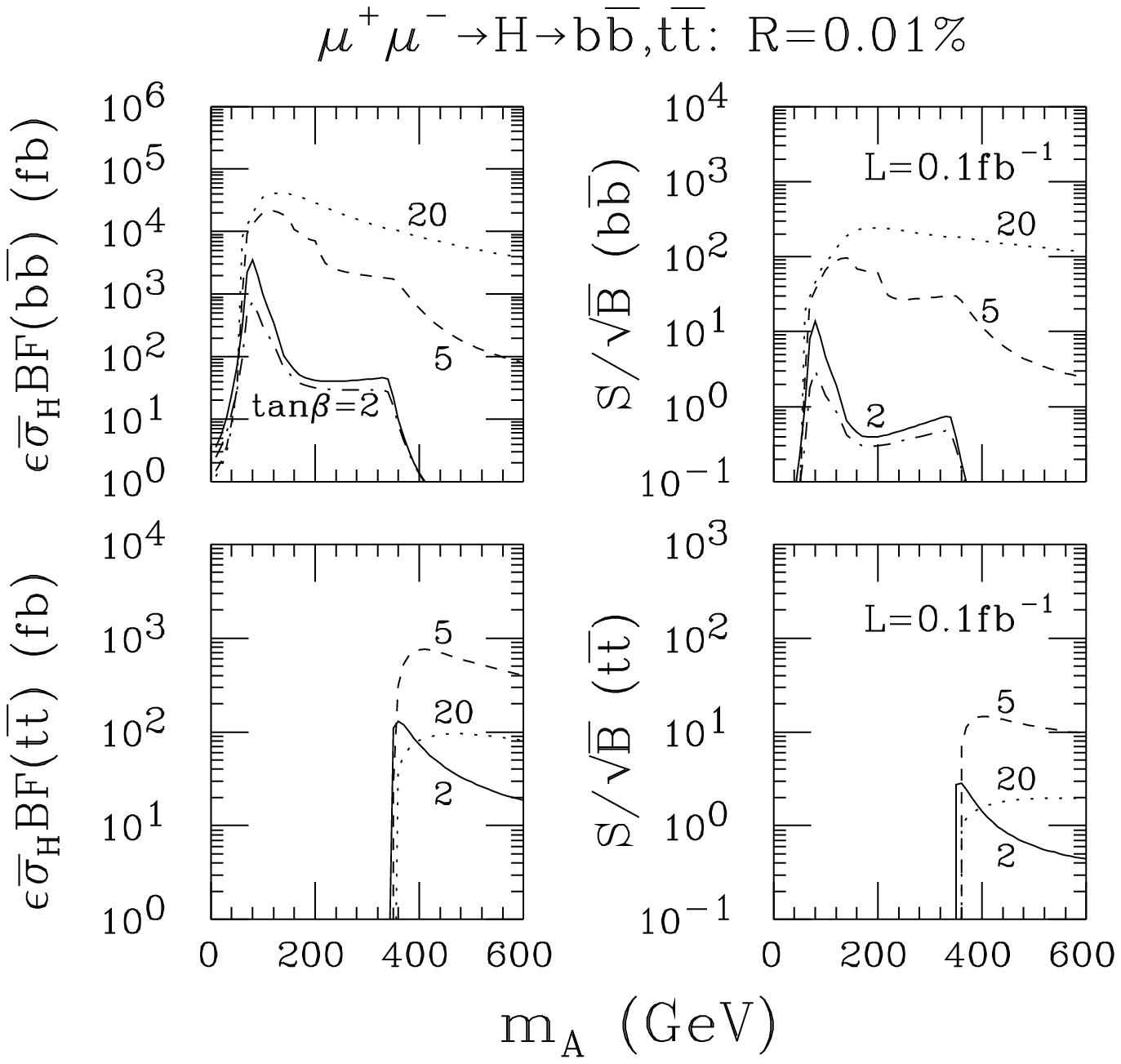,width=12.2cm}}
\begin{minipage}{12.5cm}       %%%% reduces width of caption to 12.5cm
\caption{Plot of $\eps\sighhbar \br(\hh\to b\anti b,t\anti t)$
versus $\mha$ for $\tanb=2$, 5 and 20. Also shown are the  
corresponding $S/\protect\sqrt B$ values for $L=0.1\fbi$.
The inputs are specified in the caption of
Fig.~\protect\ref{susytanbsliceshl}. Also shown as the dot-dashed curve
are the $R=0.06\%$ results at $\tanb=2$ in the $b\anti b$ channel.}
\label{susytanbsliceshh}
\end{minipage}
\end{center}
\end{figure}

\begin{figure}[htbp]
\let\normalsize=\captsize   %%%% changes the font to "\small"
\begin{center}
\centerline{\psfig{file=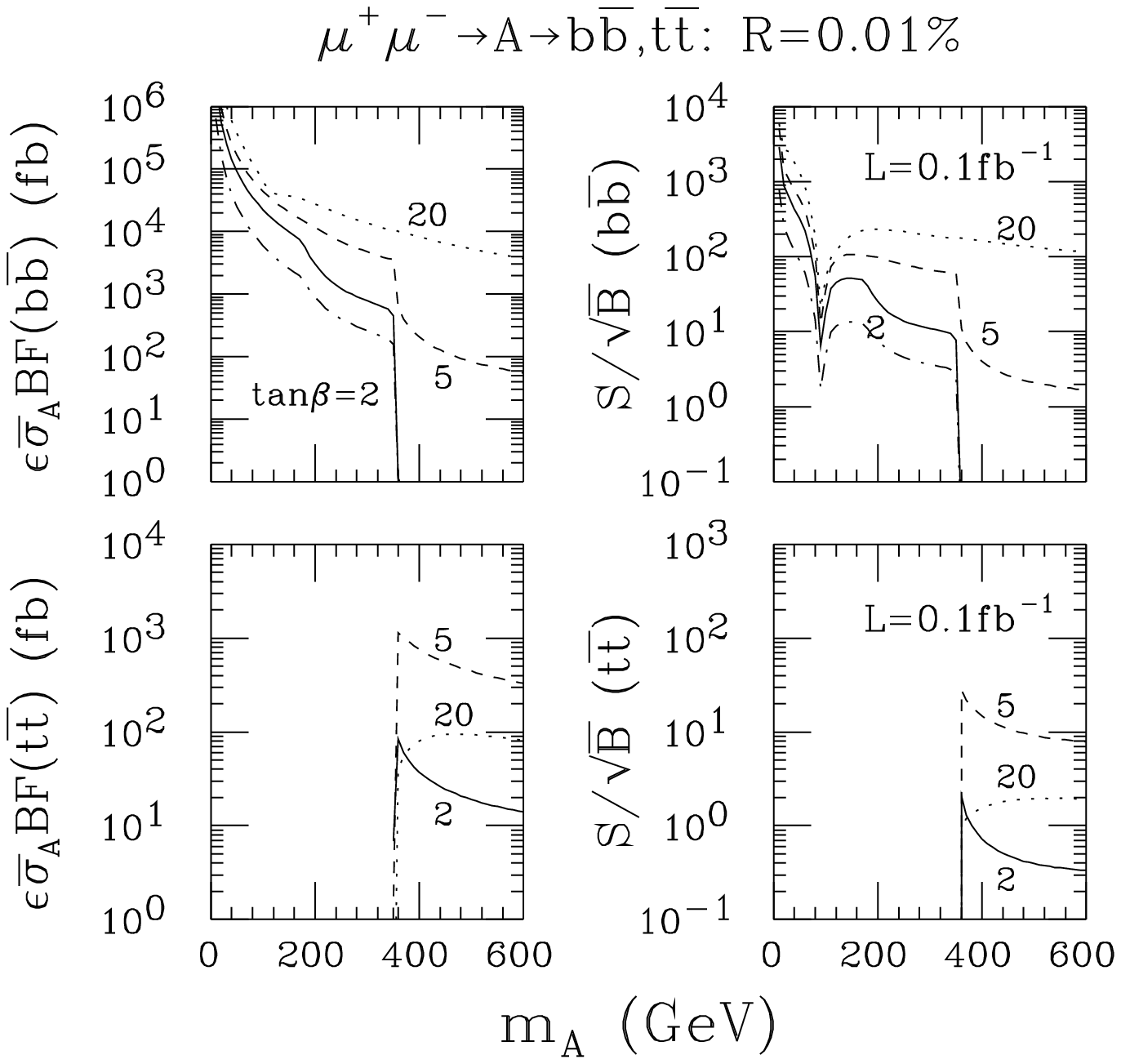,width=12.2cm}}
\begin{minipage}{12.5cm}       %%%% reduces width of caption to 12.5cm
\caption{Plot of $\eps\sighabar \br(\ha\to b\anti b,t\anti t)$ versus
$\mha$ for $\tanb=2$, 5 and 20. Also shown are the corresponding
$S/\protect\sqrt B$ values for $L=0.1\fbi$.
The inputs are specified in the caption of
Fig.~\protect\ref{susytanbsliceshl}.
Also shown as the dot-dashed curve
are the $R=0.06\%$ results at $\tanb=2$ in the $b\anti b$ channel.}
\label{susytanbslicesha}
\end{minipage}
\end{center}
\end{figure}

We first consider fixed $\tanb$ values of $2$, $5$, and $20$,
and compute $\eps\sighbar \br(\h\to b\anti b,t\anti t)$
for $\h=\hl,\hh,\ha$ as a function of $\mha$. (The corresponding
$\hl$ and $\hh$ masses can be found in Fig.~\ref{rcmasses}.)
Our results for $R=0.01\%$
appear in Figs.~\ref{susytanbsliceshl}, \ref{susytanbsliceshh},
and \ref{susytanbslicesha}.  Also shown in these figures are
the corresponding $S/\sqrt B$ values assuming an integrated luminosity
of $L=0.1\fbi$; results for other $L$ possibilities are easily
obtained by using $S/\sqrt B \propto 1/\sqrt L$.
These figures also include (dot-dashed) curves
for $R=0.06\%$ in the $b\anti b$ channel at $\tanb=2$.

Figure~\ref{susytanbsliceshl} shows that
the $\hl$ can be detected at the $5\sigma$ statistical level
with just $L=0.1\fbi$ for essentially all of parameter space,
if $R=0.01\%$.
Only for $\tanb\lsim 2$ is $\mhl$ sufficiently near
$\mz$ at large $\mha$ (for which its $\mm$ coupling is not enhanced)
that more luminosity may be required.  At low $\mha$, the
$\hl$ is not SM-like and has highly enhanced $\mm$ and $b\anti b$
couplings.  It is also no longer extremely narrow, and
is produced with a very high rate implying that high statistics
studies of its properties would be possible.
The $R=0.06\%$ $\tanb=2$ curve illustrates the large loss in  
observability that
occurs for non-optimal resolution when the $\hl$
is SM-like at large $\mha$ and has a very small width.

Results for $\eps\sighbar \br(\h\to b\anti b,t\anti t)$ for $\h=\hh$
and $\h=\ha$ are displayed in Figs.~\ref{susytanbsliceshh}
and \ref{susytanbslicesha}, respectively, along with the  
corresponding $L=0.1\fbi$ $S/\sqrt B$ values. For a luminosity 
of $L=0.01\fbi$, the $S/\sqrt B$ values of the figures
should be reduced by a factor of 0.32. For $L=0.3$, multiply by 1.7.
This range of luminosities will be that which arises when
we consider searching for the $\hh$ and $\ha$ by scanning in $\rts$.
The dot-dashed curves illustrate the fact that $R=0.06\%$
resolution does not cause a large loss in observability
relative to $R=0.01\%$ in the case of the $\ha$ and, especially, the  
$\hh$; the largest effect is for the $\tanb=2$
case in the $b\anti b$ channel. For $\tanb=5$ and 20, and
for all $t\anti t$ curves, the results for $R=0.06\%$ are virtually
indistinguishable from those for $R=0.01\%$.

\begin{figure}[htbp]
\let\normalsize=\captsize   %%%% changes the font to "\small"
\begin{center}
\centerline{\psfig{file=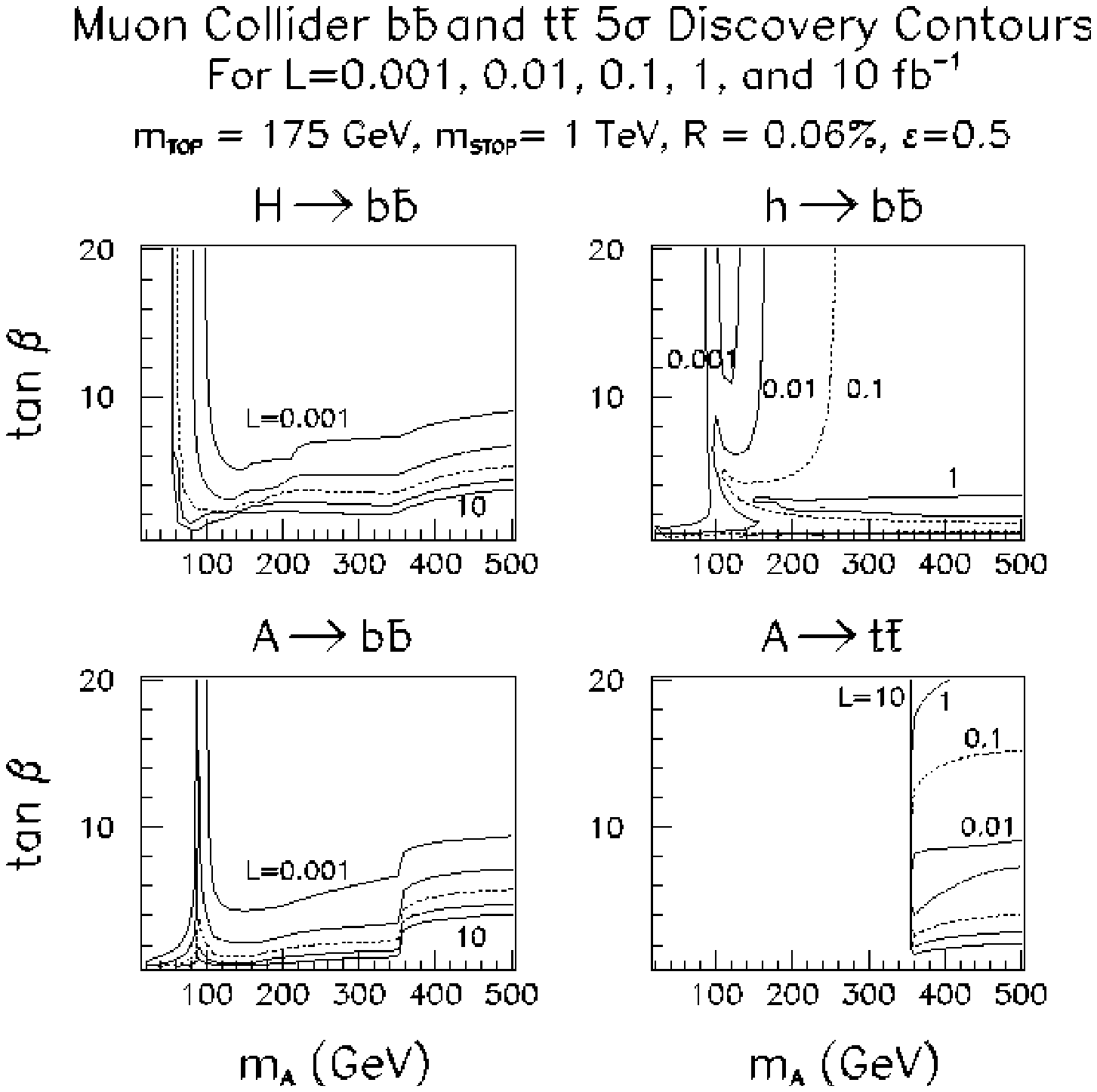,width=12.2cm}}
\begin{minipage}{12.5cm}       %%%% reduces width of caption to 12.5cm
\caption{Contours in $(\mha,\tanb)$ parameter space
of the luminosity required for $5\sigma$ Higgs signals.
Contours for $L=0.001$, 0.01, 0.1, 1, and $10\fbi$
are given. For $\ha\to t\anti t$, $L=0.001\fbi$ does not yield
a $5\sigma$ signal and no corresponding contour appears.
For $\hl\to b\anti b$, $L=10\fbi$ yields a $5\sigma$
signal for all of parameter space, and so only $L=0.001-1\fbi$
contours appear. The inputs are specified in the caption of
Fig.~\protect\ref{susytanbsliceshl}.}
\label{susyluminosityr06}
\end{minipage}
\end{center}
\end{figure}

An alternative picture that is especially useful for assessing the parameter
space region over which $\hl$, $\ha$ and/or $\hh$ discovery
will be possible at the $\mm$ collider is that given in
Fig.~\ref{susyluminosityr06}, for which we have taken $R=0.06\%$.
The contours in $(\mha,\tanb)$ parameter space denote the
luminosity required for a $5\sigma$ signal when $\rts$
is taken equal to the Higgs mass in question.
For the window labelled $\hh\to b\anti b$ we take $\rts=\mhh$,
for the $\hl\to b\anti b$ window we take $\rts=\mhl$, while $\rts=\mha$ for
the $\ha\rta b\anti b$ and $\ha\rta t\anti t$ contours.
The $5\sigma$ contours are for luminosities of
$L=0.001$, $0.01$, $0.1$, $1$, and $10\fbi$.
The larger the $L$ the larger the discovery region.
In the case of $\ha\to t\anti t$, $5\sigma$ is only achieved for
the four luminosities $L=0.01,0.1,1,10\fbi$.
In the case of the $\hl$, $L=10\fbi$ always yields
a $5\sigma$ signal within the parameter space region shown.

With regard to the $\hl$, Fig.~\ref{susyluminosityr06} shows
that for $R=0.06\%$
and luminosities somewhat less than $1\fbi$, $\hl$ could only be  
detected in the $b\anti b$
mode at large $\mha$ if $\tanb$ is sufficiently far from 1 that
$\mhl$ is not near $\mz$.
In contrast, when $\mha$ is sufficiently small that $\mhl$ is small and
the $\hl$ is no longer SM-like, and has enhanced $\mu\mu$
and $b\anti b$ couplings, rather modest luminosity is
required for a $5\sigma$ signal at $\rts=\mhl$;
for instance, $L\lsim 0.001\fbi$
will allow detection of a signal from the $\hl$ (and the possibly
overlapping $\ha$) over most of the $\mha\lsim 100\gev$
portion of parameter space even for $R=0.06\%$.
However, we have noted that it is theoretically quite likely that
$\mha$ is large and that the $\hl$ is SM-like.  Detection
of the $\hh$ and $\ha$ then becomes of paramount interest.

\subsubsection[Detecting the $\hh$ and 
$\ha$ by scanning in $\protect\rts$]{Detecting the {\protect\boldmath$\hh$} and 
{\protect\boldmath$\ha$} by scanning in {\protect\boldmath$\protect\rts$}}
\indent

In order to discover the $\hh$
or $\ha$ in the $\gsim 250\gev$ region, we must scan
over $\rts$ values between $250\gev$ and $500\gev$ (the presumed
upper limit for the FMC). The separation between scan points
is determined by the larger of the expected widths and 
the $\rts$ resolution, $\sigrts$.  If $\tanb\gsim 2$, then
for $\mhh$ and $\mha$ near $250\gev$, the $\ha$ and $\hh$ widths
are of order $0.05-0.1\gev$.  For masses near $500\gev$, their 
widths are at least $1\gev$ (cf.\  Fig.~\ref{hhhawidths}).
Meanwhile, for $R=0.01\%$ ($R=0.06\%$), $\sigrts$ ranges
from $\sim 0.018\gev$ ($\sim 0.11\gev$) to $\sim 0.035\gev$ ($\sim .21\gev$)
as $\rts$ ranges from $250\gev$ to $500\gev$.  Thus, it is reasonable
to imagine using scan points separated by $0.1\gev$ for $\mha\sim\mhh$
near $250\gev$, rising to $1\gev$ by $\rts=500\gev$. 
It will also be important to note that the luminosity required
per point for detection of the $\ha$ and $\hh$ is less
for masses below $2\mt$ than above.  In assessing the detectability
of the $\hh$ and $\ha$ by scanning we devote
\begin{itemize}
\item $L=0.01\fbi$ to each of 1000 points separated by $0.1\gev$ between
$250$ and $350\gev$,
\item $L=0.1\fbi$ to each of 100 points separated by $0.5\gev$
between $350$ and $400\gev$,
\item and $L=0.3\fbi$ to each of 100 points
separated by $1\gev$ between $400$ and $500\gev$.
\end{itemize}
This selection of points more or less ensures that if the $\hh$ and $\ha$
are present then one of the scan points would have
$\rts\sim \mhh,\mha$ within either the $\sigrts$ resolution
or the Higgs width.
The total luminosity required for this scan would be $50\fbi$.

We now employ the $5\sigma$ contours of Fig.~\ref{susyluminosityr06} to
assess the portion of $(\mha,\tanb)$ parameter
space over which the above scan will allow us
to detect the $\hh$ and $\ha$
in the $b\anti b$ and $t\anti t$ channels. 
The $5\sigma$ luminosity
contours of interest will be the curves corresponding to $L=0.01\fbi$,
$L=0.1\fbi$ and $L=1\fbi$.  The $5\sigma$ contour for $L=0.3\fbi$  
luminosity per point, as employed in our scan procedure from $400$ to $500\gev$,
is midway between these last two curves. Fig.~\ref{susyluminosityr06} shows
that, by performing the scan in the manner outlined earlier,
one can detect the $\hh,\ha$ in the $b\anti b$ mode for all $\tanb$ values
above about $2-4$ for $\mhh,\mha\lsim 2\mt$ and above about $3-5$ for
$2\mt\lsim \mhh,\mha \lsim 500\gev$.  Meanwhile, in the $t\anti t$
mode, the $\ha\to t\anti t$ signal can be seen for $\mha\gsim 2\mt$
provided $\tanb\gsim 3$. Together, the $b\anti b$ and $t\anti t$
signals are viable for a remarkably large
portion of parameter space, which includes, in particular,
essentially all of the wedge region where the LHC lacks sensitivity
(see Fig.~\ref{lhcregionshilum}).
At worst, there would be a very small $\tanb$ window for $\mha\gsim2\mt$
between $\tanb=3$ and $\tanb=4$, for which the signal might be missed
during the above described scan and also no signal seen at the LHC.  
In practice, it might
be desirable to simply devote several years of running
to the scan in order to ensure that the $\ha$ and $\hh$
are detected if present.

The implementation of the above scan is very demanding upon
the machine design because:
\begin{itemize}
\item
several rings may be needed to have high luminosities over 
a broad range of $\sqrt s$;
\item
it must be possible {\it over this broad range of energies} to quickly 
(for example, once every hour or so
in the $250$--350\gev\ range) reset $\rts$ with an accuracy that
is a small fraction of the proposed step sizes.
\end{itemize}
It is too early to say if these demands can both be met.

Finally, we note the obvious conflict between this scan and
the desirable $\rts=\mhl$, $L=50\fbi$ study of the SM-like $\hl$.
A multi-year program will be required to accomplish both tasks.

\subsubsection[Non-$b\anti b$ final 
state modes for heavy Higgs detection]{Non-{\protect\boldmath$b\anti b$} 
final state modes for heavy Higgs detection}
\indent

The reader may note that $\rts=\mhh$ does not yield
an observable $s$-channel signal in the $b\anti b$ mode for $\mha\lsim 100\gev$.
Although the $\hh$ is SM-like in this parameter region in that it
does not have enhanced coupling to $\mu\mu$ and $b\anti b$,
its decays are dominated by $\hl\hl$ and, for $\mha\lsim 60\gev$,
$\ha\ha$ pairs; $Z\ha$ decays also enter
for small enough $\mha$.  This means that the $\hh$
total width is quite large, in particular much larger
than the $\rts$ spread.  The large total width also implies
that $\br(\hh\to\mu\mu)$ is small.  Equation~(\ref{broadwidthsigma})  
then shows
that the production rate for the $\hh$ will be
small, and that the rate in the $b\anti b$ final
state will be further suppressed by the small value of $\br(\hh\to  
b\anti b)$.
The only possible channels for observation of the $\hh$ in the
$\mha\lsim 100\gev$ region are $\hl\hl,\ha\ha,Z\ha$.
As we discuss below, these could prove to be viable.

\begin{figure}[htbp]
\let\normalsize=\captsize   %%%% changes the font to "\small"
\begin{center}
\centerline{\psfig{file=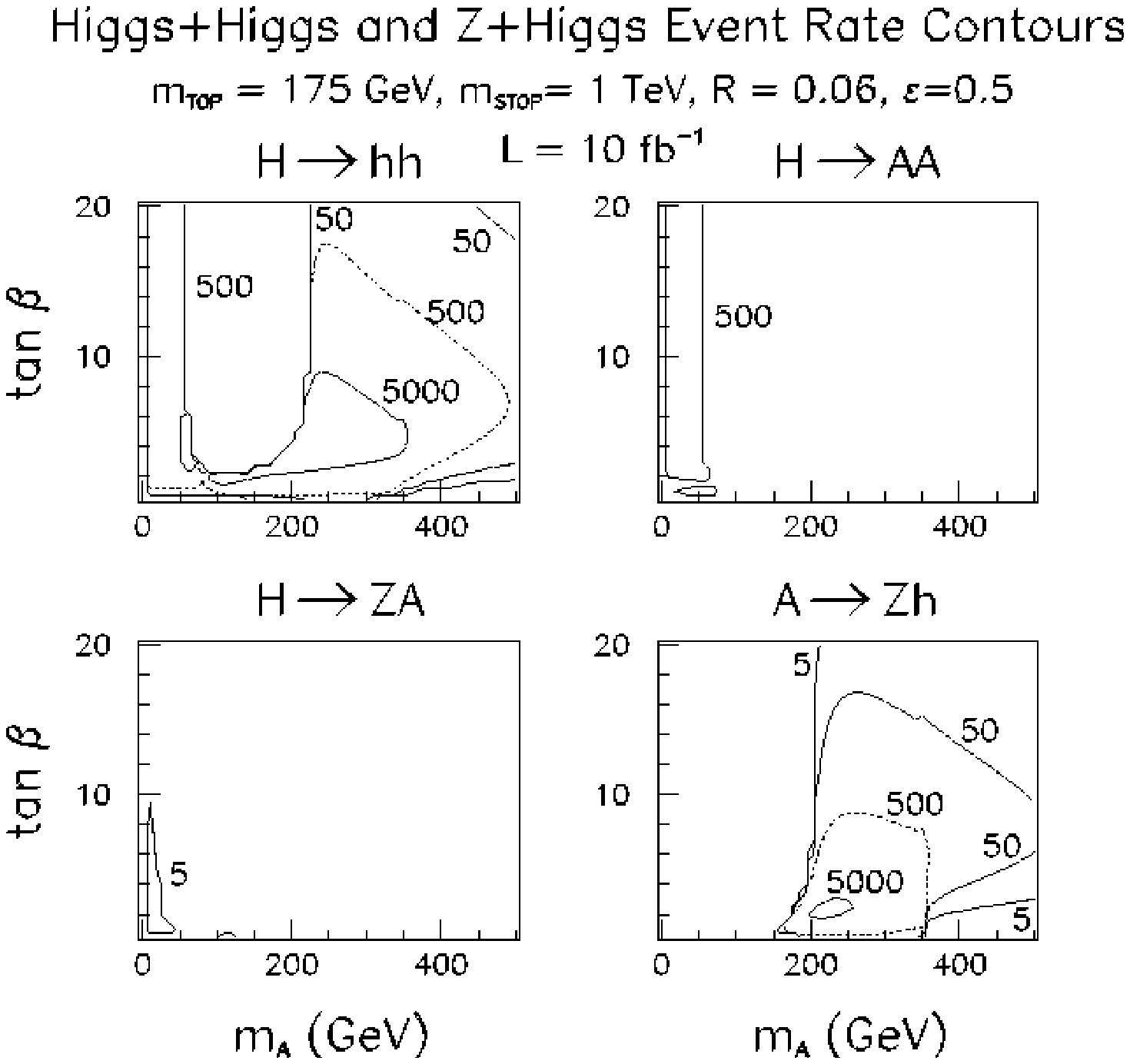,width=12.2cm}}
\begin{minipage}{12.5cm}       %%%% reduces width of caption to 12.5cm
\caption{Event rate contours for $\hh\to\hl\hl$, $\hh\to\ha\ha$,
$\hh\to Z\ha$ and $\ha\to Z\hl$ in $(\mha,\tanb)$ parameter space
for integrated luminosity $L=10\fbi$. Contours for 5, 50, 500 and  
5000 events are shown in the first and last cases. There
are 500 or more $\hh\to \ha\ha$ events if $\mha\protect\lsim 60\gev$
and $\tanb\protect\gsim 2$, but $\hh\to Z\ha$ barely
reaches the 5 event level. Two-loop/RGE-improved radiative
corrections to Higgs masses, mixing angles and self-couplings
are included, taking $\mt=175\gev$, $\mstop=1\tev$ and neglecting
squark mixing.}
\label{rateshhzh}
\end{minipage}
\end{center}
\end{figure}

The full set of
channels to be considered are
\begin{equation}
\hh\to\hl\hl,\qquad \hh\to \ha\ha,\qquad \hh\to Z\ha,\qquad \ha\to  
Z\hl.
\end{equation}
The $\hl\hl,\ha\ha$ final states primarily ($\sim 80\%$ of the time)
yield $4b$'s.  The $Z\ha,Z\hl$ final states yield $2j2b$ about 60\%
of the time.  In either case, we can demand that there be two
pairs of jets, each pair falling within narrow mass  
intervals.
In addition, two $b$-tags can be required.
Thus, these channels will have small background.
To illustrate the size of the signal in these channels, we present
in Fig.~\ref{rateshhzh} the $L=10\fbi$ signal rates for the above four  
modes,
assuming a net 50\% efficiency (including branching fractions
and tagging efficiencies, as well as double mass-binning).
In the $\hh\to\hl\hl$ case, at least 50 events are obtained
in essentially all but the $\mha=60-230,\tanb\gsim 2.5$ region; the  
5000
event contour is confined to a narrow region around  
$\mha=65-70,\tanb\gsim2$
and to the (disjoint) teardrop region labelled; the 50 and 500
event contours are as labelled. At least 500 events are predicted
in the $\mha\lsim 60$ region for all $\tanb$.
In the $\hh\to \ha\ha$ case, at least 500
events are obtained in the $\mha\lsim 60$ and $\tanb\gsim 2$ region.
In the $\hh\to Z\ha$ case, only the 5 event level
is achieved over even the small piece of parameter space shown.
Finally, in the $\ha\to Z\hl$ case all contours are easily identified
by the labelling. No events are expected for $\mha$
below about $200\gev$, where the $\ha\to Z\hl$ decay mode
is no longer kinematically allowed. It is kinematics that also  
dictates
the rather restricted regions at low $\mha$ for which
$\hh\to\ha\ha$ and $\hh\to Z\ha$ events occur.

In order to discuss the observability of the above signals, we need to
compute the background level, which we do not do in this report.
After $b$-tagging and mass reconstruction we believe that
backgrounds should be modest. In the absence of any explicit
calculation we can only make the following guesstimates.
Based on the event rates of Fig.~\ref{rateshhzh} it should be possible
to study the  $\hh\to\hl\hl$ channel
over a significant fraction of parameter space with $L\sim 1\fbi$.
In particular, luminosities at and above this level
could open up the $\mha\lsim 60\gev$
region for both this mode and the $\hh\to\ha\ha$ mode.
In contrast, it will obviously
require very substantial luminosity to detect
$\hh\to Z\ha$, even when not kinematically suppressed.
A viable $\ha\to Z\hl$ signal may be possible, when kinematically  
allowed,
only so long as $\mha$ and $\tanb$ are not large;  when $\mha$ is  
large
the tree-level coupling is suppressed (which suppression occurs
most rapidly at large $\tanb$) and there are too few events
for a useful signal.

Although these modes provide somewhat more challenging
signals than the $b\anti b$ channel signal, their observation
would provide tests of important Higgs couplings.
In particular, detection of the $\hh\to\hl\hl$ and $\hh\to\ha\ha$
modes would allow a direct probe of these very interesting
Higgs boson self-couplings. The procedure will be outlined
in a later section.
In general, determination of the Higgs boson self-couplings
is quite difficult at other machines. In particular,
even when a relevant branching fraction can be measured,
knowledge of the total width is required in order to extract
the partial width and coupling.  Without a $\mm$ collider,
measurement of the total width is only possible if the width
is substantially larger than the resolution
implied by final state mass reconstruction at the Higgs mass.  This is not
the case for the $\hh$ and $\ha$ unless $\tanb$ is very large.

\subsection{MSSM Higgs boson detection
using the bremsstrahlung tail spectrum}
\indent

In this section, we discuss an alternative way
of searching for the $\ha$ and $\hh$ by running the $\mm$
collider at full energy but looking for excess events
arising from the luminosity on the low-energy end of the bremsstrahlung tail
(see Appendix~C).
This latter technique proves to be somewhat competitive with
the scan technique just described, provided that excellent resolution
in reconstructing the $b\anti b$ final state mass can be achieved
and provided that large total integrated luminosity is devoted
to such running. It would have two distinct advantages over the scanning approach.
\begin{itemize}
\item It would not require the construction
of multiple rings in order to maintain high luminosity
over a broad range of $\rts$ collision energies.
\item A large number of
events in the $Z\h$ mode for the SM-like $\hl$
could be simultaneously accumulated.
\end{itemize}
As for the scan procedure, the bremsstrahlung tail technique
is viable only if the $\h\to\mm$ coupling is significantly
enhanced relative to the SM $\hsm\to\mm$ coupling; only then
is a Higgs boson with mass substantially
below $\rts$ produced at a large rate by virtue
of the bremsstrahlung tail.
Of course, once the $\hh$ and/or $\ha$ is found using the bremsstrahlung
technique, it would then be highly desirable to run the machine with
$\rts\sim\mhh,\mha$ in order to study in detail the widths
and other properties of the $\hh,\ha$.

For our study of the bremsstrahlung tail possibility,
we shall assume that the $b\bar b$ final state mass can be reconstructed to
within $\pm 5\gev$.  A full study of this mode of detection should
generate events, smear the $b$ jets using expected resolutions,
allow for semi-leptonic $b$ decays, and incorporate tagging efficiencies.
The reconstructed
mass of the $b\anti b$ final state for each event should then be
binned and one would then look for a peak over the expected background
level.  We will not perform this detailed simulation here.
Instead, we compute as a function of $\mbb$ (the central
value of the $b\anti b$ final state
mass) the number of events in the interval $[\mbb-5\gev,\mbb+5\gev]$.
In estimating the significance of any peak seen in the spectrum,
we will choose $\mbb$ at the center of the peak, and compare
the excess of events in the above interval (the signal $S$) to the
number of events expected if there is no Higgs boson present (the background
$B$). The statistical significance will be computed as $S/\sqrt B$.
In computing the number of events we assume an integrated
luminosity of $L=50\fbi$ and assume an event reconstruction and tagging
efficiency of $\eps=0.5$.  Correspondingly,
only the continuum $b\anti b$ final states from $\gam^\star,\zstar$ processes
will be included in $B$ (using also $\eps=0.5$). These latter assumptions
are the same ones employed in our other analyses.

\subsubsection{Mass peaks}
\indent

\begin{figure}[htbp]
\let\normalsize=\captsize   %%%% changes the font to "\small"
\begin{center}
\centerline{\psfig{file=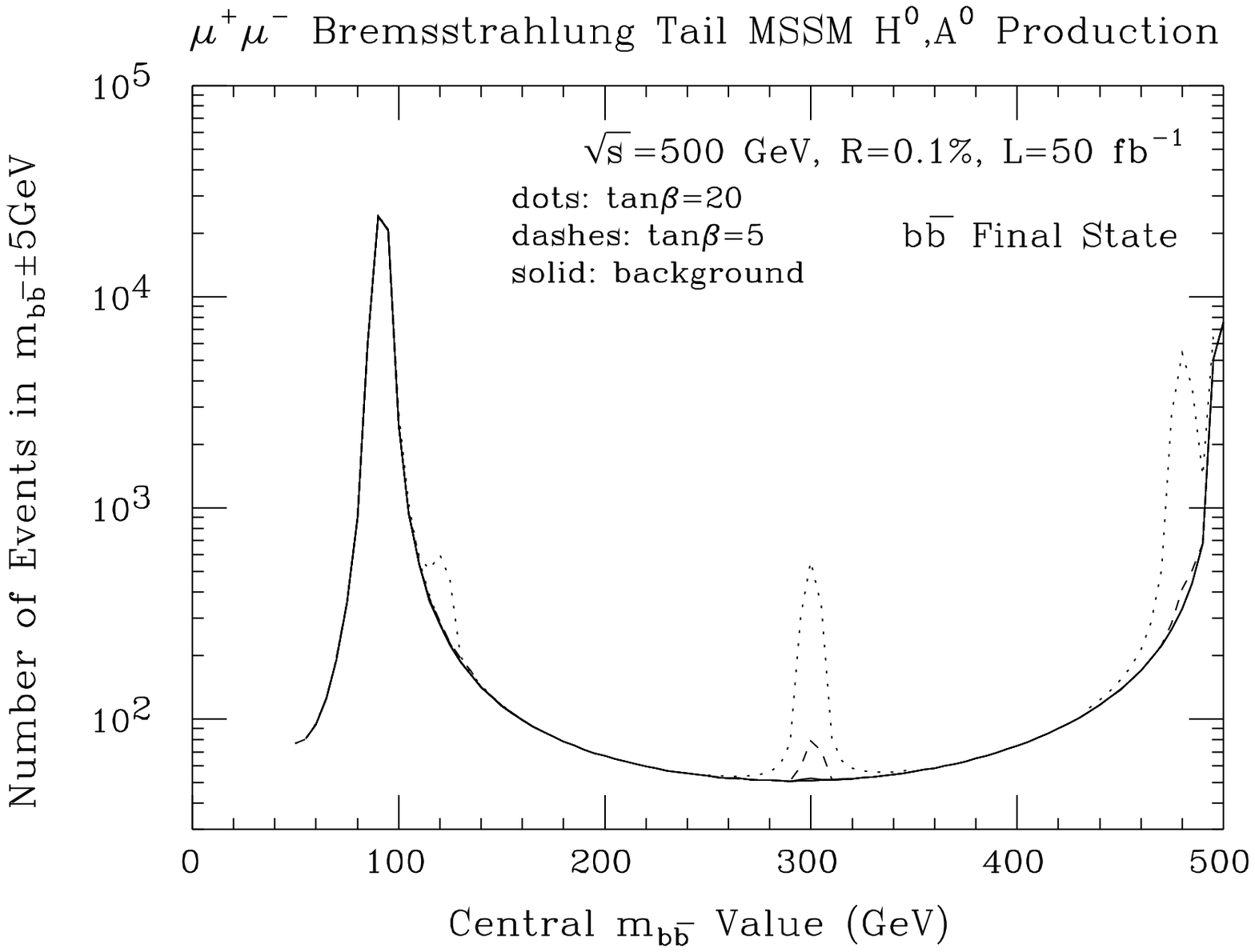,width=12.2cm}} 
\begin{minipage}{12.5cm}       %%%% reduces width of caption to 12.5cm
\caption{Taking $\protect\rts=500\gev$, integrated luminosity $L=50\fbi$,
and $R=0.1\%$, we consider the $b\anti b$ final state and
plot the number of events in the interval $[\mbb-5\gev,\mbb+5\gev]$,
as a function of the location of the central $\mbb$ value,
resulting from the low $\protect\rtshat$ bremsstrahlung tail of the  
luminosity distribution.
MSSM Higgs boson $\hh$ and $\ha$ resonances are present for
the parameter choices of $\mha=120$, $300$ and $480\gev$,
with $\tanb=5$ and $20$ in each case. Enhancements for $\mha=120$,
$300$ and $480\gev$ are visible for $\tanb=20$; $\tanb=5$ yields
visible enhancements only for $\mha=300$ and $480\gev$.
Two-loop/RGE-improved radiative corrections are included,
taking $\mt=175\gev$, $\mstop=1\tev$ and neglecting squark mixing.
SUSY decay channels are assumed to be absent.}
\label{mbbscan}
\end{minipage}
\end{center}
\end{figure}

It will be useful to first display some typical mass peaks.
In Fig.~\ref{mbbscan}, we plot the number of events in the
interval $[\mbb-5\gev,\mbb+5\gev]$ as a function of $\mbb$
for three $\mha$ choices: $\mha=120$, $300$ and $480\gev$.
In each case, results for $\tanb=5$ and $20$ are shown.
The event enhancements derive from the presence of the
$\hh$ and $\ha$ Higgs bosons.
There would be no visible
effect for the choice of $\mha=100\gev$ for any $\tanb$
value below 20. This is because all the Higgs masses are sitting
on the very large $Z$ peak and, in addition, none of the $\mm$
couplings are fully enhanced.  For the
three $\mha$ values considered in Fig.~\ref{mbbscan}, 
we observe event excesses for $\tanb=20$
in all cases. For $\tanb=5$, the $\mha=300\gev$ peak is
clear, while $\mha=480\gev$ yields a shoulder of excess events
(that is statistically significant);
nothing is visible for $\mha=120\gev$. For $\tanb\lsim 2$, 
no peaks or excesses would be
visible for any of the above $\mha$ choices.
Finally, we note that enhancements due to the $\hl$ resonance would 
not be visible, regardless of $\tanb$, for $\mha\gsim 100\gev$.

\subsubsection{Significance of signals}
\indent

\begin{figure}[htbp]
\let\normalsize=\captsize   %%%% changes the font to "\small"
\begin{center}
\centerline{\psfig{file=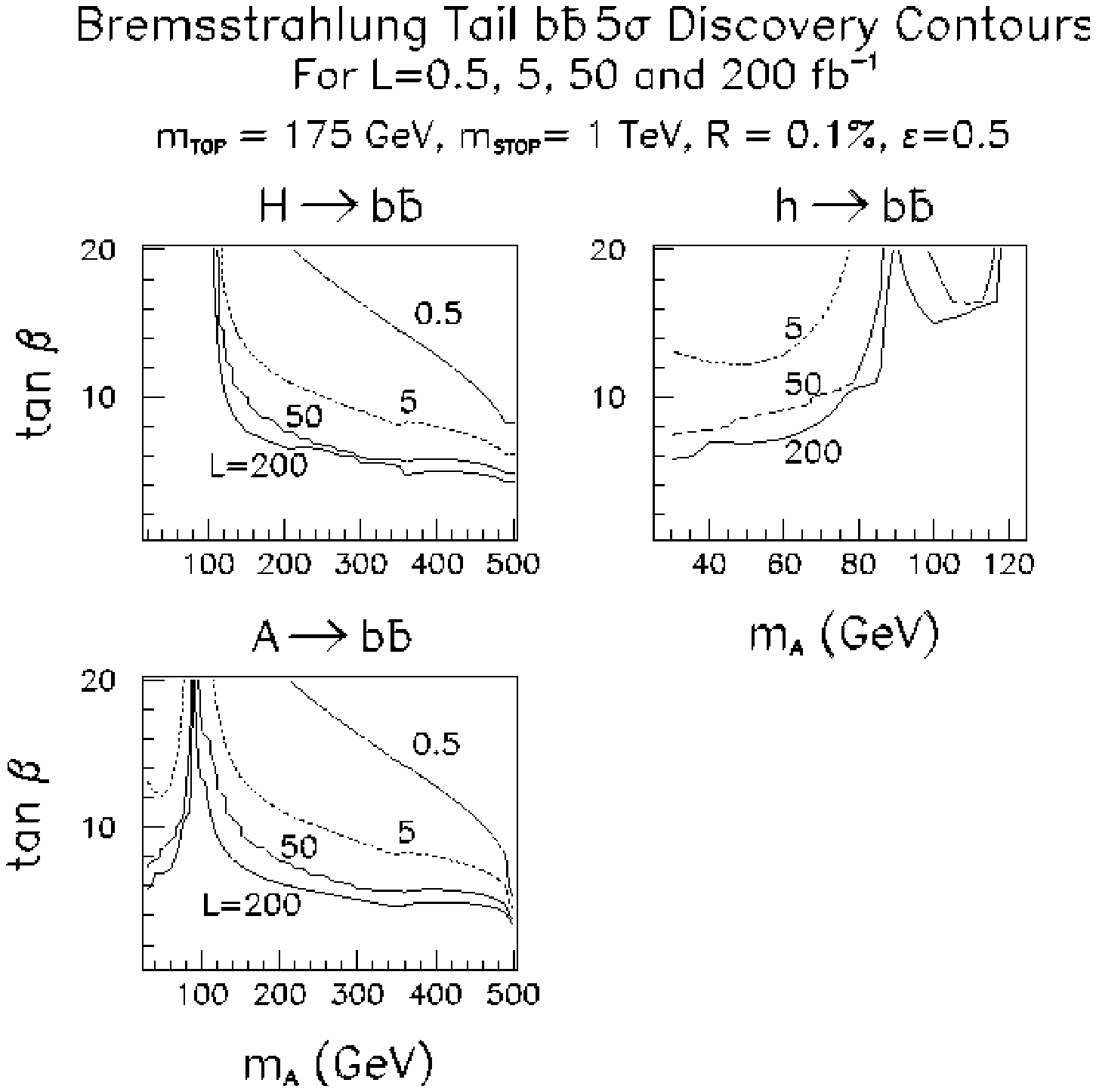,width=12.2cm}}
\begin{minipage}{12.5cm}       %%%% reduces width of caption to 12.5cm
\caption{Taking $\protect\rts=500\gev$
and $R=0.1\%$, we consider the $b\anti b$ final state and
compute the Higgs signal ($S$) and background ($B$)  rates
in the mass interval $[\mbb-5\gev,\mbb+5\gev]$,
with $\mbb=\mhh$, $\mbb=\mhl$, and $\mbb=\mha$,
resulting from the low $\protect\rtshat$ bremsstrahlung tail of the  
luminosity distribution.
$S/\protect\sqrt B=5$ contours are shown for integrated luminosities
of $L=0.5$, $5$, $50$, and $200\fbi$.
Two-loop/RGE-improved radiative corrections are included,
taking $\mt=175\gev$, $\mstop=1\tev$ and neglecting squark mixing.
SUSY decay channels are assumed to be absent.}
\label{mbbstatsig}
\end{minipage}
\end{center}
\end{figure}

We will now proceed to survey the $S/\sqrt B$ expectations.
We do this as a function of location in the $(\mha,\tanb)$
parameter space as follows.  For each choice of $(\mha,\tanb)$
we determine $\mhl$ and $\mhh$. We then compute
$S/\sqrt B$ for the three locations $\mbb=\mhl$, $\mbb=\mhh$
and $\mbb=\mha$, where $S$ and $B$ are computed by counting
events in the $\mbb\pm5\gev$ window.
Effects from overlapping Higgs resonances are included.
The $5\sigma$ discovery contours for each of these three
window locations are plotted in $(\mha,\tanb)$
parameter space for integrated luminosities of
$L=0.5$, $5$, $50$ and $200\fbi$ in Fig.~\ref{mbbstatsig},
taking $\rts=500\gev$ and $R=0.1\%$.

As expected from Fig.~\ref{mbbscan}, the window centered
at $\mbb=\mhl$ only yields a statistically significant
excess if $\tanb$ is large and $\mhl$ is not near $\mz$.
($\mhl$ near $\mz$ at high $\tanb$ corresponds to $\mha\sim 95\gev$.)
Since the $Z\hl$ mode will yield an observable signal regardless
of the $(\mha,\tanb)$ values, the bremsstrahlung tail
excess would mainly be of interest as a probe of
the $\Gamma(\hl\to\mm)$ partial width
prior to running at $\rts=\mhl$.

However, the $\pm5\gev$ intervals centered at
$\mbb=\mhh$ and $\mbb=\mha$ (which,
include events from the overlapping $\ha$ and $\hh$
resonances, respectively) yield $5\sigma$ statistical
signals for a substantial portion of parameter space if $L$ is large.
With $L=50\fbi$, a 5 sigma discovery of the $\hh$ and $\ha$
using the $\rts=500\gev$ bremsstrahlung tail is viable down to
$\tanb\gsim 6.5$ at $\mha=250\gev$ improving to
$\tanb\gsim 5$ at $480\gev$. This is not quite as far
down in $\tanb$  as can be probed for $250\lsim \mha\lsim 500\gev$
by the previously
described scan over a series of $\rts$ values using $0.01-0.3\fbi$
of luminosity at each scan point. As $\mhh,\mha$ move
closer to $\mz$, the $5\sigma$ discovery contours move to much
larger $\tanb$ values, whereas the scanning technique
would yield $5\sigma$ signals for $\tanb$ values as low
as $\tanb\sim 3-4$ all the way down to $\mha\gsim 60\gev$.

\subsubsection{Strategy: scan vs.\ maximum energy}
\indent

If $\zstar\to\hh\ha$ is not observed at a $\rts=500\gev$
$\ee$ machine and if discovery of the $\hh$ and $\ha$ in the  
$250-500\gev$
mass range is the primary goal, at the $\mm$
collider it would be a close call as to whether
it would be better to immediately embark on the $\rts$ scan
or accumulate luminosity at the maximum machine energy.
The $\rts$ scan probes $\tanb$ values that are lower
by only 1 or 2 units than the bremsstrahlung tail search.
This statement assumes that a
final state mass resolution of order $\pm 5\gev$ can be achieved
(even after including all semi-leptonic decay effects and so forth)
in the $b\anti b$ final state for the latter search.
If not, the $\rts$ scan is the preferred technique.
Thus, resolution and missing energy could become critical issues
for the detector(s) in deciding the best approach.

If an $\ee$ collider is not operational at the time a $\mm$
collider begins running, then the decision as to which approach
to choose for $\hh$ and $\ha$ discovery becomes even more delicate
unless the LHC has clearly ruled out $\mha,\mhh\lsim 250\gev$
(which it probably can do --- see Fig.~\ref{lhcregionshilum}).
Without a lower
bound on $\mha,\mhh$, the $\rts$ scan would have to be extended to lower
$\rts$, requiring more luminosity.  In contrast, by accumulating
$L=50\fbi$ at full energy, $\rts=500\gev$,
it would be possible to simultaneously either discover or rule out
$\mha,\mhh\lsim \rts/2$ for all $\tanb$
and $\rts/2\lsim\mhh,\mha,\lsim\rts$ for $\tanb\gsim 5-7$.
Note that  $\mha,\mhh\lsim\rts/2-20\gev$ can
be ruled out in the $\zstar\to\hh\h$ mode with perhaps as little as $5-10\fbi$.
For luminosities of order $10\fbi$ the bremsstrahlung tail technique
would probe $\tanb\gsim 11$ for $\mha\sim250\gev$
improving to $\tanb\gsim 6$ for $\mha\sim 500\gev$.
After accumulating the $L=5-10\fbi$, the $\mm$
collider could then be switched to the scan mode of operation
if no signal has been found.

\subsection[Detailed studies of the $\hh$ and
$\ha$]{Detailed studies of the {\protect\boldmath$\hh$} and
{\protect\boldmath$\ha$}}

\indent\indent
However the $\hh$ and $\ha$ are first detected,
one will wish to measure
the total and partial widths of the $\hh$ and $\ha$.
Once again, the $\mm$ collider can play a crucial role.
We will not give detailed estimates of what can be accomplished,
but rather confine ourselves to outlining the procedures
and strategies. The time scale and available luminosity
for implementing these
procedures depends dramatically upon whether or not
one must first discover the $\hh$ and $\ha$ by scanning or
in the bremsstrahlung tail (either of which would require a luminosity
expenditure of $L\sim 50\fbi$), 
as opposed to observing them at the LHC (typically
possible for $\tanb\lsim 3-4$ at high $\mha$) or at an $\ee$ collider
(requiring $\mha,\mhh\lsim \rts/2$).

One might presume that once
a Higgs boson with $\gamh$ larger than the rms $\rts$
spread is discovered, direct measurement
of the Higgs width would be quite straightforward with a simple
scan over several $\rts$ settings. This is indeed
the case unless there is a second
nearby Higgs boson. As it happens, the $\ha$ and $\hh$
are sufficiently degenerate in some regions of parameter space
(large $\mha$ and large $\tanb$),
see Figs.~\ref{hhhawidths} and \ref{delmhiggs}, that a measurement of the widths
of the $\ha$ and $\hh$ separately will require sorting out
two overlapping resonance bumps, which, in turn, necessitates
an appropriate scan.  Two sample possibilities were illustrated
earlier in Fig.~\ref{hhhasusyrtsscan}, where the $\hh$ and $\ha$
resonance bumps that would appear as a function of $\rts$
are illustrated for $\mha=350\gev$ in the cases $\tanb=5$ and 10.
As noted earlier, separation of the peaks and precision width
measurements are both much easier if we have excellent beam energy resolution;
we assume $R=0.01\%$.
At $\tanb=5$, we estimate that by accumulating
roughly $0.01\fbi$ at each of
3 appropriately placed $\rts$ choices near the center and on
either side of each of the two separated peaks, the widths of
the $\hh$ and $\ha$ could be measured to about 33\%;
10\% width determination would require about $0.1\fbi$ per point.
At the higher $\tanb=10$ value, one would clearly have to accumulate
data in the dip between the overlapping peaks,
near both peaks, below the double peak and above the double peak, and perform
a fit to the two Higgs resonances simultaneously. A minimum of 5 data
points would be required. Again, roughly $0.01\fbi$ per point
would be needed to determine $\gamhh$ and $\gamha$ to the 33\% level,
or $0.1\fbi$ per point for a 10\% determination.
Very large $\tanb$ values yield the worst scenarios since the $\hh$
and $\ha$ peaks are, then, simultaneously broad and very degenerate.
Determination of the individual widths would become extremely difficult.

The production rate in a given channel is proportional
to $\br(\h\to \mm)\br(\h\to X)$ (for $\sigrts\ll\gamh$), see
Eq.~(\ref{broadwidthsigma}). We then proceed as follows:
\begin{itemize}
\item
$\br(\h\to\mm)$ and $\br(\h\to b\anti b)$ can be obtained individually
if we use the type-II doublet prejudice that the $\mm$ and $b\anti b$
couplings squared are modified relative to the SM coupling by the same
factor, $f$. (A value of $\mb$ must be specified.) 
\item Given the individual branching fractions, the partial widths
can then be computed:
\begin{equation}
\Gamma(\h\to\mm,b\anti b)=\gamh \br(\h\to \mm,b\anti b)
\label{partialwidthforms}
\end{equation}
\item One can use event rates in other observable
channels, coupled with the $\br(\h\to\mm)$ determination,
to obtain results for $\br(\h\to X)$.  
\item $\gamh\times \br(\h\to X)$
then yields the partial width and coupling for any observable channel $X$.
For example, if the $\hh\to\hl\hl$ channel can be detected we
could determine the very interesting associated partial width (and, thence,
coupling) via $\Gamma(\hh\to\hl\hl)=\gamhh \br(\hh\to\hl\hl)$ or,
equivalently, 
\begin{equation}
\Gamma(\hh\to\hl\hl)={[\gamhh]^2 \br(\hh\to\mu\mu)\br(\hh\to\hl\hl)
\over \Gamma(\hh\to \mu\mu)}\,.
\label{gamhhhlhl}
\end{equation}
\end{itemize}
Of course, if $\gamh$ and $\sigrts$ are close in size,
one must avoid the approximation of
Eq.~(\ref{broadwidthsigma}), but determination of $f$ and the partial
widths and branching fractions would nevertheless be straightforward.

\subsection{Determining a Higgs boson's CP properties}

\indent\indent
A $\mupmum$ collider might well prove to be the best machine
for directly probing the CP properties of a Higgs boson that
can be produced and detected in the $s$-channel mode.
This issue has been explored in Refs.~\cite{bohguncp,atsoncp}
in the case of a general two-Higgs-doublet model.

The first possibility is to measure correlations in the
$\taup\taum$ or $t\anti t$ final states. Via such measurements,
a $\mupmum$ collider is likely to have greater sensitivity
to the Higgs boson CP properties
for $L=20\fbi$ than will the $\epem$ collider for $L=85\fbi$ (using
correlation measurements in the $Z\h$ production mode)
if $\tanb\gsim 10$ or $2\mw\lsim\mh\lsim 2\mt$.  Indeed, there
is a tendency for the $\mupmum$ CP-sensitivity to be best precisely
for parameter choices such that CP-sensitivity in the $\epem\rta Z\h$
mode is worst. Somewhat higher total luminosity ($L\sim 50\fbi$)
is generally needed in order to use these correlations to distinguish
a pure CP-odd state from a pure CP-even state.

The second possibility arises if it is possible to transversely
polarize the muon beams.  Assume that
we can have 100\% transverse polarization and that 
the $\mu^+$ transverse polarization is rotated with respect
to the $\mu^-$ transverse polarization by an angle $\phi$.  The production
cross section for a $\h$ with coupling $a+ib\gamma_5$ then behaves as
\begin{equation}
\sigma(\phi)\propto 1 - {a^2-b^2\over a^2+b^2} \cos\phi 
+ {2ab\over a^2+b^2}\sin\phi\,.
\label{cpmu}
\end{equation}
To prove that the $\h$ is a CP admixture, use the asymmetry
\begin{equation}
A_1\equiv {\sigma(\pi/2)-\sigma(-\pi/2)\over \sigma(\pi/2)+\sigma(-\pi/2)}
= {2ab\over a^2+b^2}\,.
\end{equation}
For a pure CP eigenstate, either $a$ or $b$ is zero.
For example, in the MSSM  the Higgs sector is CP-conserving;
$b=0$ for the CP-even $\hl$ and $\hh$,
while $a=0$ for the CP-odd $\ha$.
In such cases, it is necessary to employ a different 
asymmetry than that discussed in Ref.~\cite{atsoncp}. The quantity
\begin{equation}
A_2\equiv {\sigma(\pi)-\sigma(-\pi) \over \sigma(\pi)+\sigma(-\pi)}
= {a^2-b^2\over a^2+b^2}
\end{equation}
is $+1$ or $-1$ for a CP-even or CP-odd $\h$, respectively.
Background processes in the final states where
a Higgs boson can be most easily observed ({\it e.g.} $b\anti b$)
can dilute these asymmetries
substantially. Whether or not they will prove useful
depends even more 
upon the very uncertain ability to transversely polarize the muon
beams, especially while maintaining high luminosity.

Note that longitudinally polarized beams are not useful for studying
the CP properties of a Higgs produced in the $s$-channel.
Regardless of the values of $a$ and $b$
in the $\h$ coupling, the cross section is simply proportional
to $1-\lam_{\mu^+}\lam_{\mu^-}$ (the $\lam$'s
being the helicities), and is only non-zero for $LR$ or $RL$ transitions,
up to corrections of order $m_\mu^2/\mh^2$.

\section{Summary and Conclusion}

\indent\indent
A $\mm$ collider would be a remarkably powerful machine for
probing Higgs physics using direct $s$-channel production,
and thus ultimately for
finding the underlying theory of the scalar sector. In this report we have
concentrated on the procedures and
machine requirements for direct measurement of the  
properties of a Higgs boson.

\subsection{SM-like Higgs boson}

\indent\indent
We expect that a SM-like $\h$ (which nominally includes the $\hl$ 
of the MSSM) will first be detected either at the LHC or in the $Z\h$ mode
at an $\ee$ collider. If not, it would be most advantageous
to expend a small amount of luminosity at full machine energy to discover
it in the $Z\h$ mode at the $\mm$ collider. Once $\mh$ is approximately known,
a $\mm$ collider can zero-in on $\rts\simeq\mh$
for detailed studies of a SM-like Higgs boson 
provided $\mh\lsim 2\mw$ (as is the case for the $\hl$
of the MSSM). 
The mass can be measured to a fraction of an MeV
for $\mhsm\lsim 130\gev$.

Crucial to a model-independent determination of all
the properties of the Higgs boson at the $\mm$ collider is
the ability to make a direct precision measurement 
of its total width, which is very narrow for a SM-like $\h$ when $\mh<2\mw$.
The proposed method (described in Appendix~C) relies on measuring the 
ratio of the central peak 
cross section to the cross section on the wings of the peak,
a ratio that is determined by $\gamh$ alone.  
Once $\gamh$ is measured, determinations
of the crucial $\mm$ and $b\anti b$ couplings
are possible.  The precision for $\gamh$ and the $\mm$ and $b\anti b$
partial widths/couplings achieved for total integrated luminosity of 
$L=50\fbi$ and an excellent beam resolution of $R=0.01\%$
would be sufficient to distinguish the MSSM $\hl$ from
the SM $\hsm$ at the $3\sigma$ statistical level
for values of the parameter $\mha$ as large as $\sim 400\gev$
provided that $\mhl$ is not in the range $80\lsim\mhl\lsim  
100\gev$ (\ie\ near $\mz$).  No other accelerator or combination of
accelerators has the potential of seeing the $\hl$ vs. $\hsm$
differences at this level of precision out to such large $\mha$  
values.
For a SM-like Higgs with $\mh\gsim 200\gev$, the
event rate is too low for detection in the $s$-channel.

Machine requirements for the precision studies are:
\begin{itemize}
\item High luminosity ${\cal L}\gsim 2\times 10^{33} {\rm cm}^{-2} {\rm
s}^{-1}$ at $\rts\sim \mh$.
\item Excellent beam energy resolution of $R=0.01\%$.
\item Ability to adjust the machine energy $\rts$ accurately (to one part
in a million) and quickly (once an hour in the intial scan
to precisely determine $\mh$) over a $\rts$ interval of several GeV.
\end{itemize}

\subsection{Non-SM-like Higgs bosons}

\indent\indent
For other Higgs bosons with weak $WW,ZZ$ couplings (such as the $\hh$ and
$\ha$ of the MSSM), but enhanced $\mm$ and $b\anti b$ couplings,
discovery in $s$-channel collisions at the $\mm$ collider
is typically possible.  There are three possible techniques.
In order to compare these techniques it is reasonable to suppose
that the $\hh$ and $\ha$
have been excluded for $\mhh,\mha\lsim \rts/2$
via the $\zstar\to\hh\ha$ mode at an $\ee$ collider
running with $\rts\sim 500\gev$.

\begin{itemize}
\item[a)] Scan method\\
In this approach, a scan for the $\hh$ and $\ha$ of the MSSM
would be made over a sequence
of $\rts$ values all the way out
to the maximal $\rts$ value achievable at the $\mm$ collider.
Assuming that $L=50\fbi$ is devoted to the scan
and that both the $\ee$ and the $\mm$ colliders
have maximal energies of order $500\gev$, discovery 
via the scan would be robust for 
$250\lsim m_{\hh,\ha}\lsim500\gev$ if $\tan\beta\gsim3$ to 4.
Fortuitously, the domain
$250\lsim\mhh,\mha\lsim 500\gev,\tanb\lsim 3$, in which
much more luminosity would clearly be required
for discovery at the $\mm$ collider, is a parameter region 
where the $\hh$ and $\ha$ are likely to
be accessible at the LHC for accumulated luminosity of $300\fbi$
per detector (ATLAS+CMS), as illustrated in  Fig.~\ref{lhcregionshilum}.
There is, nonetheless, a small window, $3\lsim\tanb\lsim 4$,
at large $\mha$ (between about 400 and $500\gev$)
for which the LHC and the $\mm$ collider might both miss seeing
the $\hh$ and $\ha$ unless higher luminosities are accumulated.

In order that the required $L=50\fbi$ can be optimally distributed over the
full $250-500\gev$ scan range in the course of a year or two of  
running, it would be necessary to design the storage ring or rings
so that it would be possible to adjust $\rts$ quickly and accurately 
(to within a small fraction of the step size, which must be
$\lsim 0.1\gev$ in some mass ranges) while maintaining the full luminosity.

\item[b)] Bremsstrahlung tail method\\
In this technique, 
the $\ha$ and $\hh$ search is made while running the $\mm$
collider at full energy, looking for excess events
arising from the luminosity at the
low-energy end of the bremsstrahlung tail.
This approach is competitive with
the scan technique if  the $b\anti b$ final state mass can be  
reconstructed with excellent resolution (roughly $\pm5$~GeV, 
including all detector effects and semi-leptonic $b$ decays).
The lower $\tanb$ limits for $5\sigma$ signals are about
one to two units higher than for the scan technique in the $\mha=250-480\gev$
range. Thus the bremsstrahlung search 
leaves a larger gap between the upper limit in $\tanb$
for which $\hh,\ha$ discovery would
be possible at the LHC ($\tanb\lsim 3-4$ at high $\mha$) and the lower
limit for which the $\hh,\ha$ would be detected at the $\mm$
collider ($\tanb\gsim 5-7$) than would the scan technique.

The bremsstrahlung technique has the advantage of not requiring
that high luminosity be maintained
over a broad range of $\rts$ collision energies while being able
to step quickly and accurately in $\rts$, but detector
costs associated with  the very demanding resolution 
in the $b\anti b$ invariant mass might be high. 

\item[c)] Pair production\\
It may well be possible to
build a $\mm$ collider with $\rts$ substantially
above $500\gev$.  If a $\rts\geq 1\tev$ machine
with high luminosity were built instead of a $500\gev$ collider,
it could discover the $\hh,\ha$ for $\mhh,\mha\geq 500\gev$
in the pair production mode.  
\end{itemize}

If the $\hh,\ha$ have already been discovered, either 
\begin{itemize}
\item with $\mhh,\mha\lsim 250\gev$
in the $\zstar\to\hh\ha$ mode at an $\ee$ collider, or
\item with $\mhh,\mha\lsim 2\tev$ in the $\zstar\to\hh\ha$
mode at a 4 TeV $\mm$ collider, or
\item
with $\mhh,\mha\lsim 500\gev$ at the LHC
(if $\tanb\lsim 3-4$ or $\tanb\gsim 8-20$),
\end{itemize} 
scanning over a broad energy range would not be necessary at the 
$\mm$ collider. By constructing
a single appropriate storage ring and devoting
full luminosity to accumulating events at $\rts\simeq \mha,\mhh$,
detailed studies of the total widths and partial widths of the $\ha$
and $\hh$ would then be possible at the $\mm$ collider
{\it for all $\tanb$ values above 1}.

\subsection{Summary of machine and detector requirements}
\indent

We re-emphasize the crucial machine and detector characteristics for
detection and study of both SM-like Higgs bosons
and non-SM-like Higgs bosons.
\begin{itemize}
\item
High luminosity, ${\cal L}\gsim 2\times 10^{33} {\rm cm}^{-2} {\rm
s}^{-1}$, is required at any $\rts$ where a Higgs boson is
known to exist and throughout any range of energy over which we
must scan to detect a Higgs boson. 
\item
A machine design such that beamstrahlung is small compared to
the effects of bremsstrahlung (included in our studies) is highly
desirable for scan searches and precision studies. However,
significant beamstrahlung might improve the ability
to discover Higgs bosons using the low-energy tail of the
luminosity spectrum.
\item
An extremely precise beam energy, $R\sim 0.01\%$, will
be needed for precision studies of a narrow-width SM-like Higgs boson.
Such precise resolution is also extremely helpful in the zeroing-in scan
for a very narrow SM-like and is not harmful for discovering a Higgs boson
with broad width. Precision measurements of the non-SM-like $\hh$
and $\ha$ widths and separation of these two resonances when
they overlap becomes difficult if $R$ is substantially larger
than $0.01\%$.
\item
To zero-in on $\rts\simeq\mh$ for a narrow-width SM-like Higgs boson 
requires being able to rapidly set $\rts$
with an accuracy that is small compared to the beam resolution $R$,
for $\rts$ values within about
a few GeV of the (approximately known) value of $\mh$.
To discover the $\hh$ and $\ha$ by scanning requires being able
to rapidly set $\rts$ with an accuracy that is small compared to
their widths over a $\rts$ interval of order several hundred GeV.
\item
To measure $\gamh$ for a SM-like $\h$ to $\pm 10\%$,
it must be possible to  set $\rts$ with an
accuracy of order 1 part in $10^6$ over $\rts$ values 
in an interval  several times $R\mh$, \ie\
over an interval of tens of MeV.
This (and the accuracy for the mass measurements) 
requires a machine design
that allows quick spin rotation measurements
of a polarized muon in the storage ring.
\item
If both muon beams can be polarized and the polarization ($P$) maintained  
through the cooling and acceleration process, the significance of the  
$s$-channel Higgs signal can be significantly enhanced provided
the factor by which the luminosity is decreased is 
less than $(1+P^2)/(1-P^2)$.
\item 
To detect non-SM-like Higgs bosons with enhanced $\mm$ couplings
in the bremsstrahlung luminosity tail when the machine is run
at full energy, one needs excellent mass resolution ($\sim\pm 5\gev$) in the 
$b\anti b$ final state mass as reconstructed in the detector.

\end{itemize}

%\section{Conclusion}
%\indent

In conclusion, if a Higgs bosons is discovered at the LHC
and/or an $\ee$ collider, construction of a
$\mm$ collider with $\rts$ covering the range of masses observed will become
almost mandatory purely on the basis of $s$-channel Higgs physics.
There are many other motivations for building a
$\mm$ collider, especially one with $\rts\gsim 2\tev$,
based on other types of new physics that could be probed.    
The physics motivations for a high-energy $\mm$ collider will
be treated elsewhere~\cite{mupmumreport}.

\bigskip\goodbreak
\begin{center}
{\bf\uppercase{Acknowledgments}}
\end{center}

We thank D.~Cline for initially stimulating our interest in the physics of
$\mu^+\mu^-$ colliders. We thank R.~Palmer for helpful information about
the expected machine characteristics.
We are grateful for conversations with and private communications from
T.~Barklow, S.~Dawson, K.~Fujii, H.~Haber, G.~Jackson, D.~Miller, D.~Neuffer, 
F.~Paige, Z.~Parsa,  M.~Peskin, R.J.N.~Phillips and B.~Willis.
This work was supported in part by the U.S.~Department of Energy  
under Grants No.~DE-FG02-95ER40896, No.~DE-FG03-91ER40674 and  
No.~DE-FG02-91ER40661. 
Further support was provided
by the University of Wisconsin Research
Committee, with funds granted by the Wisconsin Alumni Research  
Foundation, and by the Davis Institute for High Energy Physics.

\newpage
%\begin{center}
%{\bf\uppercase{Appendices}}
%\end{center}

\addcontentsline{toc}{section}{Appendices}

\appendix
\section{Effects of bremsstrahlung}

\indent\indent
Soft photon radiation is an
important effect that must be taken into account when
considering the ultimate resolution in $\rtshat$
(where $\shat=(p_{\mu^+}+p_{\mu^-})^2$ is the invariant energy
squared of a given collision)
and peak luminosity that can be
achieved at an $\ee$ or $\mm$ collider.  In an often discussed  
approximation
\cite{softphoton} the small Gaussian spread in $\rtshat$
about the nominal central machine energy, $\rts$,
resulting from purely machine effects is ignored, and the energy
spread resulting from soft photon radiation is computed starting
from $d{\cal L}/d\rtshat\propto \delta(\rtshat -\rts)$,
where $\rts$ is the nominal machine energy.
For many types of physics, this is an entirely adequate approximation  
since
the Gaussian spread is much smaller than the structure of the  
physical
cross section.  However, there are physical processes with $\rtshat$
structure that is much narrower than the expected Gaussian spread;
production of a Higgs boson with very narrow width is a case in  
point.
In this case, it is important to assess the distortion of the  
machine-level
Gaussian shape due to soft photon radiation.  Here we give the
necessary formalism and derive an extremely accurate approximation
to the exact result that is useful for numerical  
investigations.

We start from the basic machine-level Gaussian form:
\begin{equation}
{d{\cal L}_0\over d\rtshat}= { e^{-(\rtshat-\rts)^2/2
\sigma_{\tiny\rts}^2}\over
\sqrt{2\pi}\sigrts}\equiv G(\rtshat,\rts,\sigrts)
\label{gaussian}
\end{equation}
 and $\sigrts$
where is the resolution in $\rtshat$.
This form results from the convolution of two Gaussians
for the individual beams.  In general,
\begin{equation}
{d{\cal L}\over d\rtshat}=\int dE_1 dE_2 f(E_1)f(E_2)
\delta(\rtshat-\sqrt{4E_1E_2})
\label{meld}
\end{equation}
where the $f(E)$'s give the probability for finding an electron or  
muon
of energy $E$;  ${d{\cal L}_0\over d\rtshat}$
of Eq.~(\ref{gaussian}) results when the $f$'s are
standard Gaussian forms centered at $E_0\equiv\rts/2$ with  
resolutions
$\sigrtsprime$, $G(E_i,E_0,\sigrtsprime)$;
$\sigrts$ in Eq.~(\ref{gaussian})
is then given by $\sigrts=\sqrt 2 \sigrtsprime$.
However, $E_1$ and $E_2$ are degraded by soft photon radiation, so  
that
the actual probability for finding energy $E$ in any one beam is  
given by
\begin{equation}
f(E)=\int_0^1 {dz\over z} D(z) G(E/z,E_0,\sigrtsprime)
\label{convolution}
\end{equation}
where $D(z)$ is the probability
distribution for finding an electron or muon with fraction $z$
of its initial energy after the soft-photon radiation.
Substituting into Eq.~(\ref{meld}), we obtain
\begin{eqnarray}
{d{\cal L}\over d\rtshat} &=&\int dE_1 dE_2 \int {dz_1\over z_1}  
{dz_2\over
z_2}
D(z_1)D(z_2) G(E_1/z_1,E_0,\sigrtsprime) \nonumber \\
&  \times & G(E_2/z_2,E_0,\sigrtsprime)
\left[{1\over \sqrt{z_1z_2}}
\delta\left(\sqrt{\shat\over z_1z_2}-\sqrt{4{E_1\over z_1}{E_2\over
z_2}}\right)
\right]
\label{formi}
\end{eqnarray}
where we have rewritten $\delta(\rtshat-\sqrt{4E_1E_2})$ in a useful  
form. 
Changing integration variables to $E_1/z_1$
and $E_2/z_2$, we see immediately that the $E_i$ integrations
including the $\delta$ function reproduce $d{\cal L}_0/d \rtsprime$
evaluated at $\sprime=\shat/(1-x)$, where $1-x=z_1z_2$.
Introducing $\int dx \delta(1-x-z_1z_2)=1$ under the integral
in Eq.~(\ref{formi}) we obtain
\begin{equation}
{d{\cal L}\over d\rtshat}=\int {dx\over \sqrt{1-x}}
\left.{d{\cal L}_0\over d\rtsprime}\right|_{\sprime=
{\shat\over 1-x}} {\cal D}(x)
\label{formii}
\end{equation}
where
\begin{equation}
{\cal D}(x)=\int dz_1\,dz_2\, D(z_1)D(z_2) \delta(1-x-z_1z_2)\,.
\label{ddef}
\end{equation}
In the approximation of Ref.~\cite{softphoton}, ${\cal D}$
takes the form
\begin{equation}
{\cal D}(x)=C\left[\beta x^{\beta-1} \left(1+ {3\over 4}\beta\right)
-\beta \left(1-{x\over 2}\right)\right]\,,
\label{dform}
\end{equation}
where $C\equiv 1+{2\alpha\over \pi}(\pi^2/6-1/4)$ and
${\cal D}$ has an implicit energy dependence coming from
\begin{equation}
\beta\equiv {2\alpha\over\pi}(\log {\sprime\over m^2} -1)
\label{betadef}
\end{equation}
with $\sprime =\shat/(1-x)$; in what follows we replace $\sprime$ in this formula for $\beta$
by $\shat$, the error made in doing so being extremely small.
Typical values of $\beta$ for a $\mm$ collider are
$\beta\simeq0.0632$ at $\rts=100\gev$ and $0.0792$ at $\rts=500\gev$.

To compute a given cross section, one must fold
this result for the luminosity with the physical cross section:
\begin{equation}
\overline{\sigma}=\int d\rtshat {d{\cal L}\over d\rtshat}  
\sigma(\rtshat)
\,,
\label{sigbar}
\end{equation}
where the luminosity is that found from the convolution of  
Eq.~(\ref{formii}).
This is a numerically intensive operation in a number of cases of  
particular
interest.  Thus, it is useful to develop an analytic approximation to
$d{\cal L}/d\rtshat$ in Eq.~(\ref{formii}). This we have done by  
performing
an expansion. Writing $x\equiv\epsilon/(1+\epsilon)$, defining
\begin{equation}
\rtshat\equiv \mu+\rts, \quad a\equiv
{\sqrt 2\mu\over\sigrts}\sqrt{1+{\mu\over \rts}}, \quad
\rho\equiv {2\sqrt 2 \sigrts\over \rts\sqrt{1+{\mu\over\rts}}}\,,
\label{variousdefs}
\end{equation}
and changing variables to $y\equiv\epsilon/\rho$, the exponential in
the Gaussian $d{\cal L}_0/d\rtsprime$ in Eq.~(\ref{formii})
takes the form
\begin{equation}
exp[-\mu^2/(2\sigma_{\tiny\rts}^2)]
\times exp[-y^2-ay+\rho y^3/2+{\cal O}(y^4)]\,.
\label{exponential}
\end{equation}
The first term in Eq.~(\ref{exponential}) is the `standard' Gaussian
component. In the remaining part of the exponential,
the strongly convergent quadratic
component, $exp[-y^2]$, and the very small size of $\rho$
guarantee that $x$ takes on values of order $\rho$ in ${\cal D}(x)$
and allows a convenient expansion, including
$exp[\rho y^3/2]\simeq 1+\rho y^3/2$. The most important component
of the result derives from
$\rho^\beta \beta (1+{3\over 4}\beta)\int  
dy\,exp[-y^2-ay]\,y^{\beta-1}$
which is easily expressed in terms of degenerate hypergeometric  
functions,
$_1F_1$.  Keeping some other smaller terms as well, we find the  
result:
\begin{eqnarray}
{\frac{{d{\cal L}(\rtshat)\over d\rtshat}}{{d{\cal L}_0(\rtshat)\over  
d\rtshat
}}}  &  = &
  {C\over 2}\,\Biggl[ {{- \beta\,{e^{{{{a^2}}\over 4}}}\,
           {\sqrt{\pi }}\,\rho\,\left( 1 - {\rm Erf}({a\over 2})  
\right)
             }}
 +      {1\over 2}{{\left( 1 + {{3\,\beta}\over 4} \right) \,\beta\,
         {{\rho}^{1 + \beta}}\,}} \nonumber \\
& \times & \Biggl\{
         \left( \Gamma({{3 + \beta}\over 2})\,
            {_1F_1}({{3 + \beta}\over 2},{1\over 2},
             {{{a^2}}\over 4}) -
           a\,\Gamma({{4 + \beta}\over 2})\,
            {_1F_1}({{4 + \beta}\over 2},{3\over 2},
             {{{a^2}}\over 4}) \right) \nonumber \\
&  &  +     \left( \Gamma({{1 + \beta}\over 2})\,
            {_1F_1}({{1 + \beta}\over 2},{1\over 2},
             {{{a^2}}\over 4}) -
           a\,\Gamma({{2 + \beta}\over 2})\,
            {_1F_1}({{2 + \beta}\over 2},{3\over 2},
             {{{a^2}}\over 4}) \right) \Biggr\} \nonumber \\
& + &
     {{\left( 1 + {{3\,\beta}\over 4} \right) \,\beta\,
         {{\rho}^{\beta}}\,
         \left( \Gamma({{\beta}\over 2})\,
            {_1F_1}({{\beta}\over 2},{1\over 2},
             {{{a^2}}\over 4}) -
           a\,\Gamma({{1 + \beta}\over 2})\,
            {_1F_1}({{1 + \beta}\over 2},{3\over 2},
             {{{a^2}}\over 4}) \right) }}  \Biggr]\,. \nonumber \\
& &
\label{analyticapprox}
\end{eqnarray}
The numerically most important term outlined above appears last, the  
others
being quite small corrections thereto. From it we see
the crucial dependence on $\rho^\beta$. This factor decreases (albeit  
very
slowly because of the small size of $\beta$)  with increasingly small
$\sigrts$. This non-negligible loss of peak luminosity will be  
quantified
below.

The result of Eq.~(\ref{analyticapprox})
is extremely accurate for $|\mu/\sigrts|\leq 10$ and even for  
$\mu/\sigrts=-20$
deviates by only about 3\% from a precise numerical evaluation of
the integral. As will be illustrated, the effective
luminosity remains approximately Gaussian in shape aside
from a long low-energy tail.  The effective width at half maximum
of this approximately Gaussian peak is little altered, even for
$R\equiv\sigrtsprime/(\rts/2)$ values as small as $0.0001$ (desirable
to measure the precise mass and width of a very
narrow Higgs boson). As already noted, the most
important effect of the soft-photon radiation is to reduce
the effective peak luminosity height. The peak height ratio
is that obtained by setting $\mu=0$ (\ie\ $a=0$) in  
Eq.~(\ref{analyticapprox}).
The functional form simplifies significantly, and we obtain:
\begin{eqnarray}
\left.{\frac{{d{\cal L}(\rtshat)\over d\rtshat}}
{{d{\cal L}_0(\rtshat)\over d\rtshat}}}\right|_{\rtshat=\rts}
&  = & C\Biggl\{
\rhoo^\betao\Gamma\left(1+{\betao\over 2}\right)
\left(1+{3\over 4}\betao\right) -\betao\rhoo{\sqrt\pi\over 2}
\nonumber \\
& + & {\betao\over 4}\left(1+{3\over 4}\betao\right)\rhoo^{\betao+1}
\left[\Gamma\left( {\betao+3\over  
2}\right)+\Gamma\left({\betao+1\over
2}\right)\right]\Biggr\}\,,
\label{peakratio}
\end{eqnarray}
where $\rhoo$ and $\betao$ are the values of $\rho$ and $\beta$
at $\rtshat=\rts$. In Eq.~(\ref{peakratio}) the first term is
the most important one.

\begin{figure}[htbp]
\let\normalsize=\captsize   %%%% changes the font to "\small"
\begin{center}
\centerline{\psfig{file=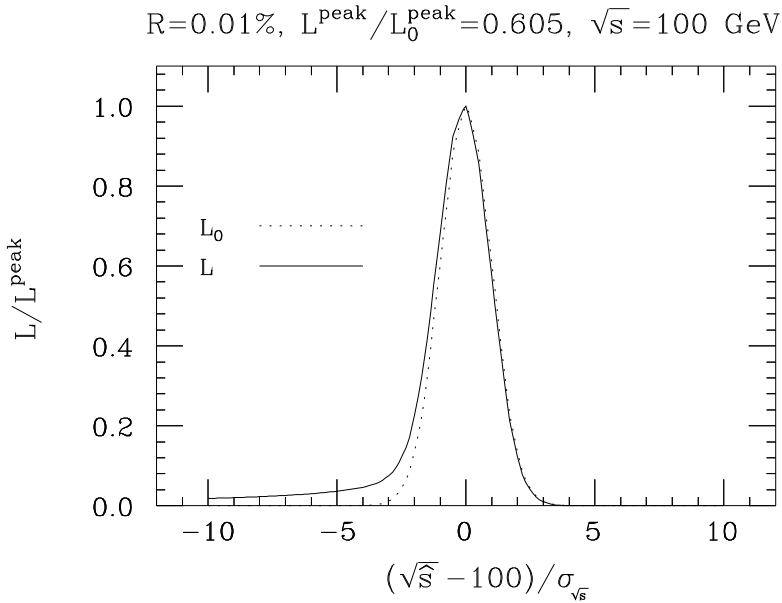,width=10.5cm}}
\begin{minipage}{12.5cm}       %%%% reduces width of caption to 12.5cm
\bigskip
\caption{{\baselineskip=0pt
$d{\cal L}/d\protect\rtshat$ relative to its peak value at
$\protect\rtshat=\protect\rts$
is plotted before and after soft-photon radiation.
We have taken $\protect\rts=100\gev$ and  $R=0.01\%$. The ratio
of peak height after including soft-photon radiation to
that before is 0.605.}}
\label{bremdistfig}
\end{minipage}
\end{center}
\end{figure}

We now illustrate the effects of the soft-photon radiation
for a $\mm$ collider. We first
present in Fig.~\ref{bremdistfig} a plot of $d{\cal L}/d\rtshat$
divided by its value at $\rtshat=\rts$ as
a function of $(\rtshat-\rts)/\sigrts$ in the case where the beam  
resolution
is 0.01\% (that is $\sigrtsprime=0.0001 \rts/2$), taking  
$\rts=100\gev$.
Results before (${\cal L}_0$) and after (${\cal L}$) including
soft-photon radiation are shown by
the dotted and solid curves, respectively.  The ratio of peak
heights in this case is 0.605, \ie\ roughly 40\% of the peak
luminosity is lost to soft-photon radiation. As promised,
the peak remains close to the original Gaussian shape within
$\pm 2\sigrts$ of the central peak, with a long tail extending
to low $\rtshat$ values.

\begin{figure}[htbp]
\let\normalsize=\captsize   %%%% changes the font to "\small"
\begin{center}
\centerline{\psfig{file=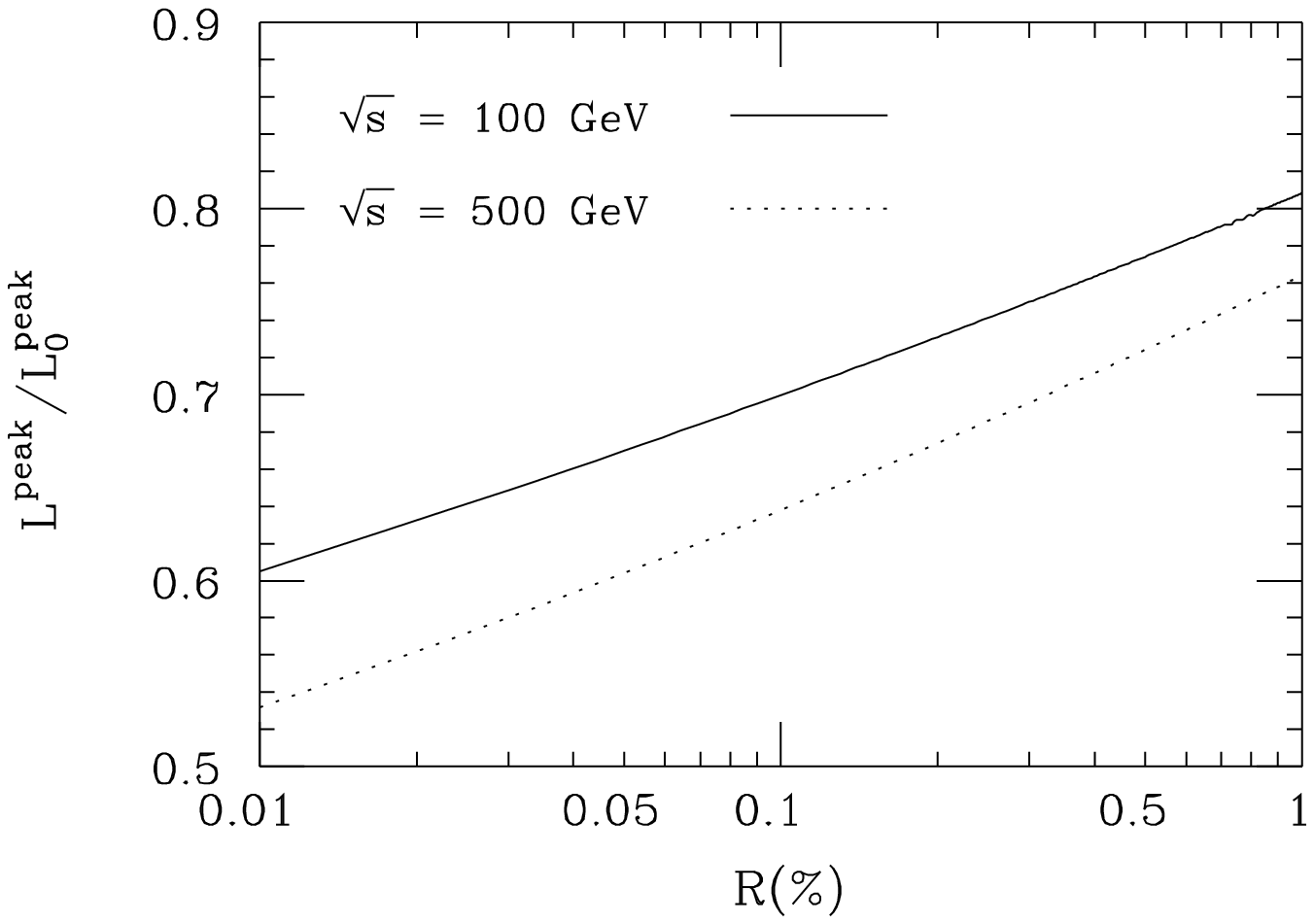,width=11.5cm}}
\begin{minipage}{12.5cm}       %%%% reduces width of caption to 12.5cm
\bigskip
\caption{{\baselineskip=0pt
$\left.{d{\cal L}\over d\protect\rtshat}/
{d{\cal L}_0\over  
d\protect\rtshat}\right|_{\protect\rtshat=\protect\rts}$
as a function of $R$ for $\protect\rts=100$ and 500 GeV.}}
\label{brempeakfig}
\end{minipage}
\end{center}
\end{figure}

To further quantify the loss of peak luminosity
that would be critical in searching for and studying a
narrow Higgs boson resonance, we consider
${d{\cal L}\over d \rtshat}/
{d{\cal L}_0\over d\rtshat}|_{\protect\rtshat=\rts}$ as a function
of $\rts$ and $R$ in Fig.~\ref{brempeakfig}.
 From Fig.~\ref{brempeakfig} one sees that the loss of peak  
luminosity
decreases as the beam resolution becomes poorer, but increases
as $\rts$ increases (due to the increase of $\beta$ with increasing  
$\rtshat$,
see Eq.~(\ref{betadef}) and below).

The results presented in the figures were obtained by direct  
numerical
integration.  However, the results from the approximate formulas,
Eqs.~(\ref{analyticapprox}) and (\ref{peakratio}), are essentially
indistinguishable from those presented.

\begin{figure}[htbp]
\let\normalsize=\captsize   %%%% changes the font to "\small"
\begin{center}
\centerline{\psfig{file=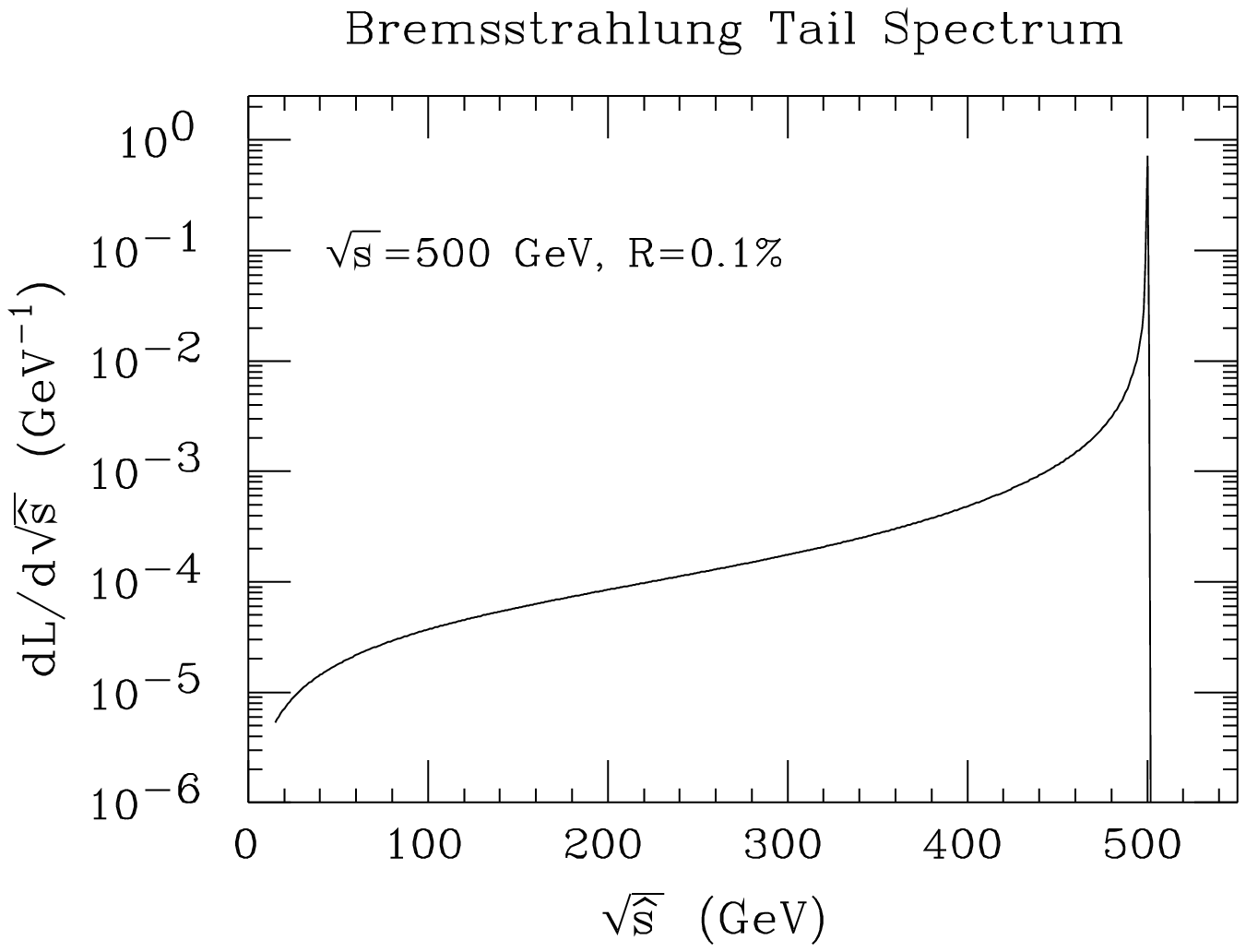,width=11.5cm}}
\begin{minipage}{12.5cm}       %%%% reduces width of caption to 12.5cm
\bigskip
\caption{{\baselineskip=0pt
${d{\cal L}\over d\protect\rtshat}$ as a function of  
$\protect\rtshat$
for $R=0.1\%$ and $\protect\rts=500\gev$. The integral under the  
curve
is normalized to 1.}}
\label{bremtail}
\end{minipage}
\end{center}
\end{figure}

Finally, we present in Fig.~\ref{bremtail}
the full bremsstrahlung tail distribution $d{\cal L}/d\rtshat$
for the case of $R=0.1\%$ and $\rts=500\gev$. The integral
$\int d\rtshat [d{\cal L}/d\rtshat]$ is normalized to 1.
The low-$\rtshat$ tail is quite independent of $R$.  Only near
the Gaussian peak region is there significant dependence of the
spectrum on $R$. It is this spectrum that we have employed
in discussing detection of Higgs bosons with enhanced $\mm$
coupling using events from the bremsstrahlung tail when
the $\mm$ collider is run at full energy of $\rts=500\gev$.

\section{The {\protect\boldmath$\mu^+\mu^- \to h \to W\wstarp,Z\zstarp$}
modes}
\indent

Whether we are above or below $2W$ or $2Z$ threshold, we include
in our event rates an efficiency of 50\% (after branching
fractions) for isolating and reconstucting the final states
of interest. Further, we search
for the optimum choice of $z_0$ such that (in the center of
mass) events with $|\cos\theta|>z_0$ are discarded.
This is primarily useful for discarding the large number of
forward/backward $W$'s or $Z$'s from the continuum backgrounds
for $\mh$ significantly larger than $2\mw$ or $2\mz$.

\bigskip\goodbreak
\noindent
\underline{\bf (a). {\protect\boldmath$\mh < 2\mw$} }

We consider only those final states where the mass of the real $W$ or $Z$
can be reconstructed, thereby excluding $W\wstar\to 2\ell2\nu$ and
$Z\zstar\to 4\nu$. For Higgs masses below $2\mw$ or $2\mz$, 
we cannot use a mass
constraint on the virtual boson to help isolate the $WW^\star$ or  
$ZZ^\star$ final state. Consequently, 
the pure QCD background to the $4j$ final state
is very substantial, and a significant 4-jet Higgs signal is very  
difficult to obtain in this region. Thus,  only the
mixed hadronic/leptonic modes ($\ell\nu 2j$ for $WW^\star$ and $2\ell 2j$
and $2\nu2j$ for $ZZ^\star$) and the visible purely leptonic  
$ZZ^\star$ modes ($4\ell$ and $2\ell2\nu$) can be employed.
The effective branching fractions ($\br$) for these final states are
\begin{equation}
\brwweff\equiv \br(\wp\wm\rta \ell\nu 2j)=2 (2/9) (2/3) \sim 0.3
\label{bfwweff}
\end{equation}
and
\begin{equation}
\begin{array} {lc}
\brzzeff&\equiv  \br(ZZ\rta 4\ell+2\ell2j+2\nu2j+2\ell2\nu) \hfill\\
\phantom{\brzzeff}&= (0.067)^2+2[(0.067)(0.699)+(0.2)(0.699)+
(0.067)(0.2)] \hfill\\
\phantom{\brzzeff}&\sim  0.42\;.\hfill
\label{bfzzeff}
\end{array}
\end{equation}

\begin{figure}[htbp]
\let\normalsize=\captsize   %%%% changes the font to "\small"
\begin{center}
\centerline{\psfig{file=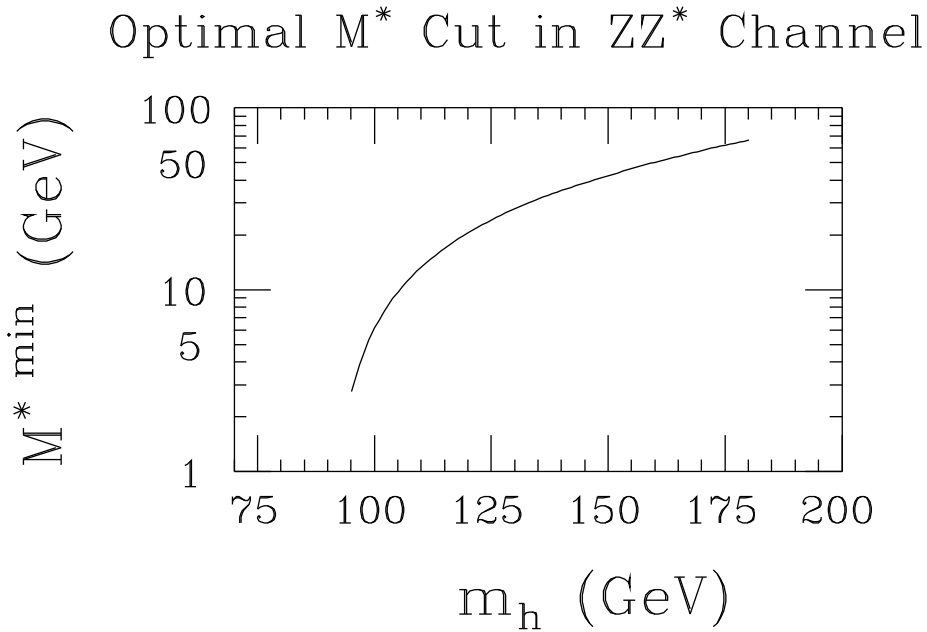,width=12.2cm}}
\begin{minipage}{12.5cm}       %%%% reduces width of caption to 12.5cm
\caption{The optimal choice for $\mstarmin$
in the $Z2j$, $Z2\ell$ and $Z2\nu$ final states is given
as a function of $\mh$ for a SM-like $\h$.}
\label{mstarcut}
\end{minipage}
\end{center}
\end{figure}

For virtual weak boson decays to leptons, the only backgrounds derive from the
$W\ell\nu$, $Z2\ell$ and $Z2\nu$ processes.
For virtual weak boson decays to jets,
the background processes are $W2j$ and $Z2j$.
In the $\wstar$ cases, the contribution to the cross
section from small $\ell\nu$ or $2j$ masses is well-behaved and small.
Our procedure is to accept events from both the Higgs and background
regardless of the virtual mass.
For the $Z2\ell$ and $Z2j$ processes,  the $2\ell$ or $2j$ can
arise from a $\zstar$ {\it or} $\gamma^\star$.  The virtual photon  
exchange causes a strong singularity and cross section growth for small  
$2\ell$ and $2j$ virtual masses.  In the $Z2\nu$ channel, especially for $\mh$
near $\mz$, there are likely to be significant backgrounds
associated with limited detector acceptance (\eg\ a three-jet
event could produce two jets with mass near $\mz$ and the third
jet could be soft or disappear down the beam pipe) and detector
fluctuations and uncertainties.  Hence, it is safer to
require a minimum virtuality on the virtual $2\nu$.
We have adopted the procedure of imposing a uniform cutoff,  
$\mstarmin$, on the invariant mass of the virtual $\zstar$ in all channels
--- a search was performed to determine
the optimal choice for $\mstarmin$ that maximizes the statistical
significance of the Higgs signal.  The resulting values
for $\mstarmin$ as a function of Higgs mass are presented in
Fig.~\ref{mstarcut}. For the plotted
$\mstarmin$ values one retains about 40\% to 50\% of the signal
at the lower values of $\mh$, rising to 85\% by $\mh=175\gev$.

\bigskip\goodbreak
\noindent
\underline{\bf (b). {\protect\boldmath$\mh > 2\mz$} }

Above $2W$ and $2Z$ thresholds, the only case for which mass  
reconstruction of both $W$'s ($Z$'s) is not possible is the 
$2\ell2\nu$ ($4\nu$)  final state.
In all other final states Higgs mass reconstruction is possible
from the final state particles (\eg\ in the four-jet final states or
the $2\ell2j$ and $4\ell$ $ZZ$ final states) or from the extremely
precise determination of the incoming total momentum plus a reconstruction
of one of the $W$'s or $Z$'s to determine the other $W$ or $Z$ (\eg\
in the $\ell\nu 2j$ $WW$ final state and the $2\nu 2j$ and $2\nu  
2\ell$ $ZZ$ final states). Since both $W$'s or $Z$'s can
be reconstructed above $2W$ or $2Z$ threshold, we can
employ the $4j$ modes in addition to those listed in Eqs.~(\ref{bfwweff})
and (\ref{bfzzeff}); the pure QCD $4j$ background can be eliminated by
requiring two equal mass pairs (within $\pm 5\gev$ of mass $\mw$ or $\mz$).
The only significant background
is then that from true continuum $WW$ and $ZZ$ production.
%
%As noted earlier, cuts on $p_T^i$ (for all $i$) and
%on $\Delta R_{ij}$ and $M_{ij}$ (for all $i\neq j$) must be imposed
%to avoid standard infrared and collinear singularities. For  $p_T^{\rm
%min}=5\gev$, $\Delta R^{\rm min}=0.04$ and $M_{\rm pair}^{\rm  min}=20\gev$
%we find: a) that the efficiency for retaining the irreducible
%continuum background and Higgs signal in the $WW$ and $ZZ$ channels
%exceeds 90\%; and b) that the QCD $4j$ background is reduced to a  level
%much smaller than that of the irreducible backgrounds.
%
Noting that we have already
included a general 50\% efficiency factor for cuts and  
reconstruction, a safe estimate in the above-threshold regions for
the luminosity required for $5\sigma$ observation in the $4j$  
channels can be obtained from the $WW$ and $ZZ$ curves of Fig.~\ref{smrates}
by dividing by $[\br(W\rta jj)^2/\brwweff]^{1/2}\approx 1.2$
and $[\br(Z\rta jj)^2/\brzzeff]^{1/2}\approx 1.1$, respectively.

%In the above-threshold region, the only significant background  (after
%simple and highly efficient cuts that require two equal
%mass systems with mass near $\mw$ or $\mz$, in the $WW$ and $ZZ$
%cases, respectively)
%is that from true continuum $WW$ or $ZZ$ pair production.

\section{Three-point determination of
{\protect\boldmath$\mhsm$} and {\protect\boldmath$\gamhsm$}}

\indent\indent
The procedure is as follows.
We perform three measurements.
At $\rts_1=\mhsm+\sigrts d$ (where $d$
is not known ahead of time, and will be determined
by the procedure) we employ a luminosity $L_1$ and
measure the total rate $N_1=S_1+B_1$. Next, we perform measurements
at $\rts_2=\rts_1-\nsigrts\sigrts$ and
$\rts_3=\rts_1+\nsigrts\sigrts$, yielding $N_2=S_2+B_2$ and  
$N_3=S_3+B_3$
events, respectively, employing luminosities of
$L_2=\rho_2L_1$ and $L_3=\rho_3L_1$, with $\rho_{2,3}>1$ ---
$\nsigrts\sim 2$ and $\rho_2=\rho_3\sim 2.5$
are good choices for maximizing sensitivity and minimizing the
error in determining $d$ (\ie\ $\mhsm$) and $\gamhsm$.
We then define the ratios
$r_2\equiv (S_2/\rho_2)/S_1$ and $r_3\equiv (S_3/\rho_3)/S_1$.
Obviously, the ratios $r_2$ and $r_3$ are determined by $d$ (\ie\
by $\mhsm$) and $\gamhsm$:
$r_i=r_i(d,\gamhsm)$. Conversely, we have implicitly
$d=d(r_2,r_3)$ and $\gamhsm=\gamhsm(r_2,r_3)$. Determining the  
statistical
errors $\Delta\mhsm$ and $\Delta\gamhsm$ is then simply a matter of
computing the partial derivatives of $d$ and $\gamhsm$
with respect to the $r_{2,3}$
and the errors on the ratios $r_{2,3}$ implied by statistics.

Assuming precise knowledge of the background level
$B\equiv B_1=B_2/\rho_2=B_3/\rho_3$,
the experimental error for either of the ratios is given by
\begin{equation}
\Delta r_i={\sqrt{r_i}\over \sqrt{\rho_i} \sqrt{S_1}}\times
\sqrt{(1+r_i\rho_i)+B/S_1(1/r_i+\rho_i r_i)}\,.
\label{dris}
\end{equation}
 The errors in the
experimental determination of $d$ and $\gamhsm$ are given
by quadrature:
\begin{eqnarray}
\Delta d & = & \left[({\partial d\over\partial r_2})^2(\Delta r_2)^2
         +({\partial d\over\partial r_3})^2(\Delta  
r_3)^2\right]^{1/2} \\
\Delta\gamhsm &=&\left[({\partial \gamhsm\over\partial r_2})^2(\Delta  
r_2)^2
         +({\partial \gamhsm\over\partial r_3})^2(\Delta  
r_3)^2\right]^{1/2}
\label{mgamerrors}
\end{eqnarray}
In practice, we compute the above partial derivatives
by first computing
\begin{equation}
{\cal M}\equiv \pmatrix{{\partial r_2\over\partial d} & {\partial
r_2\over\partial\gamhsm} \cr
                        {\partial r_3\over\partial d} & {\partial
r_3\over\partial\gamhsm} \cr }
\label{matrixdef}
\end{equation}
and then inverting; \eg\ ${\partial d \over \partial r_2}=({\cal
M}^{-1})_{11}$.  This is performed numerically, and the $\Delta  
r_i$'s
from Eq.~(\ref{dris}) are then inserted in Eq.~(\ref{mgamerrors}).

The above procedure is convenient in order to determine the  
luminosity
required for $\Delta\gamhsm/\gamhsm=1/3$ as a function of $\mhsm$
for several possible choices of beam resolution.
We have explicitly verified the accuracy and correctness
of the procedure in several specific cases as follows.
For a given $\mhsm$ and $\gamhsm$ we compute the event numbers  
$N_{1,2,3}$
and thence the ratios $r_i$ ($i=2,3$) and their statistical errors.
We then vary $\mhsm$ and $\gamhsm$ relative to their original values
and determine $\Delta\mhsm$ and $\Delta\gamhsm$ as those shifts
which result in $\Delta \chi^2=1$.  Good agreement with
the results from the above procedure is obtained.

\end{document}